\documentclass[preprint,prd,superscriptaddress,tightenlines,nofootinbib,eqsecnum]{revtex4-1}

\usepackage{amsmath}
\usepackage{amsfonts}
\usepackage{amssymb}
\usepackage{bm}
\usepackage{color}
\usepackage[colorlinks,pdfborder={0 0 0},plainpages=false]{hyperref}
\usepackage[format=plain,justification=centerlast,size=small]{caption}
\usepackage{subcaption} 
\usepackage{graphicx}

\definecolor{CiteColor}{rgb}{0,0.5,0}
\hypersetup{citecolor=CiteColor}
\definecolor{RefColor}{rgb}{0.55,0,0}
\hypersetup{linkcolor=RefColor}

\newcommand{\beq}{\begin{equation}}
\newcommand{\eeq}{\end{equation}}
\newcommand{\be}{\begin{equation}}
\newcommand{\ee}{\end{equation}}
\newcommand{\ba}{\begin{eqnarray}}
\newcommand{\ea}{\end{eqnarray}}
\newcommand{\ud}{\mathrm{d}}

\newcommand{\calO}{\mathcal{O}}
\newcommand{\calP}{\mathcal{P}}
\newcommand{\calQ}{\mathcal{Q}}
\newcommand{\calU}{\mathcal{U}}
\newcommand{\calV}{\mathcal{V}}
\renewcommand{\c}{\,,}
\newcommand{\p}{\,.}
\newcommand{\f}{\frac}
\newcommand{\Tau}{\mathcal{T}}
\newcommand{\mrm}{\mathrm}
\renewcommand{\max}{\text{max}}
\renewcommand{\min}{\text{min}}
\newcommand{\Omr}{{\Omega}_r}
\newcommand{\Omph}{{\Omega}_\phi}
\newcommand{\en}{\mathcal{E}}
\newcommand{\ang}{\mathcal{L}}

\allowdisplaybreaks

\begin{document}

\title{Comparison Between Self-Force and Post-Newtonian Dynamics: Beyond Circular Orbits}

\author{Sarp Akcay}
\affiliation{School of Mathematical Sciences and Complex \& Adaptive Systems Laboratory, University College Dublin, Belfield, Dublin 4, Ireland}
\author{Alexandre Le Tiec}
\affiliation{Laboratoire Univers et Th\'eories, Observatoire de Paris, CNRS, Universit\'e Paris Diderot, 92190 Meudon, France}
\author{Leor Barack}
\affiliation{School of Mathematics, University of Southampton, Southampton, SO17 1BJ, United Kingdom}
\author{Norichika Sago}
\affiliation{Faculty of Arts and Science, Kyushu University, Fukuoka 819-0395, Japan}
\author{Niels Warburton}
\affiliation{MIT Kavli Institute for Astrophysics and Space Research, Massachusetts Institute of Technology, Cambridge, MA 02139, USA}

\begin{abstract}
The gravitational self-force (GSF) and post-Newtonian (PN) schemes are complementary approximation methods for modelling the dynamics of compact binary systems. Comparison of their results in an overlapping domain of validity provides a crucial test for both methods, and can be used to enhance their accuracy, e.g.\ via the determination of previously unknown PN parameters. Here, for the first time, we extend such comparisons to noncircular orbits---specifically, to a system of two nonspinning objects in a bound (eccentric) orbit. To enable the comparison we use a certain orbital-averaged quantity $\langle U \rangle $ that generalizes Detweiler's redshift invariant. The functional relationship $\langle U \rangle(\Omega_r,\Omega_\phi)$, where $\Omega_r$ and $\Omega_\phi$ are the frequencies of the radial and azimuthal motions, is an invariant characteristic of the conservative dynamics. We compute $\langle U \rangle(\Omega_r,\Omega_\phi)$ numerically through linear order in the mass ratio $q$, using a GSF code which is based on a frequency-domain treatment of the linearized Einstein equations in the Lorenz gauge. We also derive $\langle U \rangle(\Omega_r,\Omega_\phi)$ analytically through 3PN order, for an arbitrary $q$, using the known near-zone 3PN metric and the generalized quasi-Keplerian representation of the motion. We demonstrate that the $\mathcal{O}(q)$ piece of the analytical PN prediction is perfectly consistent with the numerical GSF results, and we use the latter to estimate yet unknown pieces of the 4PN expression at $\mathcal{O}(q)$.
\end{abstract}

\date{\today}

\maketitle

\section{Introduction}

With the Advanced LIGO observatories scheduled to start science runs in 2015 \cite{Aa.al.13}, the next few years are likely to see first direct detections of gravitational waves from astrophysical sources. Prime targets are inspiralling and coalescing  binary systems of neutron stars and/or black holes, with predicted rates that may be as high as a few dozen per observation year \cite{Ab.al.10}. Theoretical templates of the gravitational waveforms must be developed to enable detection and interpretation of the weak signals \cite{FiCh.93}. The parameter space of these waveforms is too large for numerical relativity simulations to cover sufficiently well. Instead, the community has been seeking semi-analytical models that can be informed by a judiciously chosen set of numerical relativity templates. A leading framework is the effective one-body (EOB) model, where the two-body relativistic dynamics is mapped onto a model of (non)geodesic motion in an effective spacetime \cite{BuDa.99,BuDa.00,Ta.al.14,DaNa.14}. EOB waveforms will play a crucial role in searches based on matched filtering, and there is an important need to refine the model, particularly in the strong-field regime \cite{BiDa2.14,BiDa3.14,BiDa4.14}.

One avenue of refinement is provided by the gravitational self-force (GSF) method, a perturbative scheme based on an expansion in the mass ratio of the binary \cite{Ba.09,Po.al.11,Th.11}. The GSF approach is complementary to the post-Newtonian (PN) approximation, a weak-field/small-velocity expansion valid for arbitrary mass ratios \cite{Bl.14}. Recently, there has been much activity in attempt to ``synergize'' the two schemes. The goal of such cross-cultural studies is three-fold: to test the two independent approximation schemes---GSF and PN---and help delineate their respective domains of validity; to determine yet-unknown high-order expansion terms in both approaches (hence improving both approximations); and to help calibrate the EOB model across the entire inspiral parameter space.

To facilitate such studies requires the identification of concrete gauge-invariant physical quantities that can be computed using both approaches. A first such quantity was identified by Detweiler in 2008 within the GSF framework \cite{De.08}: the so-called ``redshift'' variable, defined for strictly circular orbits when dissipation is ignored. The functional relation between the redshift and the orbital frequency is a gauge-invariant diagnostic of the conservative sector of the binary dynamics. Detweiler made the first successful comparison with the PN prediction at 2PN order \cite{De.08}. This comparison was later extended by Blanchet \textit{et al.} to 3PN order and to even higher orders \cite{Bl.al.10,Bl.al2.10,Sh.al.14,Bl.al.14,Bl.al2.14}. The calculation of the redshift through linear order in the mass ratio was subsequently confirmed by several other GSF computations in different gauges \cite{Sa.al.08,Sh.al.11}, which provided an internal consistency check for the GSF formalism.

Soon after, Barack and Sago considered two more such ``conservative'' invariant quantities, namely the frequency of the innermost stable circular orbit (ISCO), and the rate of periastron advance \cite{BaSa.10,BaSa.11}. These results led to a plethora of comparisons between PN, GSF and numerical relativity \cite{Fa.11,Fa2.11,Le.al.11,Le.al2.12,Le.al.13,Le2.14} and the subsequent refinement of EOB theory \cite{Da.10,Ba.al.10,Ba.al.12,Ak.al.12}. More recently, the geodetic spin precession along circular orbits was computed by Dolan {\it et al.} through linear order in the mass ratio, and the numerical results successfully compared to a 3PN-accurate prediction \cite{Do.al.14}. The results allowed a numerical prediction of the (hitherto unknown) 4PN expression for the spin precession. This was later confirmed analytically by Bini and Damour \cite{BiDa3.14}, who also proceeded to obtain all PN terms up to the 8.5PN order, at linear order in the mass ratio. Dolan {\it et al.} \cite{Do.al.15} then presented a computation of the leading post-geodesic corrections to certain tidal invariants defined along the orbit. The PN series for these tidal invariants were also computed analytically up to 7.5PN order in Ref.~\cite{BiDa4.14}, still at linear order in the mass ratio.

All synergistic work so far has focused on {\em circular} orbits, for simplicity. Here, for the first time, we extend this program to orbits of arbitrary eccentricity. There are several reasons to do so. First, eccentricity provides more ``handle'' on the strong-field dynamics, giving access to new degrees of freedom in the EOB formulation. Second, although most Advanced LIGO binaries would have completely circularized by the time they enter the observable frequency band, there are scenarios where eccentricity effects could become observable and would give access to much interesting physics \cite{We.03,OL.al.09,Th2.11,AnPe.12,KoLe.12,Sa.al.14}. Third, eccentric inspirals in the extreme-mass-ratio regime will be key sources for a future mHz-band detector in space \cite{GaPo.13,Am.al.13,Am.al2.13,Am.al.14,Ba.al.14}.

A gauge-invariant quantity for eccentric orbits, suitable for synergistic studies, was introduced by Barack and Sago in Ref.~\cite{BaSa.11} (henceforth BS2011). This quantity is a straightforward generalization of Detweiler's redshift, obtained by averaging the time component of the particle's four-velocity with respect to proper time over one epicyclic period of the motion. In other words, it is the ratio between the period measured in the coordinate time of a static observer at infinity and the proper-time period. This ``averaged redshift,'' denoted here $\langle U \rangle $, is defined with the dissipative piece of the GSF ignored. The functional relationship between  $\langle U \rangle $ and the two invariant frequencies that characterize the motion is a gauge-invariant diagnostic of the conservative eccentric-orbit dynamics. BS2011 calculated $\langle U \rangle $ (numerically) through linear order in the mass ratio for a sample of strong-field orbits, but they stopped short of attempting a calculation in a weaker-field regime where a meaningful comparison with PN results might be possible. The method of BS2011, which is based on a time-domain numerical integration of the relevant field equations, was best suited for tackling strong-field orbits, and its performance deteriorated fast with increasing orbital radius because of the longer evolution time required. 

Here we extend the range of BS2011's calculation into the weaker-field regime, derive a 3PN-accurate formula for $\langle U \rangle $, valid for any mass ratio, and compare between the numerical GSF results and the analytical PN prediction in the small mass-ratio limit. This is the first such comparison for noncircular orbits. It shows a good agreement for large and medium separations, and allows us to assess the performance of the PN expansion all the way down to the innermost stable orbit. Moreover, we are also able, through fits to the numerical GSF data, to extract some information about the 4PN approximation.   

Our numerical GSF calculation improves on that of BS2011 in both accuracy and weak-field reach. This improvement is achieved in two ways. First, our computation is based on the {\it frequency}-domain approach of Akcay {\it et al.} \cite{Ak.al.13}, in which the field equations are reduced to ordinary differential equations. This offers significant computational saving, particularly at lower eccentricities ($e \lesssim 0.4$). Second, we have found a way to significantly simplify the expression given in BS2011 for $\langle U \rangle $ as a function of the two orbital frequencies. The new form requires a simpler type of numerical input, which can be obtained at greater accuracy. 

This paper is organized as follows. In Sec.~\ref{sec:GSF_Delta_U} we review relevant results for bound motion in Schwarzschild spacetime and for the redshift as defined for circular orbits. We then extend the definition to eccentric orbits and obtain a simple expression for the generalized redshift $\langle U \rangle $ in terms of calculable perturbative quantities. Section \ref{sec:num_results} discusses the numerics and sources of error, and displays a sample of numerical results for $\langle U \rangle $. In Sec.~\ref{sec:PN_Delta_U} we perform a detailed derivation of the PN expression for $\langle U \rangle $ through 3PN order. Our calculations rely crucially on the known 3PN near-zone metric and the 3PN quasi-Keplerian representation of the motion. The numerical GSF and analytical PN results are compared in Sec.~\ref{sec:comparison}. In Appendix \ref{app:proof} we establish the equivalence between our simplified formulation of $\langle U \rangle $ and that of BS2011. Appendix~\ref{app:test-mass} derives some useful PN formulas valid in the test-mass limit.

Table~\ref{table:notation} summarizes some of our notation, for easy reference. In the GSF context, we denote the mass of the background Schwarzschild geometry by $m_2$ and the mass of the orbiting particle by $m_1$, with the assumption that $q \equiv m_1/m_2\ll 1$. In the PN context, the two particles of masses $m_1$ and $m_2$ have an arbitrary mass ratio $q$. We will set $G=c=1$, except in Sec.~\ref{sec:PN_Delta_U} where we keep these constants explicit in PN expressions. We use a metric signature $(-,+,+,+)$. 

\begin{table}[h]
	\centering
	\begin{tabular}{lcl}
		\toprule            
		$m_1$					& & particle's mass \\
		$m_2$					& & black hole's mass \\
		$m = m_1 + m_2$			& & total mass \\
		$q = m_1/m_2$			& & mass ratio \\
		$\nu = m_1 m_2/m^2$		& & symmetric mass ratio \\
		$\Delta = (m_2-m_1)/m$	& & reduced mass difference \\
		$\Omr$					& & radial (epicyclic) frequency \\
		$\Omph$					& & average azimuthal frequency \\
		\botrule
	\end{tabular}
	\caption{Important symbols.}
	\label{table:notation}
\end{table}
%

\section{Generalized redshift: Formulation in self-force approach}\label{sec:GSF_Delta_U}

\subsection{Bound geodesic orbits in Schwarzschild spacetime}\label{sec:Schwarzschild_geodesics}

We first review relevant results for bound geodesic motion in Schwarzschild spacetime. We consider a test particle of mass $m_1$ moving on a bound (timelike) geodesic orbit in the Schwarzschild spacetime of a black hole of mass $m_2$. Using Schwarzschild coordinates $\{ t,r,\theta,\phi\}$, we label the position of the particle by $x_{\rm p}^\alpha (\tau_0) = \left( t_{\rm p}(\tau_0), r_{\rm p}(\tau_0), \theta_{\rm p}(\tau_0), \phi_{\rm p}(\tau_0) \right)$, with four-velocity $u^{\alpha}_0 \equiv \ud x^\alpha_{\rm p} / \ud \tau_0$,
where $\tau_0$ is a proper-time parameter along the geodesic, and the label `0' indicates normalization with respect to the background (Schwarzschild) metric $ g^0_{\alpha\beta}$, i.e.,  $ g^0_{\alpha\beta} u^{\alpha}_0 u^\beta_0 = -1 $. Without loss of generality, we confine the motion to lie in the equatorial plane, i.e., $\theta_{\rm p} = \pi/2 $, such that $ u^\theta_0 = 0$. We parameterize the geodesics by the two constants of motion: the specific energy $\en \equiv -u_{0t} $ and specific angular momentum $ \ang\equiv u_{0\phi} $, where $ u_{0\alpha} = g^0_{\alpha \beta} u^\beta_0 $.

The geodesic equation of motion is given by $u_0^\beta \nabla_{0\beta} \, u_0^\alpha = 0$, where $\nabla_{0\beta}$ is the covariant derivative compatible with the background metric $ g^0_{\alpha\beta}$. For the above setup, this gives
\begin{subequations}
	\begin{align}
		\frac{\ud t_{\rm p}}{\ud \tau_0} &= \f{\en}{f(r_{\rm p})} \c \label{eq:dt0_dtau0} \\
		\frac{\ud \phi_{\rm p}}{\ud \tau_0} &= \f{\ang}{r_{\rm p}^2} \c \label{eq:dphi0_dtau0} \\
		\left( \frac{\ud r_{\rm p}}{\ud \tau_0} \right)^2 &= \en^2 - V_{\mrm{eff}}(r_{\rm p}; \ang^2 ) \c \label{eq:dr0_dtau0}
	\end{align}
\end{subequations}
where $ f(r) \equiv 1-2m_2/r $, and $V_{\mrm{eff}}(r; \ang^2 ) \equiv f(r) \left(1 + \ang^2 /  r^2 \right)$ is an effective potential for the radial motion. Bound (eccentric) geodesics exist for $2\sqrt{2}/3 < \en < 1$ and $\ang > 2 \sqrt{3} \, m_2$. For a given bound geodesic, the radial distance $r_{\rm p}(\tau_0)$ is confined to a finite range $ 2m_2<r_{\mrm{min}} \le r_{\rm p}(\tau_0) \le r_{\mrm{max}}<\infty $ with $ r_{\mrm{min}}, r_{\mrm{max}} $ denoting periastron and apastron radii, respectively. These two turning-point radii can be mapped bijectively to $\{\en, \ang\}$. Thus the pair $ \{r_{\mrm{min}}, r_{\mrm{max}}\} $ can also parameterize the family of bound geodesics.

Another such pair is given by the dimensionless ``semi-latus rectum'' $p$ and the ``eccentricity'' $e$, defined by
\be\label{eq:p0_e0}
p \equiv \f{2 \, r_\max \, r_\min}{m_2 \, (r_\max + r_\min )} \c \qquad e \equiv \f{r_\max - r_\min}{r_\max + r_\min } \p
\ee
These relations can be inverted to yield the Keplerian-like formulas
\be\label{eq:r_min_max}
r_\max = \f{p \, m_2}{1-e} \c \qquad r_\min = \f{ p \, m_2}{1+e} \p
\ee
We can further express the specific energy and angular momentum in terms of $p$ and $e$ by solving the equation $ \en^2 = V_{\mrm{eff}}(r;\ang^2) $ at $r=\{r_\min,r_\max\}$. Using Eqs.~(\ref{eq:r_min_max}), this yields
\be\label{eq:E0_L0}
\en = \left[\f{(p -2- 2e)(p-2+2e)}{p (p - 3- e^2)}\right]^{1/2} , \qquad \ang = \f{p \, m_2}{\sqrt{p - 3- e^2}} \p
\ee

Following Darwin \cite{Da.61}, we parameterize the radial motion using the ``relativistic anomaly'' $\chi$ via
\be\label{eq:r_of_chi}
	r_{\rm p}(\chi) = \frac{p \, m_2}{1+e \cos\chi} \c
\ee
where $\chi=0$ and $\chi=\pi$ correspond to periastron and apastron passages, respectively. Using Eq.~\eqref{eq:dr0_dtau0} with Eq.~\eqref{eq:r_of_chi}, we obtain 
\be\label{eq:dtau0_dchi}
 \frac{\ud \tau_0}{\ud \chi} = \frac{m_2 \, p^{3/2}}{{(1+e\cos\chi)}^2} \, \sqrt{\frac{p-3-e^2}{p-6-2e\cos\chi}} \c
\ee
which, with the help of Eqs.~(\ref{eq:dt0_dtau0}), (\ref{eq:dphi0_dtau0}) and (\ref{eq:E0_L0}), also gives
\begin{subequations}\label{eq:dt0-dphi0}
\ba
\f{\ud t_{\rm p}}{\ud \chi} & = & \f{ m_2 \, p^2 }{(p - 2- 2e \cos{\chi}){(1+e \cos{\chi})}^2} \, \sqrt{\f{(p-2-2 e)(p-2+2e)}{p - 6- 2e \cos{\chi}}} \c \label{eq:dt0_dchi} \\
\f{\ud \phi_{\rm p}}{\ud \chi} & = & \sqrt{\f{p}{p - 6- 2 e \cos{\chi}}} \p \label{eq:dphi0_dchi}
\ea
\end{subequations}
The functions $\tau_0(\chi)$, $t_{\rm p}(\chi)$ and $\phi_{\rm p}(\chi)$ are all monotonically increasing along the orbit. The radial periods in coordinate and proper times are calculated, respectively, via
\be\label{eq:radial_period}
T_{r0} = \int_0^{2\pi} \f{\ud t_{\rm p}}{\ud \chi} \, \ud \chi \c \qquad \Tau_{r0} = \int_0^{2\pi} \f{\ud \tau_0}{\ud \chi} \, \ud \chi \c
\ee
and the accumulated azimuthal angle between successive periastron passages is 
\be\label{eq:Phi}
\Phi_0 = \int_0^{2\pi} \f{\ud \phi_{\rm p}}{\ud \chi} \, \ud \chi = 4 \sqrt{\f{p}{p-6+2 e}}\: \mrm{ellipK}\left(\f{4 e}{p - 6+2 e}\right) .
\ee
Here $\mrm{ellipK}(k) \equiv \int_0^{\pi/2} {(1-k \sin^2{\theta})}^{-1/2} \, \ud \theta$ is the complete elliptic integral of the first kind and subscripts `0' serve to distinguish the geodesic values $T_{r0}$, $\Tau_{r0}$ and $\Phi_0$ from their corresponding GSF-perturbed quantities to be introduced below. For any $(p,e)$ we have $\Phi_0 > 2 \pi $, hence the periastron advances. 

We can now define the radial and (average) azimuthal frequencies via
\be\label{eq:geod_frequencies}
\Omr \equiv \f{2\pi}{T_{r0}} \c \qquad \Omph \equiv \f{\Phi_0}{T_{r0}} \p
\ee
The pair $\{\Omr,\Omph\}$ provides a gauge-invariant parametrization of eccentric orbits. It should be noted, however, that the mapping between $(p,e)$ and $(\Omr, \Omph)$ is not bijective: there exist (infinitely many) pairs of physically distinct geodesics of different $\{p,e\}$ values but the same set of frequencies. This degeneracy, first noted in BS2011, was thoroughly studied in \cite{Wa.al.13}. The phenomenon is a feature of orbits very close to the innermost stable orbit. Since in this work we focus on less bound orbits (for the purpose of comparison with PN theory), the phenomenon of isofrequency pairing will not be relevant to us.

In the parameter space of eccentric geodesics, stable orbits are located in the region given by $ p > 6 + 2 e $. The curve $p = 6 + 2 e $ is called the {\it separatrix}. Along it both $ \Phi_0 $ and $T_{r0}$ diverge, but $\Omph $ remains finite. This gives rise to the so-called ``zoom-whirl'' behavior \cite{Cu.al.94}, where the orbiting particle zooms in from far away, whirls around the black hole many times, thus accumulating a large azimuthal phase, then zooms back out. In the limit $p\rightarrow 6+2 e $, the particle sits exactly at the peak of the effective potential and whirls infinitely on an unstable circular geodesic.

\subsection{The redshift invariant for circular orbits}\label{sec:U_generalized}

Now let the particle's mass $m_1$ be finite but small, i.e.,
\be
q \equiv m_1/m_2 \ll 1 \c
\ee 
and consider the effect of self-interaction on the motion through $\calO(q)$. Within the context of linear perturbation theory, Detweiler and Whiting \cite{DeWh.03} showed that such a particle follows a geodesic motion in a certain smooth, effective, locally-defined spacetime with metric
\be\label{eq:perturbed_metric}
g_{\alpha \beta} = g^0_{\alpha\beta} +  h^{R}_{\alpha \beta} \p
\ee
Here, $h^{R}_{\alpha \beta}$ is a certain smooth piece of the physical (retarded) metric perturbation produced by the particle. The physical perturbation itself is a solution of the linearized Einstein equation, sourced by the particle's energy-momentum, with suitable ``retarded'' boundary conditions. How $ h^{R}_{\alpha \beta} $ may be computed in practice, on a Schwarzcshild background, is discussed, for example, in Ref.~\cite{BaSa.10}.

Within linear perturbation theory, $h^{R}_{\alpha \beta}$ may be split into a dissipative piece and a conservative (time-symmetric) piece, and the effects of the two pieces may be considered separately. The conservative part of the perturbation is defined as $h^{R,{\rm cons}}_{\alpha \beta}=\frac{1}{2} \, (h^{R,{\rm ret}}_{\alpha \beta}+h^{R,{\rm adv}}_{\alpha \beta})$, where $h^{R,{\rm ret}}_{\alpha \beta}\equiv h^{R}_{\alpha \beta}$ and $h^{R,{\rm adv}}_{\alpha \beta}$ is a smooth perturbation constructed just like $h^{R,{\rm ret}}_{\alpha \beta}$ but starting with the particle's ``advanced'' metric perturbation. Replacing $h^{R,{\rm ret}}_{\alpha \beta}\to h^{R,{\rm cons}}_{\alpha \beta}$ in the effective metric (\ref{eq:perturbed_metric}) amounts to ``turning off'' the dissipation. The resulting equations of motion capture only conservative aspects of the dynamics.

In Ref.~\cite{De.08} Detweiler considered a particle in circular geodesic motion in the ``conservative'' effective spacetime 
\be
g^{\rm cons}_{\alpha \beta} = g^0_{\alpha\beta} +  h^{R,{\rm cons}}_{\alpha \beta} \p
\ee
In the absence of dissipation the orbit remains circular, and the spacetime possesses a helical Killing vector field, which, on the orbit, is proportional to the 4-velocity $u^{\alpha} = \ud x^{\alpha} / \ud \tau$. We introduce here $\tau$ as a proper-time parameter along the geodesic in the effective metric $g^{\rm cons}_{\alpha \beta}$, with $u^{\alpha}$ normalized with respect to that metric, i.e., $g^{\rm cons}_{\alpha \beta} u^{\alpha} u^{\beta} = -1$. Thanks to the helical symmetry, all components of the particle's 4-velocity are invariant under gauge transformations that respect the helical symmetry \cite{Sa.al.08}. Detweiler proposed to use the functional relationship between $u^{t}$ and $\Omega\equiv u^\phi/u^t$ as a gauge-invariant benchmark for the conservative self-force effect beyond the geodesic approximation. The frequency $\Omega$ is the circular-orbit reduction of the frequency $\Omph$ defined earlier for eccentric orbits. The quantity $u^t$ (or rather, its inverse) may be assigned a heuristic meaning of ``redshift'' (as measured in the smooth metric $h^{R,{\rm cons}}_{\alpha \beta}$ by a static asymptotic observer located along the helical symmetry axis), but it should be remembered that the true redshift, as measured in the physical metric of the particle, is, of course, divergent.

Detweiler obtained \cite{De.08}
\be
u^t(\Omega) = u^t_0(\Omega) + q \, u^t_{\rm gsf}(\Omega) \label{eq:ut_circular} \c
\ee
where $u^t_0 = {\big[1- 3 (m_2\Omega)^{2/3}\big]}^{-1/2}$ is the geodesic limit, and 
\ba\label{eq:ut_gsf_circular}
q \, u^t_{\rm gsf}  =  \frac{1}{2} u^t_0 u^{\alpha}u^{\beta} h_{\alpha\beta}^{R,{\rm cons}} 
\ea
is the $\calO(q)$ correction arising from self-interaction. Note that the correction $u^t_{\rm gsf}$ is defined for a fixed value of $\Omega$ at the background, which ensures its gauge invariance. In Ref.~\cite{De.08} and subsequent work \cite{Bl.al.10,Bl.al2.10} (see also \cite{Sh.al.14,Bl.al.14,Bl.al2.14}), Detweiler and collaborators calculated numerically the post-geodesic correction $u^t_{\rm gsf}(\Omega)$, and showed that it is consistent with corresponding PN expressions in an overlapping domain of validity. 

Detweiler's numerical results were derived using the Regge-Wheeler gauge.~An independent calculation using a direct numerical integration of the Lorenz-gauge form of the perturbation equations later recovered the same invariant relation $u^t_{\rm gsf}(\Omega)$ \cite{Sa.al.08}. This comparison highlighted a subtlety in the notion of invariance as applied to $u^t_{\rm gsf}(\Omega)$: the gauge transformation between the Lorenz-gauge metric perturbation and the Regge-Wheeler one does {\em not} leave $u^t_{\rm gsf}(\Omega)$ invariant, due to a certain minor gauge irregularity of the Lorenz-gauge metric (that was first identified in Ref.~\cite{BaLo.05} and further discussed in \cite{Sa.al.08}). Specifically, the physical metric perturbation does not vanish at infinity when expressed in the Lorenz gauge; see Eq.~(\ref{alpha}) below. While the perturbation remains helically symmetric, the transformation to an ``asymptotically flat'' gauge like Regge-Wheeler's (or the harmonic gauge of PN theory), in which Eq.\ (\ref{eq:ut_gsf_circular}) applies,  has a generator that itself does not have a helical symmetry. As a result, the transformation introduces a correction to $u^t_{\rm gsf}(\Omega)$. Denoting by $\hat h_{\alpha\beta}$ the \emph{Lorenz- gauge} metric perturbation, one finds \cite{Sa.al.08}
\ba
q \, u^t_{\rm gsf} = \frac{1}{2} u^t_0  u^{\alpha} u^{\beta} \hat h_{\alpha\beta}^{R,{\rm cons}} + \alpha {\cal E} (u_0^t)^2 \p
\label{eq:ut_gsf_circular_Lor} 
\ea
The parameter $\alpha$ is extracted from the Lorenz-gauge perturbation as prescribed in Eq.\ (\ref{alpha}) below; for a circular orbit it reads $ \alpha= q(m_2\Omega)^{2/3}u^t_0$. One must be mindful, when working in the Lorenz gauge (as we do here), to take proper account of this gauge irregularity. We shall return to this point in more detail when discussing eccentric orbits.

\subsection{The redshift invariant generalized to eccentric orbits}\label{sec:GSF_computation}

Now consider an eccentric orbit subject to the conservative effect of the GSF. In absence of dissipation, the orbit remains bound and has a constant radial period $T_r$ and a constant accumulated azimuthal phase $\Phi$ per radial period. Hence it possesses a well defined pair of frequencies $\{\Omr,\Omph\}$, defined via Eq.\ (\ref{eq:geod_frequencies}) with the subscripts `0' dropped. The functional relation between these invariant frequencies and any gauge-dependent set of parameters can be written as the sum of a ``geodesic'' term and a GSF correction; such relations were derived in explicit form in BS2011 but will not be needed here. 

The GSF-perturbed orbit is a geodesic in the effective metric $g^{\rm cons}_{\alpha \beta}=g^0_{\alpha\beta}+  h_{\alpha\beta}^{R,{\rm cons}}$, with tangent four-velocity $u^{\alpha}$ normalized in $g^{\rm cons}_{\alpha \beta}$. It is easily checked that $u^t$ is no longer gauge-invariant in a pointwise sense when the orbit is noncircular. Instead, BS2011 suggested to consider the orbital average 
\be\label{eq:U}
\langle U \rangle \equiv \langle u^t \rangle \equiv \f{1}{\Tau_r} \int_0^{\Tau_r} u^t \, \ud \tau = \f{T_r}{\Tau_r} \c
\ee
where $\Tau_r$ is the radial period measured in proper time $\tau$. BS2011 argued that $\langle U \rangle$ is invariant under gauge transformations that respect the periodicity of the orbit and are well behaved (in a certain sense) at infinity. We may split $\langle U \rangle$ in the form
\be
\langle U \rangle(\Omega_i) = {\langle U \rangle}_0(\Omega_i) + q \, \langle U \rangle_{\rm gsf}(\Omega_i) \c
\ee
where $\Omega_i \equiv \{\Omr,\Omph\}$, ${\langle U \rangle}_0$ is the geodesic limit of $\langle U \rangle$ taken with fixed $\Omega_i$, and $ q \langle U \rangle_{\rm gsf}$ is the GSF correction, defined for fixed $\Omega_i$. The functional relation $\langle U \rangle_{\rm gsf}(\Omega_i)$ is an invariant measure of the GSF effect on the eccentric orbit, and it is the quantity that we will use for our GSF--PN comparison in this paper.

The geodesic limit of $\langle U \rangle$ is given by
\be\label{eq:U0}
{\langle U \rangle}_0 = \f{T_{r0} }{\Tau_{r0} } \c
\ee
where the periods $T_{r0}$ and $\Tau_{r0}$ may be calculated via \eqref{eq:radial_period} given the parameters $p,e$ of the geodesic orbit. BS2011 describes a practical method for (numerically) inverting the relations $\Omega_i(p,e)$ in order to obtain $p(\Omega_i)$ and $e(\Omega_i)$. This method may be used in conjunction with Eqs.~\eqref{eq:radial_period} and \eqref{eq:U0} in order to compute ${\langle U \rangle}_0$ for given frequencies $\Omega_i$.

Our goal now is to express $\langle U \rangle_{\rm gsf}(\Omega_i)$ explicitly in terms of calculable perturbative quantities (the metric perturbation and/or the GSF). Since fixing $\Omega_i$ fixes $T_r$, the only contribution to $\langle U \rangle_{\rm gsf}(\Omega_i)$ comes from the $\calO(q)$ difference $\Tau_r-\Tau_{r0}$. From the normalizations $g^{0}_{\alpha \beta}u_0^{\alpha}u_0^{\beta}=-1$ and $(g^{0}_{\alpha \beta}+  h_{\alpha\beta}^{R,{\rm cons}})u^{\alpha}u^{\beta}=-1$ one obtains
\be\label{eq:dtau0_dtau}
\f{\ud \tau_0}{\ud \tau} = 1 +\f{1}{2} u_0^{\alpha}u_0^{\beta} h_{\alpha\beta}^{R,\text{cons}} \equiv 1 +\f{1}{2} h_{uu}^{R} \c
\ee
where terms of $\calO(q^2)$ and higher are omitted. Since the contraction $h_{uu}^{R}$ automatically picks out the conservative piece of $h_{\alpha\beta}^{R}$, the label `cons' becomes redundant and we have dropped it. Neglecting subleading terms in the mass ratio $q$, we now obtain
\be
\Tau_r-\Tau_{r0}= \int_{0}^{\Tau_r}\left(1-\frac{\ud\tau_0}{\ud\tau}\right)\ud\tau = - \frac{1}{2} \Tau_{r0} \langle h_{uu}^{R} \rangle \c
\ee
where the average is taken with respect to $\tau$ (or $\tau_0$) over a radial period. The $\calO(q)$ perturbation of $\langle U \rangle = T_r / \Tau_r$ at fixed $\Omega_i$ therefore reads
\be\label{Ugsf0}
q \, \langle U \rangle_{\rm gsf} = - \frac{T_{r0}}{{(\Tau_{r0})}^2} \, (\Tau_r - \Tau_{r0}) =\f{1}{2} {\langle U \rangle}_0 \langle h_{uu}^{R} \rangle .
\ee

This would be our final result for $\langle U \rangle_{\rm gsf}$ if $h_{uu}^R$ were to be calculated in a suitable ``asymptotically flat'' gauge. Our calculation, however, will be performed in the Lorenz gauge, which suffers from the aforementioned irregularity at infinity. Let us now describe this irregularity more specifically. For either circular or noncircular orbits, the Lorenz-gauge metric component $\hat h_{tt}$ tends to a finite nonzero value at $r\to\infty$ (other components are regular). This behavior is due entirely to the static piece of the mass monopole perturbation, and therefore the asymptotic value of $\hat h_{tt}$ does not depend on the angular direction even for eccentric orbits; it depends only on the orbital parameters. To remove this gauge artifact, following BS2011 we introduce the normalized time coordinate $t=(1+ \alpha)\hat t$, where $\hat t$ denotes the original Lorenz-gauge time coordinate, and $\alpha=\alpha(\Omega_i)$ is given by
\be\label{alpha}
\alpha=-\frac{1}{2}\hat h_{tt}(r\to\infty) \p
\ee
This normalization, which amounts to an $\calO(q)$ gauge transformation away from the Lorenz gauge, corrects the asymptotic behavior. Under $\hat t\to t$ we have, at leading order,
\be
\hat h_{uu}^{R} \to h_{uu}^{R} + 2\alpha g_{tt}^{0} {\langle U \rangle}_0^2 = h_{uu}^{R} - 2\alpha{\cal E} {\langle U \rangle}_0 \p
\ee
Thus, to re-express $\langle U \rangle_{\rm gsf}$ in Eq.~\eqref{Ugsf0} in terms of the Lorenz-gauge perturbation, we need simply replace  $h_{uu}^{R}\to \hat h_{uu}^{R} + 2\alpha{\cal E} {\langle U \rangle}_0$. We finally get
\be\label{Ugsf}
q \, \langle U \rangle_{\rm gsf}=\f{1}{2} {\langle U \rangle}_0 \langle \hat h_{uu}^{R} \rangle + \alpha {\cal E} {\langle U \rangle}_0^2 \p
\ee

Equation \eqref{Ugsf} is one of our main results, giving $\langle U \rangle_{\rm gsf}$ in terms of quantities directly calculable using existing GSF codes: the $R$-field $\hat h_{\alpha\beta}^{R}$ in the Lorenz gauge, and the corresponding asymptotic parameter $\alpha$. It is clear that Eq.~(\ref{Ugsf}) reduces to \eqref{eq:ut_gsf_circular} in the circular-orbit limit. As in the circular case, the expression for $\langle U \rangle_{\rm gsf}$ involves only the $R$-field along the orbit (and the parameter $\alpha$), and not the GSF itself. Our result \eqref{Ugsf} is much simpler than the one derived in BS2011 using a different procedure. In that work, certain simplifications that reduce the expression for $\langle U \rangle_{\rm gsf}$ to the form (\ref{Ugsf}) have been overlooked. In Appendix \ref{app:proof} we establish the equivalence between the two results.

\section{Numerical calculation of the generalized redshift}\label{sec:num_results}

We have used the frequency-domain computational framework of Ref.~\cite{Ak.al.13} in order to compute $\langle U \rangle_{\rm gsf}$ for a large sample of orbits, focusing primarily on obtaining weak-field data for PN comparisons. Our calculation is based on Eq.\ (\ref{Ugsf}), which takes as input the regularized Lorenz-gauge metric perturbation evaluated along the orbit (as well as the asymptotic value $\alpha$, also to be read off the Lorenz-gauge perturbation). Since the GSF correction $q \langle U \rangle_{\rm gsf}$ (defined at fixed frequencies $\Omega_i$) is of $\calO(q)$, it is sufficient to use as input the metric perturbation calculated along {\em geodesic} orbits. For convenience we shall use $p,e$ (as defined in Sec.\ \ref{sec:Schwarzschild_geodesics}), rather than $\Omega_i$, to parameterize these geodesics, and will thus express our results in the form $\langle U \rangle_{\rm gsf}=\langle U \rangle_{\rm gsf}(p,e)$. It is important to emphasize that our results refer to the GSF correction to the functional relation $\langle U \rangle_{\rm gsf}(\Omega_i)$ {\it defined for fixed invariant frequencies $\Omega_i$}, even though we use the geodesic parameters $p$ and $e$ as independent variables. These two facts should not be confused.  

\subsection{Details of numerics and sources of error}

We use the eccentric-orbit GSF code of Ref.~\cite{Ak.al.13} to obtain the metric perturbation $\hat h^R_{\alpha\beta}(\chi) $ along the geodesic orbit. This code employs a frequency-domain approach, coupled with the method of extended homogeneous solutions of Ref.~\cite{Ba.al2.08}, to compute the regularized metric perturbation $\hat h^R_{\alpha\beta}$. It then outputs $\hat h^R_{uu}(\chi) $ at 2400 evenly spaced points along the orbit, and interpolates the numerical data using {\it Mathematica}'s \texttt{Interpolation} function. In its default setting, \texttt{Interpolation} fits cubic polynomials between successive data points. Since $\hat h^{R}_{uu}(\chi)$ is very smooth this level of interpolation is sufficient for our purposes. We subsequently calculate the orbital average $ \langle \hat h^{R}_{uu} \rangle$ using \texttt{NIntegrate} with the appropriate numerical integration options/controls offered by {\it Mathematica}. The coefficient $\alpha$ is extracted, using \eqref{alpha}, from the static monopole piece of the metric perturbation, whose construction is prescribed in App.~B of \cite{Ak.al.13}. Since this piece is essentially known analytically (its computation involves the evaluation of a certain orbital integral, easily done with {\it Mathematica} at extremely high accuracy), numerical error in our calculation of $\langle U \rangle_{\rm gsf}$ comes entirely from the numerical evaluation of $\hat h^{R}_{uu}(\chi)$. Reference \cite{Ak.al.13} contains a thorough analysis of error sources for $\hat h^{R}_{uu}(\chi)$ and the GSF. Here, we briefly review two dominant sources. 

Each Fourier mode of our computation has associated with it a frequency, $\omega=\tilde{n}\Omega_r + \tilde{m}\Omega_\phi$, where $\tilde{n}$ and $\tilde{m}$ are integer harmonic numbers. The dominating source of numerical error depends on the value of $\omega$. For modes of sufficiently large frequency ($ m_2 \, \omega\gtrsim 10^{-4}$), the dominant error comes from the estimation of the contribution from the tail of uncomputed multipoles of large $l$ values. Typically, we compute the contributions to all the $l$-modes up to and including $l=15$, and estimate the remaining contribution to the mode sum by fitting numerical data to suitable power-law models of the large-$l$ behavior \cite{BaSa.07,BaSa.10,Ak.al.13}. This is a relatively well-modelled and well-controlled source of error, and it can be reduced in a straightforward manner using additional computational resources.

For modes with small frequencies, $ m_2 \, \omega \lesssim 10^{-4}$, a second source of numerical error takes over. This comes from rounding errors introduced when inverting the matrix of amplitude coefficients as part of the procedure for computing inhomogeneous solutions to the Lorenz-gauge field equations \cite{Ak.al.13}. When $\omega$ is very small, the matrix becomes nearly singular, and its inversion using machine-precision arithmetic introduces large errors. The problematic ``nearly-static'' modes occur generically in our calculation, since, given any orbital parameters, there will exist values of $\tilde{m}$ and $\tilde{n}$ in the Fourier sum for which $ m_2 \, \omega$ is very small. In practice, the sum over $\tilde{n}$ and $\tilde{m}$ is truncated once our results reach a desired accuracy. Consequently, the problem is less severe for low-eccentricity orbits, where the effective frequency band is narrow, and more severe at high eccentricity, where the broad frequency band implies a higher chance of encountering nearly static modes. Ultimately, this restricts our calculation to orbits with eccentricities of $e\lesssim0.4$. Low-$\omega$ modes are encountered also when the fundamental frequencies themselves are small, as with weak-field orbits---the main focus of the present work. Our code incorporates several methods for mitigating this small-frequency problem (see Ref.~\cite{Ak.al.13} for details), but even with these techniques employed, our current calculation appears limited to orbits with $p\lesssim130$; at larger $p$ we observe a rapid reduction in accuracy. 

The issue of nearly-static modes has been addressed in a very recent paper by Osburn {\it et al.} \cite{Os.al.14}, who proposed additional mitigation methods. These may be used to improve the performance of weak-field calculations in future work. 

\subsection{Numerical results}

Table~\ref{table:numerical_results1} displays a sample of our numerical results for $\langle U\rangle_{\text{gsf}}$. Parenthetical figures indicate estimated error bars on the last displayed decimals; for instance, $-0.0556761(1)$ stands for $-0.0556761 \pm 1\!\times \!10^{-7}$. Additional data may be made available to interested readers upon request from the authors. Some of the data shown in the table are plotted in Fig.~\ref{fig:GSF_data_all} of Sec.~\ref{sec:comparison}, where we discuss the comparison with PN results.

\begin{table}[h!]
	\begin{tabular}{@{}cclllll@{}}
	    \toprule
		$e$ & \hspace{0.3cm} & \hspace{0.5cm} $p=10$ & \hspace{0.5cm} $p=15$ & \hspace{0.5cm} $p=20$ & \hspace{0.5cm} $p=25$ \\
		\hline
		0.05 && $-0.12878023(4)$ & $-0.07751154(5)$ & $-0.0556761252(8)$ & $-0.0434829334(1)$ \\ 
		0.10 && $-0.1277540(3)$  & $-0.0768706(2)$  & $-0.05522166(7)$	 & $-0.043132423(1)$  \\ 
		0.15 && $-0.1260434(2)$  & $-0.07580395(6)$ & $-0.05446527(1)$	 & $-0.042548963(6)$  \\ 
		0.20 && $-0.123648(3)$	 & $-0.07431376(9)$ & $-0.05340854(9)$	 & $-0.041733648(3)$  \\ 
		0.25 && $-0.120567(2)$   & $-0.07240329(5)$ & $-0.0520537(3)$ 	 & $-0.04068802(6)$   \\ 
		0.30 && $-0.1168020(6)$	 & $-0.0700768(5)$  & $-0.05040377(4)$	 & $-0.0394141(2)$	  \\ 
		0.35 && $-0.112352(2)$	 & $-0.0673398(5)$  & $-0.0484623(4)$ 	 & $-0.0379143(1)$	  \\ 
		0.40 && $-0.107221(2)$	 & $-0.064199(1)$	& $-0.0462337(9)$ 	 & $-0.0361916(2) $   \\ 
		\hline
		$e$ & \hspace{0.3cm} & \hspace{0.5cm} $p=30$ & \hspace{0.5cm} $p=35$ & \hspace{0.5cm} $p=40$ & \hspace{0.5cm} $p=50$ \\
	\hline
		0.05 && $-0.0356833158(1)$  & $-0.0302606957(1)$ & $-0.0262706836(5)$	& $-0.0207905297(5)$ \\ 
		0.10 && $-0.035398479(5)$	& $-0.0300209627(8)$ & $-0.0260637929(8)$	& $-0.0206282073(6)$ \\ 
		0.15 && $-0.03492427(6)$	& $-0.029621799(3)$  & $-0.025719277(2)$	& $-0.0203578673(9)$ \\ 
		0.20 && $-0.034261480(8)$	& $-0.029063791(5)$  & $-0.025237591(2)$	& $-0.019979802(2)$  \\ 
		0.25 && $-0.03341121(1)$	& $-0.028347763(7)$  & $-0.024619373(3)$	& $-0.01949443(6)$	 \\ 
		0.30 && $-0.0323749(2)$		& $-0.0274748(4)$	 & $-0.02386545(8)$		& $-0.01890227(5)$	 \\ 
		0.35 && $-0.0311543(2)$		& $-0.0264462(2)$	 & $-0.0229768(4)$		& $-0.0182040(1)$ 	 \\ 
		0.40 && $-0.0297515(6)$		& $-0.0252634(2)$	 & $-0.0219547(5)$		& $-0.0174004(4)$	 \\ 
		\hline
		$e$ & \hspace{0.3cm} & \hspace{0.5cm} $p=60$ & \hspace{0.5cm} $p=70$ & \hspace{0.5cm} $p=80$ & \hspace{0.5cm} $p=90$ \\
		\hline
		0.05 && $-0.0172030750(2)$	& $-0.0146718447(3)$ & $-0.0127901170(3)$	& $-0.0113362814(8)$ \\ 
		0.10 && $-0.0170695612(3)$	& $-0.0145584700(2)$ & $-0.0126916092(2)$	& $-0.0112491974(8)$ \\ 
		0.15 && $-0.016847175(1)$	& $-0.0143696130(3)$ & $-0.0125275072(7)$	& $-0.0111041186(3)$ \\ 
		0.20 && $-0.016536122(4)$	& $-0.0141054255(4)$ & $-0.0122979274(7)$	& $-0.0109011371(5)$ \\ 
		0.25 && $-0.01613669(1)$	& $-0.0137661204(6)$ & $-0.012003033(2)$	& $-0.010640382(2)$  \\ 
		0.30 && $-0.01564925(2)$ 	& $-0.013351971(9)$  & $-0.011643033(7)$	& $-0.010322020(3)$  \\ 
		0.35 && $-0.01507427(8)$ 	& $-0.0128633(1)$	 & $-0.01121819(3)$		& $-0.00994625(3)$   \\ 
		0.40 && $-0.0144123(3)$ 	& $ -0.0123002(5)$	 & $-0.0107288(2)$		& $-0.0095133(2)$	 \\ 
		\hline
		$e$ & \hspace{0.3cm} & \hspace{0.5cm} $p=100$ & \hspace{0.5cm} $p=110$ & \hspace{0.5cm} $p=120$ & \hspace{0.5cm} $p=130$ \\
		\hline
		0.05 && $-0.0101792669(2)$ 	 & $-0.0092365822(2)$ 	& $-0.0084537150(1)$ & $-0.0077931956(9)$ \\ 
		0.10 && $-0.0101012344(10)$  & $-0.0091658975(1)$ 	& $-0.0083891149(2)$ & $-0.0077337155(1)$ \\ 
		0.15 && $-0.0099712296(7)$ 	 & $-0.0090481316(3)$ 	& $-0.0082814820(1)$ & $-0.0076346120(1)$ \\ 
		0.20 && $-0.0097893274(4)$ 	 & $-0.0088833462(1)$ 	& $-0.0081308692(1)$ & $-0.0074959280(1)$ \\ 
		0.25 && $-0.0095556341(8)$ 	 & $-0.0086716263(10)$  & $-0.0079373488(7)$ & $-0.0073177296(5)$ \\ 
		0.30 && $-0.009270281(4)$ 	 & $-0.008413085(2)$ 	& $-0.007701017(5)$	 & $-0.007100089(2)$  \\ 
		0.35 && $-0.00893344(2)$ 	 & $-0.00810786(1)$ 	& $-0.00742198(3)$	 & $-0.00684311(3)$   \\ 
		0.40 && $ -0.0085453(2)$ 	 & $ -0.0077561(2)$		& $-0.0071004(2)$	 & $-0.0065469(2)$	  \\ 
		\botrule
	\end{tabular}
	\caption{Numerical data for the GSF contribution $\langle U \rangle_{\text{gsf}}$ to the generalized redshift (defined with fixed invariant frequencies $\Omega_i$), for various eccentric geodesic orbits in a Schwarzschild background. The orbital parameters $e$ (eccentricity) and $p$ (semi-latus rectum) are defined in Sec.~\ref{sec:Schwarzschild_geodesics}. Parenthetical figures indicate estimated error bars on the last displayed decimals.}
	\label{table:numerical_results1}
\end{table}

As a check of our frequency-domain computation, we compare our results for $\langle U \rangle_{\mrm{gsf}}$ with those obtained in BS2011 using a time-domain method. BS2011 provided a small sample of numerical results in the range $p \le 20$ and $e \le 0.5$. The comparison is shown in Table~\ref{table:e0.4_data}. There is evidently a good agreement between the two sets of numerical results, although in some of the entries the values appear not fully consistent given the stated error bars (in all these cases the BS2011 values are smaller than ours). We have strong evidence to suggest that the source of disagreement is a slight underestimation of the magnitude of systematic error in the time-domain analysis of BS2011: We have tested the output of our frequency-domain code against accurate GSF data published in Ref.~\cite{Os.al.14}, and against yet unpublished redshift data calculated by van de Meent \cite{vdM.15} (using a very different frequency-domain method based on a semi-analytical treatment of Teukolsky's equation \cite{MeSh.15}). These comparisons strongly favor the frequency-domain data in the table. 

Also evident from the table is the fact that our code's accuracy starts to degrade for $e=0.4$; however, it still matches BS2011's results to five or six significant digits. No published numerical data exist to allow comparison beyond $p = 20$. (Reference \cite{Os.al.14} gives results for $e \le 0.7$ and $p \le 90$, but these are for the GSF components, not for $\langle U \rangle_{\mrm{gsf}}$.)

\begin{table}[h]
	\begin{tabular}{@{}lllllll@{}}
    	\toprule
  		$e$ & \hspace{0.3cm} & \qquad $p = 10$ & \hspace{0.3cm} & \qquad $p=15$ & \hspace{0.3cm} & \qquad $p = 20 $ \\
		\hline 
		$0.1$ \, Here && $-0.1277540(3)$ && $-0.0768706(2)$ && $-0.05522166(7)$ \\
		\qquad\! BS2011 && $-0.1277554(7)$ && $-0.0768709(1)$ && $-0.05522177(4)$ \\
		\hline
		$0.2$ && $-0.123648(3)$ && $-0.07431376(9)$ && $-0.05340854(9)$ \\
		&& $-0.1236493(7)$ && $-0.0743140(1)$ && $-0.05340866(4)$ \\
		\hline
		$0.3$ && $-0.1168020(6)$ && $-0.0700768(5)$ && $-0.05040377(4)$ \\
		&& $-0.1168034(6)$ && $-0.0700771(1)$ && $-0.05040388(4)$ \\
		\hline
		$0.4$ && $-0.107221(2)$ && $-0.064199(1)$ && $-0.0462337(9)$ \\
		&& $-0.1072221(5)$ && $-0.0641991(1)$ && $-0.04623383(4)$ \\
		\botrule 
	\end{tabular}
	\caption{Our frequency-domain numerical results for $\langle U \rangle_{\text{gsf}}$ and the corresponding time-domain values from BS2011. Each cell shows our result (top) in comparison to BS2011's (bottom). The relative disagreement between the two data sets is $\sim 10^{-6}$, roughly consistent with the magnitude of error bars. As discussed in the text, evidence suggests that our results are accurate to within the error bars given, whereas the magnitude of error in the time-domain data is slightly underestimated in some cases.   No time-domain data exist for $p>20$ to allow comparison in the weak-field domain.}
	\label{table:e0.4_data}
\end{table}
%

\section{Generalized redshift: post-Newtonian calculation}\label{sec:PN_Delta_U}

We shall now derive the invariant relation ${\langle U \rangle}(\Omr,\Omph)$ within the PN approximation. Our calculations will be similar in spirit to those performed by Arun \textit{et al.} \cite{Ar.al.08,Ar.al2.09}, except that we will consider the orbital average of a quantity that is related to the orbital dynamics of a binary of nonspinning compact objects, modelled as point particles, while Refs.~\cite{Ar.al.08,Ar.al2.09} calculated the orbital-averaged fluxes of energy and angular momentum radiated at infinity. Furthermore, while these fluxes are invariant under the exchange $1 \leftrightarrow 2$ of the bodies' labels, and requires knowledge of the gravitational field in the wave-zone, the generalized redshift ${\langle U \rangle}$ is a property of one particle, whose evaluation involves the near-zone metric.

\subsection{Redshift variable in standard harmonic coordinates}\label{secII}

\subsubsection{The regularized 3PN metric}

Throughout Sec.~\ref{sec:PN_Delta_U} we assume that $m_1 < m_2$, and treat $m_1$ as the ``particle'' orbiting the ``black hole'' of mass $m_2$. The redshift of the particle can be computed from the knowledge of the \textit{regularized} PN metric $g_{\alpha\beta}(y_1) \equiv g_{\alpha\beta}(t,\mathbf{y}_1)$, generated by the two bodies and evaluated at the coordinate location $\mathbf{y}_1(t)$ of the particle, as \cite{Bl.al.10}
\beq\label{def_U}
	U \equiv u_1^t = \biggl( - g_{\alpha\beta}(y_1) \, \frac{v_1^\alpha v_1^\beta}{c^2} \biggr)^{-1/2} ,
\eeq
where $v_1^\alpha = (c, \mathbf{v}_1)$, with $\mathbf{v}_1 = \ud \mathbf{y}_1 / \ud t$ the coordinate velocity of the particle. The generalized redshift will be given by the proper-time average of Eq.~\eqref{def_U} over one radial period.

The regularized PN metric $g_{\alpha\beta}(y_1)$ was itself computed up to 2.5PN order, in harmonic coordinates, in Ref.~\cite{Bl.al.98}. This calculation was then extended to 3PN order in Ref.~\cite{Bl.al.10}, partly based on existing computations of the 3PN equations of motion using Hadamard regularization \cite{BlFa.01} and dimensional regularization \cite{Bl.al.04}. Reference \cite{Bl.al.10} performed two calculations of the 3PN regularized metric, using both Hadamard and dimensional regularizations, obtaining the same metric but expressed in two different harmonic coordinate systems. The two metrics were found to differ by an infinitesimal 3PN coordinate transformation in the ``bulk,'' i.e., outside the particle's worldlines, and also by an intrinsic \textit{shift} of these worldlines. Combining Eqs.~(4.2) and (A15) of Ref.~\cite{Bl.al.10}, the 3PN-accurate expression of the regularized metric reads, in the standard harmonic coordinates corresponding to the use of Hadamard regularization,\footnote{As usual we denote by $r_{12} = \vert \mathbf{y}_1 - \mathbf{y}_2 \vert$ the coordinate separation, by $\mathbf{n}_{12} = (\mathbf{y}_1 - \mathbf{y}_2)/ r_{12}$ the unit direction from particle $2$ to particle $1$, and by $\mathbf{v}_{12} = \mathbf{v}_1 - \mathbf{v}_2$ the relative velocity, where $\mathbf{v}_a = \ud \mathbf{y}_a / \ud t$ is the 3-velocity of particle $a$. The Euclidean scalar product between two 3-vectors $\mathbf{A}$ and $\mathbf{B}$ is denoted $(A B)$. Parentheses around indices are used to indicate symmetrization, i.e., $A^{(i} B^{j)} = \frac{1}{2} ( A^i B^j + A^j B^i)$.}
\begin{subequations}\label{g1}
	\begin{align}
		g_{00}(y_1) &= -1 + \frac{2 G m_2}{c^2 r_{12}} + \frac{G m_2}{c^4 r_{12}} \biggl[ 4 v_2^2 - (n_{12}v_2)^2 - 3 \frac{G m_1}{r_{12}} - 2 \frac{G m_2}{r_{12}} \biggr] \nonumber \\ &+ \frac{G m_2}{c^6 r_{12}} \biggl[\frac{3}{4} (n_{12}v_2)^4 - 3 (n_{12}v_2)^2 v_2^2 + 4 v_2^4 + \frac{G m_2}{r_{12}} \biggl( 3 (n_{12}v_2)^2 - v_2^2 + 2 \frac{G m_2}{r_{12}} \biggr) \nonumber \\ &+ \frac{G m_1}{r_{12}} \biggl( - \frac{87}{4} (n_{12}v_1)^2 + \frac{47}{2} (n_{12}v_1)(n_{12}v_2) - \frac{55}{4} (n_{12}v_2)^2 - \frac{39}{2} (v_1 v_2) + \frac{23}{4} v_1^2 + \frac{47}{4} v_2^2 \nonumber \\ &- \frac{G m_1}{r_{12}} + \frac{17}{2} \frac{G m_2}{r_{12}} \biggr) \biggr] + \frac{G m_2}{c^8 r_{12}} \biggl[ - \frac{5}{8} (n_{12}v_2)^6 - 5 (n_{12}v_2)^2 v_2^4 + 3 (n_{12}v_2)^4 v_2^2 + 4 v_2^6 \nonumber \\ &+ \frac{G m_2}{r_{12}} \biggl( -4 (n_{12}v_2)^4 + 5 (n_{12}v_2)^2 v_2^2 - v_2^4 \biggr) + \frac{G m_1}{r_{12}} \biggl( - \frac{617}{24} (n_{12}v_1)^4 + \frac{491}{6} (n_{12}v_1)^3 (n_{12}v_2) \nonumber \\ &- \frac{225}{4} (n_{12}v_1)^2 (n_{12}v_2)^2 + \frac{41}{2} (n_{12}v_1) (n_{12}v_2)^3 + \frac{53}{8} (n_{12}v_2)^4 - \frac{79}{4} (n_{12}v_1)^2 (v_1 v_2) \nonumber \\ &+ 42 (n_{12}v_1) (n_{12}v_2) (v_1 v_2) + \frac{101}{4} (n_{12}v_2)^2 (v_1 v_2) + \frac{49}{4} (v_1 v_2)^2 - \frac{27}{8} (n_{12}v_1)^2 v_1^2 \nonumber \\ &+ \frac{23}{4} (n_{12}v_1) (n_{12}v_2) v_1^2 - \frac{273}{8} (n_{12}v_2)^2 v_1^2 - 25 (v_1 v_2) v_1^2 + \frac{39}{8} v_1^4 - \frac{305}{8} (n_{12}v_1)^2 v_2^2 \nonumber \\ &+ \frac{139}{4} (n_{12}v_1) (n_{12}v_2) v_2^2 - \frac{291}{8} (n_{12}v_2)^2 v_2^2 - 62 (v_1 v_2) v_2^2 + \frac{77}{2} v_1^2 v_2^2 + \frac{235}{8} v_2^4 \biggr) \nonumber \\ &+ \frac{G^2 m_1^2}{r_{12}^2} \biggl( \ln{\left( \frac{r_{12}}{r_0} \right)} \biggl\{ 32 (n_{12}v_{12})^2 - \frac{32}{3} v_{12}^2 \biggr\} + (n_{12}v_1)^2 \biggl\{ \frac{2789}{100} - \frac{364}{5} \ln{\left( \frac{r_{12}}{r'_1} \right)} \biggr\} \nonumber \\ &+ (n_{12}v_1) (n_{12}v_2) \biggl\{ \frac{6603}{50} + \frac{584}{5} \ln{\left( \frac{r_{12}}{r'_1} \right)} \biggr\} + (n_{12}v_2)^2 \biggl\{ - \frac{1881}{20} - 44 \ln{\left( \frac{r_{12}}{r'_1} \right)} \biggr\} \nonumber \\ &+ (v_1 v_2) \biggl\{ - \frac{3053}{150} - \frac{584}{15} \ln{\left( \frac{r_{12}}{r'_1} \right)} \biggr\} + v_1^2 \biggl\{ - \frac{1613}{100} + \frac{364}{15} \ln{\left( \frac{r_{12}}{r'_1} \right)} \biggr\} \nonumber \\ &+ v_2^2 \biggl\{ \frac{1421}{60} + \frac{44}{3} \ln{\left( \frac{r_{12}}{r'_1} \right)} \biggr\} + \frac{G m_1}{r_{12}} \biggl\{ - \frac{1091}{60} - \frac{44}{3} \ln{\left( \frac{r_{12}}{r'_1} \right)} + \frac{32}{3} \ln{\left( \frac{r_{12}}{r_0} \right)} \biggr\} \biggr) \nonumber \\ &+ \frac{G^2 m_1 m_2}{r_{12}^2} \biggl( \ln{\left( \frac{r_{12}}{r_0} \right)} \biggl\{ 32 (n_{12}v_{12})^2 - \frac{32}{3} v_{12}^2 \biggr\} + (n_{12}v_1)^2 \biggl\{ - \frac{109}{12} - \frac{141}{16} \pi^2 \biggr\} \nonumber \\ &+ (n_{12}v_1) (n_{12}v_2) \biggl\{ - \frac{197}{6} + \frac{177}{8} \pi^2 \biggr\} + (n_{12}v_2)^2 \biggl\{ \frac{391}{6} - \frac{213}{16} \pi^2 \biggr\} + (v_1 v_2) \biggl\{ \frac{812}{9} - \frac{59}{8} \pi^2 \biggr\} \nonumber \\ &+ v_1^2 \biggl\{ - \frac{299}{18} + \frac{47}{16}\pi^2 \biggr\} + v_2^2 \biggl\{ - \frac{1097}{18} + \frac{71}{16} \pi^2 \biggr\} + \frac{G m_1}{r_{12}} \biggl\{ - \frac{3571}{60} - \frac{15}{8} \pi^2 - \frac{44}{3} \ln{\left( \frac{r_{12}}{r'_1} \right)} \nonumber \\ &+ \frac{64}{3} \ln{\left( \frac{r_{12}}{r_0} \right)} \biggr\} + \frac{G m_2}{r_{12}} \biggl\{ - \frac{28141}{630} - \frac{15}{8} \pi^2 + \frac{44}{3} \ln{\left( \frac{r_{12}}{r'_2} \right)} + \frac{32}{3} \ln{\left( \frac{r_{12}}{r_0} \right)} \biggr\} \biggr) \nonumber \\ &+ \frac{G^2 m_2^2}{r_{12}^2} \biggl( - (n_{12}v_2)^2 + 2 v_2^2 - 2 \frac{G m_2}{r_{12}} \biggr) \biggr] + o(c^{-8}) \, , \\
        g_{0i}(y_1) &= - \frac{4 G m_2}{c^3 r_{12}} v_2^i + \frac{G m_2}{c^5 r_{12}} \biggl[ v_2^i \biggl( 2 (n_{12}v_2)^2 - 4 v_2^2 - 2 \frac{G m_1}{r_{12}} + \frac{G m_2}{r_{12}} \biggr) + 4 \frac{G m_1}{r_{12}} v_1^i \nonumber \\ &+ n_{12}^i \biggl( \frac{G m_1}{r_{12}} \bigl\{ 10 (n_{12}v_1) + 2 (n_{12}v_2) \bigr\} - \frac{G m_2}{r_{12}} (n_{12}v_2) \biggr) \biggr] + \frac{G m_2}{c^7 r_{12}} \biggl[ v_2^i \biggl( - \frac{3}{2} (n_{12}v_2)^4 \nonumber \\ &+ 4 (n_{12}v_2)^2 v_2^2 - 4 v_2^4 + \frac{G m_1}{r_{12}} \biggl\{ 48 (n_{12}v_1)^2 - 44 (n_{12}v_1) (n_{12}v_2) + 10 (n_{12}v_2)^2 \nonumber \\ &+ 40 (v_1 v_2) - 16 v_2^2 - 26 v_1^2 \biggr\} + \frac{G m_2}{r_{12}} \biggl\{ - 2 (n_{12}v_2)^2 + v_2^2 - 2 \frac{G m_2}{r_{12}} \biggr\} \nonumber \\ &+ \frac{G^2 m_1 m_2}{r_{12}^2} \biggl\{ \frac{95}{6} - \frac{3}{4} \pi^2 \biggr\} + \frac{G^2 m_1^2}{r_{12}^2} \biggl\{ - \frac{1013}{75} + \frac{24}{5} \ln{\left( \frac{r_{12}}{r'_1} \right)} \biggr\} \biggr) \nonumber \\ &+ v_1^i \biggl( \frac{G m_1}{r_{12}} \biggl\{ - \frac{17}{2} (n_{12}v_1)^2 - 15 (n_{12}v_1) (n_{12}v_2) + \frac{43}{2} (n_{12}v_2)^2 + 3 (v_1 v_2) \nonumber \\ &+ \frac{17}{2} v_1^2 - \frac{15}{2} v_2^2 \biggr\} + \frac{G^2 m_1 m_2}{r_{12}^2} \biggl\{ - \frac{57}{2} + \frac{3}{4} \pi^2 \biggr\} + \frac{G^2 m_1^2}{r_{12}^2} \biggl\{ \frac{1973}{75} - \frac{24}{5} \ln{\left( \frac{r_{12}}{r'_1} \right)} \biggr\} \biggr) \nonumber \\ &+ n_{12}^i \biggl( \frac{G m_1}{r_{12}} \biggl\{ \frac{21}{2} (n_{12}v_1)^3 - \frac{43}{2} (n_{12}v_1)^2 (n_{12}v_2) - \frac{29}{2} (n_{12}v_1) (n_{12}v_2)^2 + \frac{3}{2} (n_{12}v_2)^3 \nonumber \\ &- 11 (n_{12}v_1) (v_1 v_2) - 19 (n_{12}v_2) (v_1 v_2) + \frac{1}{2} (n_{12}v_1) v_1^2 + \frac{39}{2} (n_{12}v_2) v_1^2 + \frac{41}{2} (n_{12}v_1) v_2^2 \nonumber \\ &+ \frac{3}{2} (n_{12}v_2) v_2^2 \biggr\} + \frac{G m_2}{r_{12}} (n_{12}v_2) \biggl\{ 2 (n_{12}v_2)^2 - v_2^2 - 2 \frac{G m_2}{r_{12}} \biggr\} + \frac{G^2 m_1 m_2}{r_{12}^2} \biggl\{ \frac{51}{2} (n_{12}v_1) \nonumber \\ &- \frac{97}{2} (n_{12}v_2) - \frac{9}{4} \pi^2 (n_{12}v_{12}) \biggr\} + \frac{G^2 m_1^2}{r_{12}^2} \biggl\{ - \frac{2398}{25} (n_{12}v_1) + \frac{1088}{25} (n_{12}v_2) \nonumber \\ &+ \frac{72}{5} (n_{12}v_{12}) \ln{\left( \frac{r_{12}}{r'_1} \right)} \biggr\} \biggr) \biggr] + o(c^{-7}) \, , \\
        g_{ij}(y_1) &= \delta^{ij} + \frac{2 G m_2}{c^2 r_{12}} \delta^{ij} + \frac{G m_2}{c^4 r_{12}} \biggl[ \delta^{ij} \biggl( - (n_{12}v_2)^2 + \frac{G m_1}{r_{12}} + \frac{G m_2}{r_{12}} \biggr) + 4 v_2^i v_2^j \nonumber \\ &+ n_{12}^i n_{12}^j \biggl( - 8 \frac{G m_1}{r_{12}} + \frac{G m_2}{r_{12}} \biggr) \biggr] + \frac{G m_2}{c^6 r_{12}} \biggl[ \delta^{ij} \biggl( \frac{3}{4} (n_{12}v_2)^4 - v_2^2 (n_{12}v_2)^2 \nonumber \\ &- \frac{G m_2}{r_{12}} (n_{12}v_2)^2 + \frac{G m_1}{r_{12}} \biggl\{ - \frac{71}{4} (n_{12}v_{12})^2 + \frac{47}{4} v_{12}^2 - \frac{59}{5} \frac{G m_1}{r_{12}} + \frac{25}{6} \frac{G m_2}{r_{12}} \biggr\} \biggr) \nonumber \\ &+ v_2^i v_2^j \biggl( 4 v_2^2 - 2 (n_{12}v_2)^2 - 10 \frac{G m_1}{r_{12}} - \frac{G m_2}{r_{12}} \biggr) + 24 \frac{G m_1}{r_{12}} v_1^{(i} v_2^{j)} - 16 \frac{G m_1}{r_{12}} v_1^i v_1^j \nonumber \\ &+ n_{12}^i n_{12}^j \biggl( \frac{G m_1}{r_{12}} \biggl\{ - 16 (n_{12}v_1)^2 + 32 (n_{12}v_1) (n_{12}v_2) - 12 v_{12}^2 + \frac{182}{5} \frac{G m_1}{r_{12}} \biggr\} \nonumber \\ &+ \frac{G m_2}{r_{12}} \biggl\{ -2 (n_{12}v_2)^2 + 3 \frac{G m_1}{r_{12}} + 2 \frac{G m_2}{r_{12}} \biggr\} \biggr) + 40 \frac{G m_1}{r_{12}} (n_{12}v_{12}) \, n_{12}^{(i} v_1^{j)} \nonumber \\ &+ n_{12}^{(i} v_2^{j)} \biggl( \frac{G m_1}{r_{12}} \bigl\{ - 60 (n_{12}v_1) + 36 (n_{12}v_2) \bigr\} + 2 \frac{G m_2}{r_{12}} (n_{12}v_2) \biggr) \biggr] + o(c^{-6}) \, .
	\end{align}
\end{subequations}
Since we are considering only the \textit{conservative} part of the binary dynamics, we did not include in \eqref{g1} the dissipative 2.5PN radiation-reaction terms; these can be found in Eqs.~(7.6) of Ref.~\cite{Bl.al.98}. Notice the occurence at 3PN order of some logarithmic terms, containing two constants $r'_1$ and $r'_2$ (one for each body) that have the dimension of a length. These ultraviolet (UV) regularization parameters come from regularizing the self-field of point particles using the Hadamard regularization of Ref.~\cite{BlFa.01}. The constants $r'_1$ and $r'_2$ are gauge-dependent, as they can be arbitrarily changed by a coordinate transformation of the bulk metric \cite{BlFa.01,Bl.al.10} or by some shifts of the worldlines of the particles \cite{Bl.al.04}. The metric coefficient $g_{00}(y_1)$ also involves a constant $r_0$ that originates from the infrared (IR) regularization of the metric at spatial infinity, as discussed in Ref.~\cite{Bl.al.10}. This arbitrary IR scale should also disappear from the final gauge-invariant results.

We introduce the expression \eqref{g1} of the regularized 3PN metric into the definition \eqref{def_U} of the redshift, and expand in powers of $1/c$, keeping all terms up to $\calO(c^{-8})$. This gives an expression for $U$ as a function of the two masses $m_1$ and $m_2$, the coordinate separation $r_{12}$, and the scalar products $(n_{12}v_1)$, $(n_{12}v_2)$, $(v_1v_1)$, $(v_1v_2)$ and $(v_2v_2)$, as well as the regularization constants $r_0$, $r'_1$ and $r'_2$, in an arbitrary reference frame. The resulting expression is too lengthy to be displayed here.

\subsubsection{Reduction to the center-of-mass frame}

We wish to specialize the previous expression to the \textit{center-of-mass} (CM) frame, which is consistently defined at 3PN order by the vanishing of the center-of-mass integral deduced from the 3PN binary equations of motion \cite{deA.al.01}. This condition yields expressions for the individual positions $\mathbf{y}_a$ and velocities $\mathbf{v}_a$ relatively to the CM in terms of the relative position $\mathbf{y} \equiv \mathbf{y}_1 - \mathbf{y}_2$ and relative velocity $\mathbf{v} \equiv \mathbf{v}_1 - \mathbf{v}_2$ \cite{BlIy.03}. Since these results play an important role in our algebraic manipulations, we recall here the expressions for the functional relationships $\mathbf{y}_a[\mathbf{y},\mathbf{v}]$ in the harmonic gauge that was used to derive the regularized metric \eqref{g1}. Thus,
\begin{subequations}\label{y_a}
	\begin{align}
		\mathbf{y}_1 &= \left[ X_2 + \nu \, (X_1 - X_2) \, \calP \right] \mathbf{y} + \nu \, (X_1 - X_2) \, \calQ \; \mathbf{v}+ o(c^{-6}) \, , \\
		\mathbf{y}_2 &= \left[ - X_1 + \nu \, (X_1 - X_2) \, \calP \right] \mathbf{y} + \nu \, (X_1 - X_2) \, \calQ \; \mathbf{v} + o(c^{-6}) \, ,
	\end{align}
\end{subequations}
where $\nu \equiv m_1 m_2 / m^2$ is the symmetric mass ratio and $X_a \equiv m_a / m$, with $m \equiv m_1 + m_2$ the total mass of the binary. The coefficients $\calP$ and $\calQ$ depend on the parameters $m$ and $\nu$, the separation $r \equiv \vert \mathbf{y} \vert$, the relative velocity squared $v^2 \equiv (vv)$, and the radial velocity $\dot{r} \equiv (nv)$. They explicitly read \cite{BlIy.03}
\begin{subequations}\label{P-Q}
	\begin{align}
		\calP &= \frac{1}{c^2} \, \bigg\{ \frac{v^2}{2} - \frac{Gm}{2r} \bigg\} \nonumber\\ &+ \frac{1}{c^4} \, \bigg\{ v^4 \biggl( \frac{3}{8} - \frac{3}{2} \nu \biggr) + \frac{Gm}{r} v^2 \biggl( \frac{19}{8} + \frac{3}{2} \nu \biggr) + \frac{Gm}{r} \dot{r}^2 \biggl( - \frac{1}{8} + \frac{3}{4} \nu \biggr) + \frac{G^2 m^2}{r^2} \biggl( \frac{7}{4} - \frac{\nu}{2} \biggr) \bigg\} \nonumber\\ &+ \frac{1}{c^6} \, \bigg\{ v^6 \bigg( \frac{5}{16} - \frac{11}{4} \nu + 6 \nu^2 \bigg) +  \frac{Gm}{r} v^4 \bigg( \frac{53}{16} - 7 \nu - \frac{15}{2} \nu^2 \bigg) + \frac{G m}{r} \dot{r}^4 \bigg( \frac{1}{16} - \frac{5}{8} \nu + \frac{21}{16} \nu^2 \bigg) \nonumber \\ &\qquad + \frac{G m}{r} v^2 \dot{r}^2 \bigg( - \frac{5}{16} + \frac{21}{16} \nu - \frac{11}{2} \nu^2 \bigg) + \frac{G^2 m^2}{r^2} v^2 \biggl( \frac{101}{12} - \frac{33}{8} \nu + 3 \nu^2 \biggr) \nonumber \\ &\qquad + \frac{G^2 m^2}{r^2} \dot{r}^2 \biggl( - \frac{7}{3} + \frac{73}{8} \nu + 4 \nu^2 \biggr) + \frac{G^3 m^3}{r^3} \biggl( - \frac{14351}{1260} + \frac{22}{3} \ln{\bigg( \frac{r}{r''_0} \bigg)} + \frac{\nu}{8} - \frac{\nu^2}{2} \biggr) \bigg\} \, , \label{calP} \\
		\calQ &= - \frac{7}{4} \frac{G m \, \dot{r}}{c^4} + \frac{G m \, \dot{r}}{c^6} \, \bigg\{ v^2 \biggl( - \frac{15}{8} + \frac{21}{4} \nu \biggr) + \dot{r}^2 \biggl( \frac{5}{12} - \frac{19}{24} \nu \biggr) - \frac{G m}{r} \biggl( \frac{235}{24} + \frac{21}{4} \nu \biggr) \bigg\} \, .
	\end{align}
\end{subequations}
Again, we did not include the radiation-reaction contributions at 2.5PN order. A logarithmic term contributes at 3PN order in \eqref{calP}; it involves a particular combination $r''_0$ of the gauge constants $r'_1$ and $r'_2$, that is defined by
\beq
	(X_1 - X_2) \ln{r''_0} = X_1^2 \ln{r'_1} - X_2^2 \ln{r'_2} \, .
\eeq

By computing the time derivatives of Eqs.~\eqref{y_a}--\eqref{P-Q} and by applying, where necessary, an iterative order-reduction of all accelerations by means of the CM equations of motion given in Eqs.~(3.9)--(3.10) of Ref.~\cite{BlIy.03}, we obtain expressions analogous to \eqref{y_a}--\eqref{P-Q} for the particle's individual velocities $\mathbf{v}_a$ as functions of the relative variables $\mathbf{y}$ and $\mathbf{v}$. Replacing the positions and velocities by their CM expressions $\mathbf{y}_a[\mathbf{y},\mathbf{v}]$ and $\mathbf{v}_a[\mathbf{y},\mathbf{v}]$ yields 3PN-accurate expressions for the scalar products $(n_{12}v_1)$, $(n_{12}v_2)$, $(v_1v_1)$, $(v_1v_2)$, $(v_2v_2)$ as functions of $r$, $\dot{r}$ and $v^2$. Finally, the CM expression for the redshift $U[r,\dot{r},v^2]$ in standard harmonic coordinates takes the form
\beq\label{U_CM}
	U = 1 + \frac{1}{c^2} \, U_\text{N} + \frac{1}{c^4} \, U_\text{1PN} + \frac{1}{c^6} \, U_\text{2PN} + \frac{1}{c^8} \, U_\text{3PN} + o(c^{-8}) \, , 
\eeq
where the various PN contributions read
\begin{subequations}\label{U_CMbis}
	\begin{align}
		U_\text{N} &= \left( \frac{1}{4} + \frac{\Delta}{4} - \frac{\nu}{2} \right) v^2 + \left( \frac{1}{2} + \frac{\Delta}{2} \right) \frac{G m}{r} \, , \label{U_Newt} \\
		U_\text{1PN} &=  \left( \frac{3}{16} + \frac{3}{16} \Delta - \nu - \frac{5}{8} \Delta \, \nu + \frac{11}{8} \nu^2 \right) v^4 + \left( \frac{5}{4} + \frac{5}{4} \Delta - \frac{\nu}{2} - \nu^2 \right) \frac{G m}{r} v^2 \nonumber \\ &+ \left( \frac{\Delta}{2} - \nu \right) \nu \, \frac{G m}{r} \dot{r}^2 + \left( \frac{1}{4} + \frac{\Delta}{4} - 2 \nu \right) \frac{G^2 m^2}{r^2} \, , \\
		U_\text{2PN} &= \left( \frac{5}{32} + \frac{5}{32} \Delta - \frac{3}{2} \nu - \frac{19}{16} \Delta \, \nu + \frac{157}{32} \nu^2 + \frac{75}{32} \Delta \, \nu^2 - \frac{85}{16} \nu^3 \right) v^6 \nonumber \\ &+ \left( \frac{27}{16} + \frac{27}{16} \Delta - \frac{37}{8} \nu - \frac{13}{4} \Delta \, \nu + \frac{5}{16} \nu^2 - \frac{19}{16} \Delta \, \nu^2 + \frac{11}{2} \nu^3 \right) \frac{G m}{r} v^4 \nonumber \\ &+ \left( - \frac{5}{8} + \frac{7}{8} \Delta - \frac{11}{8} \nu - \frac{23}{8} \Delta \, \nu + 6 \nu^2 \right) \nu \, \frac{G m}{r} \dot{r}^2 v^2 \nonumber \\ &+ \left( \frac{23}{8} + \frac{23}{8} \Delta - 2 \nu - \frac{17}{8} \Delta \, \nu + \frac{59}{8} \nu^2 - \frac{1}{4} \Delta \, \nu^2 - \frac{3}{2} \nu^3 \right) \frac{G^2 m^2}{r^2} v^2 \nonumber \\ &+ \left( \frac{1}{4} + \frac{\Delta}{4} - 5 \nu + \frac{19}{8} \Delta \, \nu - \frac{55}{8} \nu^2 + \frac{5}{4} \Delta \, \nu^2 - \frac{9}{2} \nu^3 \right) \frac{G^2 m^2}{r^2} \dot{r}^2 \nonumber \\ &+ \left( - \frac{3}{8} \Delta + \frac{9}{16} \nu + \frac{9}{16} \Delta \, \nu - \frac{3}{2} \nu^2 \right) \nu \, \frac{G m}{r} \dot{r}^4 + \left( \frac{1}{4} + \frac{\Delta}{4} - \frac{9}{8} \nu - \frac{1}{8} \Delta \, \nu \right) \frac{G^3 m^3}{r^3} \, , \\
		U_\text{3PN} &= \left( \frac{35}{256} + \frac{35}{256} \Delta - 2 \nu - \frac{221}{128} \Delta \, \nu + \frac{705}{64} \nu^2 + \frac{939}{128} \Delta \, \nu^2 - \frac{855}{32} \nu^3 - \frac{665}{64} \Delta \nu^3 +  \frac{3059}{128} \nu^4 \right) v^8 \nonumber \\ &+ \left( \frac{65}{32} + \frac{65}{32} \Delta - \frac{49}{4} \nu - \frac{163}{16} \Delta \, \nu + \frac{169}{8} \nu^2 + \frac{45}{4} \Delta \, \nu^2 + \frac{57}{8} \nu^3 + \frac{145}{16} \Delta \, \nu^3 - \frac{255}{8} \nu^4 \right) \frac{G m}{r} v^6 \nonumber \\ &+ \left( - \frac{61}{32} + \frac{19}{32} \Delta + \frac{131}{32} \nu - \frac{229}{32} \Delta \, \nu + \frac{251}{16} \nu^2 + \frac{67}{4} \Delta \, \nu^2 - \frac{289}{8} \nu^3 \right) \nu \, \frac{G m}{r} \dot{r}^2 v^4 \nonumber \\ &+ \left( \frac{231}{32} + \frac{231}{32} \Delta - \frac{19}{2} \nu - \frac{39}{4} \Delta \, \nu + \frac{297}{16} \nu^2 + \frac{129}{16} \Delta \, \nu^2 - \frac{597}{16} \nu^3 - \Delta \, \nu^3 + \frac{55}{4} \nu^4 \right) \frac{G^2 m^2}{r^2} v^4 \nonumber \\ &+ \left( \frac{15}{32} - \frac{33}{32} \Delta + \frac{9}{2} \Delta \, \nu - 9 \nu^2 - \frac{87}{16} \Delta \, \nu^2 + \frac{111}{8} \nu^3 \right) \nu \, \frac{G m}{r} \dot{r}^4 v^2 \nonumber \\ &+ \left( \frac{3}{8} + \frac{3}{8} \Delta - \frac{323}{32} \nu + \frac{283}{32} \Delta \, \nu - \frac{667}{32} \nu^2 - \frac{617}{32} \Delta \, \nu^2 + \frac{725}{16} \nu^3 - \frac{27}{2} \Delta \, \nu^3 + \frac{161}{4} \nu^4 \right) \frac{G^2 m^2}{r^2} \dot{r}^2 v^2 \nonumber \\ &+ \left( \frac{35}{8} + \frac{35}{8} \Delta + \biggl[ - \frac{39731}{2520} + \frac{71}{64} \pi^2 \biggr] \nu + \biggl[ - \frac{23761}{1260} + \frac{71}{64} \pi^2 \biggr] \Delta \, \nu - \frac{64933}{5040} \nu^2 + \frac{35}{16} \Delta \, \nu^2 \right. \nonumber \\ &\left. \qquad + \; 12 \nu^3 - \frac{1}{2} \Delta \, \nu^3 - 2 \nu^4 - \frac{16}{3} \nu \, \ln{\left( \frac{r}{r_0} \right)} + \frac{11}{3} \, (1 - \Delta + \nu - \Delta \, \nu) \, \nu \, \ln{\left( \frac{r}{r'_1} \right)} \right. \nonumber \\ &\left. \qquad + \; \frac{11}{3} \, (- 1 - \Delta + 3 \nu + \Delta \, \nu) \, \nu \, \ln{\left( \frac{r}{r'_2} \right)} \right) \frac{G^3 m^3}{r^3} v^2 \nonumber \\ &+ \left( \frac{5}{16} \Delta - \frac{15}{32} \nu - \frac{35}{32} \Delta \, \nu + \frac{35}{16} \nu^2 + \frac{5}{8} \Delta \, \nu^2 - \frac{15}{8} \nu^3 \right) \nu \, \frac{G m}{r} \dot{r}^6 \nonumber \\ &+ \left( - \frac{425}{48}  - \frac{73}{6} \Delta + \frac{177}{16} \nu + \frac{\Delta \, \nu}{8} - \frac{2}{3} \nu^2 + \frac{3}{2} \Delta \, \nu^2 - 9 \nu^3 \right) \nu \, \frac{G^2 m^2}{r^2} \dot{r}^4 \nonumber \\ &+ \left( \frac{5}{4} + \frac{5}{4} \Delta - \biggl[ \frac{17917}{420} + \frac{213}{64} \pi^2 \biggr] \nu + \biggl[ \frac{5767}{840} - \frac{213}{64} \pi^2 \biggr] \Delta \, \nu + \frac{122333}{1680} \nu^2 + \frac{7}{16} \Delta \, \nu^2 \right. \nonumber \\ &\left. \qquad - \; 16 \nu^3 + \frac{5}{2} \Delta \, \nu^3 - 14 \nu^4 + 16 \nu \, \ln{\left( \frac{r}{r_0} \right)} + 11 \, (- 1 + \Delta - \nu + \Delta \, \nu) \, \nu \, \ln{\left( \frac{r}{r'_1} \right)} \right. \nonumber \\ &\left. \qquad + \; 11 \, (1 + \Delta - 3 \nu - \Delta \, \nu) \, \nu \, \ln{\left( \frac{r}{r'_2} \right)} \right) \frac{G^3 m^3}{r^3} \dot{r}^2 \nonumber \\ &+ \left( \frac{3}{16} + \frac{3}{16} \Delta - \biggl[ \frac{67853}{5040} + \frac{15}{32} \pi^2 \biggr] \nu - \biggl[ \frac{20141}{5040} + \frac{15}{32} \pi^2 \biggr] \Delta \, \nu - \frac{656}{315} \nu^2 + \frac{16}{3} \nu \ln{\left( \frac{r}{r_0} \right)} \right. \nonumber \\ &\left. \qquad + \; \frac{11}{3} \, (- 1 + \Delta) \, \nu \, \ln{\left( \frac{r}{r'_1} \right)} + \frac{11}{3} \, (1 + \Delta - 2 \nu) \, \nu \, \ln{\left( \frac{r}{r'_2} \right)} \right) \frac{G^4 m^4}{r^4} \, . \label{U_3PN}
	\end{align}
\end{subequations}
Here, $\Delta \equiv (m_2 - m_1) / m = X_2 - X_1 = \sqrt{1-4\nu}$ denotes the reduced mass difference, so that the test-mass limit of particle $1$ corresponds to $\nu \to 0$. Since the redshift \eqref{def_U} is a property of particle $1$, the expressions \eqref{U_CMbis} are not symmetric by exchange $1 \leftrightarrow 2$ of the bodies' labels. The redshift of particle $2$ is simply obtained by setting $\Delta \to - \Delta$ in Eqs.~\eqref{U_CMbis}. As expected, the regularization constants $r_0$, $r'_1$ and $r'_2$ that enter the expression \eqref{g1} of the regularized 3PN metric appear in the CM expression \eqref{U_CM}--\eqref{U_CMbis} for the redshift. In Sec.~\ref{sec:averaging} we will check that the orbital averaging cancels out the dependance on these arbitrary length scales.

\subsection{Redshift variable in alternative coordinates}

In the previous section we obtained an expression for the redshift variable in the standard harmonic (SH) coordinate system, namely the coordinate system in which the 3PN equations of motion were originally derived \cite{BlFa.01,BlIy.03}. These coordinates are such that the equations of motion involve some gauge-dependent logarithmic terms at 3PN order. Importantly, these logarithms prevent the use of the 3PN quasi-Keplerian representation of the binary motion (reviewed in Sec. IV C below), thus impeding the averaging of the redshift over an orbit. Therefore, it is useful to have the expression for the redshift in a modified harmonic (MH) coordinate system, without logarithmic terms in the equations of motion, such as the one used in Refs.~\cite{Ar.al.08,Ar.al2.09}. Alternatively, we shall use ADM-type coordinates, which are also free of such logarithms at 3PN order in the equations of motion. Both the MH coordinates and the ADM coordinates are suitable for a 3PN quasi-Keplerian parametrization of the motion \cite{Me.al2.04}. This will require us to re-express the redshift in terms of the variables $r$, $\dot{r}$ and $v^2$ in these alternative coordinate systems.

\subsubsection{Modified harmonic coordinates}

The trajectories $\mathbf{y}'_a(t)$ of the particles in MH coordinates are related to their counterparts $\mathbf{y}_a(t)$ in SH coordinates by some 3PN shifts $\bm{\xi}_a(t)$ of the worldlines induced by a coordinate transformation in the ``bulk,'' namely $\mathbf{y}'_a = \mathbf{y}_a + \bm{\xi}_a$ \cite{BlFa.01}. Therefore, in the CM frame, the MH coordinate separation $\mathbf{y}'$ is related to the SH coordinate separation $\mathbf{y}$ through $\mathbf{y}' = \mathbf{y} + \bm{\xi}$, where the relative shift $\bm{\xi} \equiv \bm{\xi}_1 - \bm{\xi}_2$ is given by \cite{Ar.al.08}
\beq\label{xi}
	\bm{\xi}_\text{(SH $\to$ MH)} = - \frac{22}{3} \frac{G^3 m^3 \nu}{c^6 r^2} \ln{\left( \frac{r}{r'_0} \right)} \, \mathbf{n} + o(c^{-6}) \, ,
\eeq
with $\mathbf{n} \equiv \mathbf{y} / r$ the unit direction pointing from particle $2$ to particle $1$. Following \cite{Ar.al.08}, we introduced the ``logarithmic barycenter'' $r'_0$ of the constants $r'_1$ and $r'_2$, (not to be confused with the IR constant $r_0$):
\beq
	\ln{r'_0} \equiv X_1 \ln{r'_1} + X_2 \ln{r'_2} \, .
\eeq

The expression $U'[r,\dot{r},v^2]$ for the redshift in MH coordinates can then be deduced from the formula for $U[r,\dot{r},v^2]$ in SH coordinates by means of the \textit{functional} equality $U' = U + \delta_\xi U$, where
\beq\label{dxiU}
	\delta_\xi U = - \frac{\partial U}{\partial r} \, \delta_\xi r - \frac{\partial U}{\partial \dot{r}} \, \delta_\xi \dot{r} - \frac{\partial U}{\partial v^2} \, \delta_\xi v^2 + \calO(\xi^2) \, ,
\eeq
with
\begin{subequations}\label{deltas}
	\begin{align}
		\delta_\xi r &= (n \xi) + \calO(\xi^2) \, , \\
		\delta_\xi \dot{r} &= ( n \dot{\xi}) + \frac{(v \xi)}{r} - \frac{\dot{r}}{r} \, (n \xi) + \calO(\xi^2) \, , \\
		\delta_\xi v^2 &= 2 (v \dot{\xi}) + \calO(\xi^2) \, .
	\end{align}
\end{subequations}
Since the relative shift \eqref{xi} comes at 3PN order, the nonlinear terms $\calO(\xi^2)$ in Eqs.~\eqref{dxiU} and \eqref{deltas} contribute at leading 6PN order, and can thus be neglected. Plugging the expression \eqref{xi} into Eqs.~\eqref{deltas}, we find the explicit expressions
\begin{subequations}\label{deltas_bis}
	\begin{align}
		(\delta_\xi r)_\text{(SH $\to$ MH)} &= - \frac{22}{3} \frac{G^3 m^3 \nu}{c^6 r^2} \ln{\left( \frac{r}{r'_0} \right)} , \\
		(\delta_\xi \dot{r})_\text{(SH $\to$ MH)} &= - \frac{22}{3} \frac{G^3 m^3 \nu}{c^6 r^3} \left\{ \dot{r} - 2 \dot{r} \ln{\left( \frac{r}{r'_0} \right)} \right\} , \\
		(\delta_\xi v^2)_\text{(SH $\to$ MH)} &= - \frac{44}{3} \frac{G^3 m^3 \nu}{c^6 r^3} \left\{ \dot{r}^2 + \left( v^2 - 3 \dot{r}^2 \right) \ln{\left( \frac{r}{r'_0} \right)} \right\} .
	\end{align}
\end{subequations}

In order to compute the change $\delta_\xi U$ in the redshift induced by the relative shift \eqref{xi}, we only require the Newtonian expression for $U[r,\dot{r},v^2]$, which is given by Eq.~\eqref{U_Newt}. Combined with \eqref{dxiU} and \eqref{deltas_bis}, this gives
\begin{align}\label{dxiU_MH}
	(\delta_\xi &U)_\text{(SH $\to$ MH)} = \frac{11}{3} \frac{G^3 m^3 \nu}{c^8 r^3} \bigg\{ \left( 1 + \Delta - 2\nu \right) \dot{r}^2 + \left( \left( 1 + \Delta \right) \left( v^2 - 3 \dot{r}^2 \right) - \frac{2Gm}{r} \right) \nu \, \ln{\left( \frac{r}{r'_1} \right)} \nonumber \\ &\quad + \left( \left( 1 + \Delta - 3 \nu - \Delta \, \nu \right) \left( v^2 - 3 \dot{r}^2 \right) - \frac{Gm}{r} \left( 1 + \Delta - 2 \nu \right)  \right) \ln{\left( \frac{r}{r'_2} \right)} \biggr\} + o(c^{-8}) \, .
\end{align}
Adding the above shift to the formula \eqref{U_CM}--\eqref{U_CMbis} for the redshift in SH coordinates yields the expression for the redshift in MH coordinates. Since $U' = U + \delta_\xi U$ is a functional equality, the resulting MH redshift is expressed as a function of the ``dummy'' variables $r$, $\dot{r}$ and $v^2$.

Adding together Eqs.~\eqref{U_3PN} and \eqref{dxiU_MH}, we find that the UV regularization constant $r'_2$ disappears from the expression for $U'[r,\dot{r},v^2]$ in MH coordinates. However the UV and IR constants $r'_1$ and $r_0$ remain and enter the result through the logarithmic contributions
\beq\label{Uln}
	{\big[U'\big]}_\text{log} = \frac{11}{3} \frac{G^3 m^3 \nu}{c^8 r^3} \left( v^2 - 3 \dot{r}^2 - \frac{G m}{r} \right) \left\{ \left( 1 - \Delta + 2 \nu \right) \ln{\left( \frac{r}{r'_1} \right)} - \frac{16}{11} \ln{\left( \frac{r}{r_0} \right)} \right\} .
\eeq
For circular orbits, such that $\dot{r} = 0$ and $v^2 = Gm / r + \calO(c^{-2})$, these logarithmic contributions cancel out. We will see that the factor $(v^2 - 3 \dot{r}^2 - Gm/r)/r^3$ vanishes when averaged over one radial period, such that the constants $r_0$ and $r'_1$ will cancel out from the final, gauge-invariant result for the orbital-averaged redshift, as expected.

\subsubsection{ADM-type coordinates}\label{subsec:ADM}

Similarly, the individual trajectories $\mathbf{y}'_a(t)$ of the particles in ADM coordinates are related to the trajectories $\mathbf{y}_a(t)$ in SH coordinates by some shifts $\bm{\xi}_a(t)$ of the worldlines: $\mathbf{y}'_a = \mathbf{y}_a + \bm{\xi}_a$ \cite{Da.al2.01,deA.al.01}. In the CM frame, the ADM coordinate separation $\mathbf{y}'$ is related to the SH coordinate separation $\mathbf{y}$ through $\mathbf{y}' = \mathbf{y} + \bm{\xi}$, where the relative shift $\bm{\xi} = \bm{\xi}_1 - \bm{\xi}_2$ reads \cite{BlIy.03,Ar.al.08}
\begin{align}\label{xiADM}
	\bm{\xi}&_\text{(SH $\to$ ADM)} = \frac{G m}{c^4} \left\{ \left[ - \frac{5}{8} \nu \, v^2 + \frac{\nu}{8} \, \dot{r}^2 - \frac{Gm}{r} \biggl( \frac{1}{4} + 3 \nu \biggr) \right] \mathbf{n} + \frac{9}{4} \nu \, \dot{r} \, \mathbf{v} \right\} \nonumber \\ &+ \frac{G m \nu}{c^6} \, \Biggl\{ \left[ v^4 \biggl( - \frac{1}{2} + \frac{11}{8} \nu \biggr) + \dot{r}^2 v^2 \biggl( \frac{5}{16} - \frac{15}{16} \nu \biggr) + \dot{r}^4 \biggl( - \frac{1}{16} + \frac{5}{16} \nu \biggr) - \frac{Gm}{r} v^2 \biggl( \frac{451}{48} + \frac{3}{8} \nu \biggr) \right. \nonumber \\ &\qquad\qquad\quad + \left. \frac{Gm}{r} \dot{r}^2 \biggl( \frac{161}{48} - \frac{5}{2} \nu \biggr) + \frac{G^2 m^2}{r^2} \biggl( \frac{2773}{280} + \frac{21}{32} \pi^2 - \frac{22}{3} \ln{\biggl( \frac{r}{r'_0} \biggr)} \biggr) \right] \mathbf{n} \nonumber \nonumber \\ &\qquad\quad\, + \left[ v^2 \biggl( \frac{17}{8} - \frac{21}{4} \nu \biggr) + \dot{r}^2 \biggl( - \frac{5}{12} + \frac{29}{24} \, \nu \biggr) + \frac{Gm}{r} \biggl( \frac{43}{3} + 5 \nu \biggr) \right] \dot{r} \, \mathbf{v} \Biggr\} + o(c^{-6}) \, ,
\end{align}
from which the authors of Ref.~\cite{Ar.al.08} deduced, using Eqs.~\eqref{deltas}, the transformation of variables that we need to compute the redshift in ADM coordinates:\footnote{The remainder $\calO(\xi^2)$ in Eqs.~\eqref{dxiU} and \eqref{deltas} is of order 4PN, which is still negligible in the transformation to ADM coordinates.}
\begin{subequations}\label{delta_bis}
	\begin{align}
		(\delta_\xi &r)_\text{(SH $\to$ ADM)} = \frac{Gm}{c^4} \left\{ \frac{5}{8} \nu \, v^2 - \frac{19}{8} \nu \, \dot{r}^2 + \frac{Gm}{r} \biggl( \frac{1}{4} + 3\nu \biggr) \right\} \nonumber \\ &+ \frac{G m \nu}{c^6} \left\{ v^4 \biggl( \frac{1}{2} - \frac{11}{8} \nu \biggr) + \dot{r}^2 v^2 \biggl( - \frac{39}{16} + \frac{99}{16} \nu \biggr) + \dot{r}^4 \bigg( \frac{23}{48} - \frac{73}{48} \nu \biggr) + \frac{Gm}{r} v^2 \biggl( \frac{451}{48} + \frac{3}{8} \nu \biggr) \right. \nonumber \\ &\qquad - \left. \frac{Gm}{r} \dot{r}^2 \biggl( \frac{283}{16} + \frac{5}{2} \nu \biggr) + \frac{G^2 m^2}{r^2} \biggl( - \frac{2773}{280} + \frac{22}{3} \ln{\biggl( \frac{r}{r'_0} \biggr)} - \frac{21}{32} \pi^2 \biggr) \right\} , \\
		(\delta_\xi &\dot{r})_\text{(SH $\to$ ADM)} = \frac{Gm}{c^4 r} \dot{r} \left\{ - \frac{19}{4} \nu \, v^2 + \frac{19}{4} \nu \, \dot{r}^2 + \frac{Gm}{r} \biggl( - \frac{1}{4} + \frac{\nu}{2} \biggr) \right\} \nonumber \\ &+ \frac{G m \nu}{c^6 r} \dot{r} \left\{ v^4 \biggl( - \frac{39}{8} + \frac{99}{8} \nu \biggr) + \dot{r}^2 v^2 \biggl( \frac{163}{24} - \frac{443}{24} \nu \biggr) + \dot{r}^4 \biggl( - \frac{23}{12} + \frac{73}{12} \nu \biggr) \right. \nonumber \\ &\qquad - \frac{Gm}{r} v^2 \biggl( \frac{1603}{48} + \frac{17}{4} \nu \biggr) + \frac{Gm}{r} \dot{r}^2 \biggl( \frac{1777}{48} + \frac{131}{24} \nu \biggr) \nonumber \\ &\qquad + \left. \frac{G^2 m^2}{r^2} \biggl( \frac{3121}{105} - \frac{44}{3} \ln{\biggl( \frac{r}{r'_0} \biggr)} + \frac{21}{16} \pi^2 - \frac{11}{4} \nu \biggr) \right\} , \\
		(\delta_\xi &v^2)_\text{(SH $\to$ ADM)} = \frac{Gm}{c^4 r} \left\{ - \frac{13}{4} \nu \, v^4 +  \frac{5}{2} \nu \, \dot{r}^2 v^2 + \frac{3}{4} \nu \, \dot{r}^4 + \frac{Gm}{r} v^2 \biggl( \frac{1}{2} + \frac{21}{2} \nu \biggr) - \frac{Gm}{r} \dot{r}^2 \biggl( 1 + \frac{19}{2} \nu \biggr) \right\} \nonumber \\ &+ \frac{G m \nu }{c^6 r} \left\{ v^6 \biggl( - \frac{13}{4} + \frac{31}{4} \nu \biggr) + \dot{r}^2 v^4 \biggl( \frac{31}{8} - \frac{75}{8} \nu \biggr) + \dot{r}^4 v^2 \biggl( - \frac{3}{2} \nu \biggr) + \dot{r}^6 \biggl( - \frac{5}{8} + \frac{25}{8} \nu \biggr) \right. \nonumber \\ &\qquad - \frac{Gm}{r} v^4 \biggl( \frac{9}{8} + \frac{25}{4} \nu \biggr) + \frac{Gm}{r} \dot{r}^2 v^2 \biggl( - \frac{131}{8} + \frac{121}{4} \nu \biggr) + \frac{Gm}{r} \dot{r}^4 \biggl( \frac{99}{4} - \frac{259}{12} \nu \biggr) \nonumber \\ &\qquad + \frac{G^2 m^2}{r^2} v^2 \biggl( - \frac{3839}{420} + \frac{44}{3} \ln{\biggl( \frac{r}{r'_0} \biggr)} - \frac{21}{16} \pi^2 + \nu \biggr) \nonumber \\ &\qquad + \left. \frac{G^2 m^2}{r^2} \dot{r}^2 \biggl( \frac{28807}{420} - 44 \ln{\biggl( \frac{r}{r'_0} \biggr)} + \frac{63}{16} \pi^2 - \frac{13}{2} \nu \biggr) \right\} .
	\end{align}
\end{subequations}

The expression $U'[r,\dot{r},v^2]$ for the redshift in ADM coordinates can then be deduced from the result \eqref{U_CM}--\eqref{U_CMbis} in SH coordinates via the functional equality $U' = U + \delta_\xi U$. Using the expressions \eqref{dxiU} and \eqref{delta_bis}, the SH redshift is found to be modified by 2PN and 3PN corrections that read
\begin{align}\label{dxiU_ADM}
	(\delta_\xi &U)_\text{(SH $\to$ ADM)} = \frac{G m}{c^6 r} \, \biggl\{ (1 + \Delta - 2\nu) \, \nu \, \biggl( - \frac{13}{16} v^4 + \frac{5}{8} v^2 \dot{r}^2 + \frac{3}{16} \dot{r}^4 \biggr) \nonumber \\ &\qquad + \frac{Gm}{r} v^2 \biggl( \frac{1}{8} + \frac{\Delta}{8} + \frac{33}{16} \nu + \frac{37}{16} \Delta \, \nu - \frac{21}{4} \nu^2 \biggr) \nonumber \\ &\qquad - \frac{Gm}{r} \dot{r}^2 \biggl( \frac{1}{4} + \frac{\Delta}{4} + \frac{11}{16} \nu + \frac{19}{16} \Delta \, \nu - \frac{19}{4} \nu^2 \biggr) - \frac{G^2 m^2}{r^2} \biggl( \frac{1}{8} + \frac{\Delta}{8} + \frac{3}{2} \nu + \frac{3}{2} \Delta \, \nu \biggr) \biggl\} \nonumber \\ &+ \frac{G m \nu}{c^8 r} \, \biggl\{ - v^6 \biggl( \frac{65}{32} + \frac{65}{32} \Delta - \frac{161}{16} \nu - 6 \Delta \, \nu + \frac{205}{16} \nu^2 \biggr) \nonumber \\ &\qquad + v^4 \dot{r}^2 \biggl( \frac{61}{32} + \frac{61}{32} \Delta - \frac{297}{32} \nu - \frac{175}{32} \Delta \, \nu + \frac{185}{16} \nu^2 \biggr) \nonumber \\ &\qquad + v^2 \dot{r}^4 \biggl( \frac{9}{32} + \frac{9}{32} \Delta - \frac{15}{8} \nu - \frac{21}{16} \Delta \, \nu + \frac{45}{16} \nu^2 \biggr) \nonumber \\ &\qquad - \dot{r}^6 \biggl( \frac{5}{32} + \frac{5}{32} \Delta - \frac{35}{32} \nu - \frac{25}{32} \Delta \, \nu + \frac{25}{16} \nu^2 \biggr) \nonumber \\ &\qquad + \frac{Gm}{r} \dot{r}^4 \biggl( \frac{661}{96} + \frac{661}{96} \Delta - \frac{1669}{96} \nu + \frac{125}{96} \Delta \, \nu - \frac{11}{6} \nu^2 \biggr) \biggl\} \nonumber \\ &+ \frac{G^2 m^2}{c^8 r^2} \, \biggl\{ v^4 \biggl( \frac{3}{16} + \frac{3}{16} \Delta - \frac{39}{16} \nu - \frac{33}{16} \Delta \, \nu - 18 \nu^2 - 14 \Delta \, \nu^2 + \frac{287}{8} \nu^3 \biggr) \nonumber \\ &\qquad + v^2 \dot{r}^2 \biggl( - \frac{3}{8} - \frac{3}{8} \Delta + \frac{53}{32} \nu + \frac{29}{32} \Delta \, \nu + \frac{847}{32} \nu^2 + \frac{361}{32} \Delta \, \nu^2 - 36 \nu^3 \biggr) \nonumber \\ &\qquad + \frac{Gm}{r} v^2 \biggl( \frac{5}{16} + \frac{5}{16} \Delta + \biggl[ \frac{2189}{1120} - \frac{21}{64} \pi^2 \biggr] \nu + \biggl[ \frac{2329}{1120} - \frac{21}{64} \pi^2 \biggr] \Delta \, \nu + \biggl[ \frac{5263}{1680} + \frac{21}{32} \pi^2 \biggr] \nu^2 \nonumber \\&\qquad\qquad + \frac{\Delta \, \nu^2}{16} - 8 \nu^3 + \frac{11}{3} \, (1 + \Delta) \, \nu^2 \ln{\bigg( \frac{r}{r'_1} \biggr)} + \frac{11}{3} \, (1 + \Delta - 3 \nu - \Delta \, \nu) \, \nu \ln{\bigg( \frac{r}{r'_2} \biggr)} \biggr) \nonumber \\ &\qquad - \frac{Gm}{r} \dot{r}^2 \biggl( \frac{5}{4} + \frac{5}{4} \Delta - \biggl[ \frac{53099}{3360} + \frac{63}{64} \pi^2 \biggr] \nu - \biggl[ \frac{50159}{3360} + \frac{63}{64} \pi^2 \biggr] \Delta \, \nu + \biggl[ \frac{15821}{420} + \frac{63}{32} \pi^2 \biggr] \nu^2 \nonumber \\&\qquad\qquad + \frac{11}{8} \Delta \, \nu^2 - \frac{59}{4} \nu^3 + 11 \, (1 + \Delta) \, \nu^2 \ln{\bigg( \frac{r}{r'_1} \biggr)} + 11 \, (1 + \Delta - 3 \nu - \Delta \, \nu) \, \nu \ln{\bigg( \frac{r}{r'_2} \biggr)} \biggr) \nonumber \\ &\qquad + \frac{G^2 m^2}{r^2} \biggl( - \frac{1}{8} - \frac{\Delta}{8} + \biggl[ \frac{2493}{560} + \frac{21}{64} \pi^2 \biggr] \nu + \biggl[ \frac{1933}{560} + \frac{21}{64} \pi^2 \biggr] \Delta \, \nu + 12 \nu^2 \nonumber \\&\qquad\qquad - \frac{22}{3} \, \nu^2 \ln{\bigg( \frac{r}{r'_1} \biggr)} - \frac{11}{3} \, (1 + \Delta - 2\nu) \, \nu \ln{\bigg( \frac{r}{r'_2} \biggr)} \biggr) \biggr\} + o(c^{-8}) \, .
\end{align}
Although the additional contribution \eqref{dxiU_ADM} in ADM coordinates is more involved than its counterpart \eqref{dxiU_MH} in MH coordinates, they share the same logarithmic terms. Thus, adding Eqs.~\eqref{U_3PN} and \eqref{dxiU_ADM} we find that the UV regularization constant $r'_2$ disappears from the expression for $U'[r,\dot{r},v^2]$ in ADM coordinates, while the constants $r_0$ and $r'_1$ remain and enter the final result through the logarithmic terms \eqref{Uln}.

\subsection{The generalized quasi-Keplerian representation}\label{sec:gen_QK}

Before we discuss the orbital averaging of the redshift in Sec. IV D, we must summarize the 3PN generalized quasi-Keplerian (QK) representation of the motion of Memmesheimer \textit{et al.} \cite{Me.al2.04}. Indeed, since averaging over one radial period is most conveniently performed using an explicit solution of the equations of motion, the generalized QK representation is an essential input for our 3PN calculation. The QK representation was originally introduced by Damour and Deruelle \cite{DaDe.85} to account for the leading-order 1PN general relativistic effects in the timing formula of the Hulse-Taylor binary pulsar. It was later extended at 2PN order in Refs.~\cite{DaSc.88,ScWe.93,We.95}, in ADM coordinates, and more recently at 3PN order \cite{Me.al2.04} in both ADM and harmonic coordinates.

We first introduce the mean anomaly
\beq\label{l_t}
	\ell \equiv \Omega_r \, (t - t_\text{per}) \, ,
\eeq
where $t_\text{per}$ is the coordinate time at a periastron passage and $\Omega_r = 2 \pi / T_r$ is the radial frequency (also known as the mean motion $n$), i.e., the frequency associated with the periodicity $T_r$ of the radial motion. The mean anomaly simply maps one radial period $t \in [t_\text{per},t_\text{per}+T_r)$ to the trigonometric interval $\ell \in [0,2\pi)$. We then adopt a parametric description of the binary's motion in polar coordinates, in the CM frame, in terms of the eccentric anomaly $u \in [0,2\pi)$. At 3PN order, this parametrization reads
\begin{subequations}\label{paramQK}
	\begin{align}
		r(u) &= a_r \left( 1 - e_r \cos{u} \right) , \label{r_u} \\
		\ell(u) &= u - e_t \sin{u} + f_t \, \sin{V} + g_t \left( V - u \right) + i_t \sin{2V} + h_t \sin{3V} \, , \label{l_u} \\
		\phi(u) &= \phi_\text{per} + K \left( V + f_\phi \sin{2V} + g_\phi \sin{3V} + i_\phi \sin{4V} + h_\phi \sin{5V} \right) ,
	\end{align}
\end{subequations}
where $\phi_\text{per}$ is the value of the orbital phase when $t = t_\text{per}$, at a periastron passage, $K \equiv 1 + k$ is the fractional angle of advance of the periastron per orbital revolution, such that the angle of return to periastron is given by $\Phi = 2\pi K$ (equivalent to $\Delta \Phi = 2\pi \, k$), and the true anomaly $V$ is defined by
\beq\label{V_u}
	V(u) = u + 2 \arctan{\left(\frac{\beta_\phi \sin{u}}{1 - \beta_\phi \cos{u}}\right)} \, ,
\eeq
with $\beta_\phi \equiv \left[ 1 - (1-e_\phi^2)^{1/2} \right] / e_\phi$. Equations \eqref{l_t}--\eqref{V_u} provide a 3PN-accurate generalization of the usual Keplerian representation of the Newtonian motion.\footnote{In the Newtonian limit, $a_r$ is the semi-major axis, the three eccentricities coincide ($e_t = e_r = e_\phi \equiv e$), an eccentric orbit does not precess ($K = 1$), and $f_t = g_t = i_t = h_t = f_\phi = g_\phi = i_\phi = h_\phi = 0$.}

The previous generalized QK representation is complete only once the orbital elements $\Omega_r$, $K$, $a_r$, $e_t$, $e_r$, $e_\phi$, $f_t$, $g_t$, $i_t$, $h_t$, $f_\phi$, $g_\phi$, $i_\phi$ and $h_\phi$ have been related to the first integrals of the motion, namely the binding energy $E$ and the orbital angular momentum $J$, both per reduced mass $\mu = m_1 m_2 / m$. Following Ref.~\cite{Ar.al.08}, we shall instead make use of the convenient, dimensionless, coordinate-invariant quantities\footnote{For circular orbits, we have the well-known Newtonian limits $\varepsilon \sim v^2 / c^2 \sim G m / (r c^2)$ and $j \sim 1$.}
\beq
	\varepsilon \equiv - \frac{2 E}{c^2} \, , \qquad j \equiv - \frac{2 E J^2}{(G m)^2} \, ,
\eeq
such that $\varepsilon > 0$ and $j > 0$ for a generic bound eccentric orbit (since $E < 0$ for such orbits). Notice the PN scalings $\varepsilon = \calO(c^{-2})$ and $j = \calO(c^0)$. Therefore, we shall consider expansions in powers of the PN parameter $\varepsilon$, with coefficients depending on $j$ and $\nu$. In ADM coordinates, the 3PN-accurate expressions for the orbital elements read \cite{Me.al2.04,Ar.al.08}
\begin{subequations}\label{orb_elts_ADM}
	\begin{align}
		\Omega_r^\text{ADM} &= \frac{\varepsilon^{3/2} c^3}{G m} \bigg\{ 1 + \frac{\varepsilon}{8} \bigl[ - 15 + \nu \bigr] + \frac{\varepsilon^{2}}{128} \biggl[ 555 + 30 \nu + 11 \nu^2 + \frac{192}{j^{1/2}} \bigl( - 5 + 2 \nu \bigr) \biggr] \nonumber \\ &\qquad + \frac{\varepsilon^{3}}{3072} \biggl[ - 29385 - 4995 \nu - 315 \nu^2 + 135 \nu^3 + \frac{5760}{j^{1/2}} \bigl( 17 - 9 \nu + 2 \nu^2 \bigr) \nonumber \\ &\qquad\qquad\qquad\! - \frac{16}{j^{3/2}} \bigl( 10080 - 13952 \nu + 123 \pi^2 \nu + 1440 \nu^2 \bigr) \biggr] + o(\varepsilon^3) \bigg\} \, , \label{n} \\
		K^\text{ADM} &= 1 + \frac{3\varepsilon}{j} + \frac{\varepsilon^2}{4} \biggl[ \frac{3}{j} \bigl( - 5 + 2 \nu \bigr) + \frac{15}{j^2} \bigl( 7 - 2 \nu \bigr) \biggr] \nonumber \\ &\qquad + \frac{\varepsilon^{3}}{128} \biggl[ \frac{24}{j} \bigl( 5 - 5 \nu + 4 \nu^2 \bigr) - \frac{1}{j^2} \bigl( 10080 - 13952 \nu + 123 \pi^2 \nu + 1440 \nu^2 \bigr) \nonumber \\ &\qquad\qquad\quad\; + \frac{5}{j^3} \bigl( 7392 - 8000 \nu + 123 \pi^2 \nu + 336 \nu^2 \bigr) \biggr] + o(\varepsilon^3) \, , \label{K} \\
		a_r^\text{ADM} &= \frac{G m}{c^2 \varepsilon} \biggl\{ 1 + \frac{\varepsilon}{4} \bigl[ - 7 + \nu \bigr] + \frac{\varepsilon^2}{16} \bigg[ 1 + 10 \nu + \nu^2 + \frac{1}{j} \bigl( - 68 + 44 \nu \bigr) \bigg] \nonumber \\ &\qquad + \frac{\varepsilon^{3}}{192} \biggl[ 3 - 9 \nu - 6 \nu^2 + 3 \nu^3 + \frac{1}{j} \bigl( 864 - 2212 \nu - 3 \pi^2 \nu + 432 \nu^2 \bigr) \nonumber \\ &\qquad\qquad\quad\; - \frac{1}{j^2} \bigl( 6432 - 13488 \nu + 240 \pi^2 \nu + 768 \nu^2 \bigr) \biggr] + o(\varepsilon^3) \biggr\} \, , \label{a_r} \\
		e_t^\text{ADM} &= \Biggl\{ 1 - j + \frac{\varepsilon}{4} \bigl[ - 8 + 8 \nu + j \bigl( 17 - 7 \nu \bigr) \bigr] \nonumber \\ &\qquad + \frac{\varepsilon^2}{8} \biggl[ 8 + 4 \nu + 20 \nu^2 - \frac{24}{j^{1/2}} \bigl( 5 - 2 \nu \bigr) + 24 j^{1/2} \bigl( 5 - 2 \nu \bigr) \nonumber \\ &\qquad\qquad\quad - j \bigl( 112 - 47 \nu + 16 \nu^2 \bigr) + \frac{4}{j} \bigl( 17 - 11 \nu \bigr) \biggr] \nonumber\\ &\qquad + \frac{\varepsilon^{3}}{192} \biggl[ 24 \bigl( - 2 + 5 \nu \bigr) \bigl( - 23 + 10 \nu + 4 \nu^2 \bigr) + 15 j \bigl( 528 - 200 \nu + 77 \nu^2 - 24 \nu^3 \bigr) \nonumber \\ &\qquad\qquad\quad\; - 72 j^{1/2} \bigl( 265 - 193 \nu + 46 \nu^2 \bigr) - \frac{2}{j} \bigl( 6732 - 12508 \nu + 117 \pi^2 \nu + 2004 \nu^2 \bigr) \nonumber \\ &\qquad\qquad\quad\; + \frac{2}{j^{1/2}} \bigl( 16380 - 19964 \nu + 123 \pi^2 \nu + 3240 \nu^2 \bigr) \nonumber \\ &\qquad\qquad\quad\; - \frac{2}{j^{3/2}} \bigl( 10080 - 13952 \nu + 123 \pi^2 \nu + 1440 \nu^2 \bigr) \nonumber \\ &\qquad\qquad\quad\; + \frac{96}{j^2} \bigl( 134 - 281 \nu + 5 \pi^2 \nu + 16 \nu^2 \bigr) \biggr] + o(\varepsilon^3) \Biggr\}^{1/2} , \label{e_t_ADM} \\
		e_r^\text{ADM} &= \Biggl\{ 1 - j + \frac{\varepsilon}{4} \bigl[ 24 - 4 \nu + 5 j \bigl( - 3 + \nu \bigr) \bigr] \nonumber \\ &\qquad	+ \frac{\varepsilon^2}{8} \biggl[ 52 + 2 \nu + 2 \nu^2 - j \bigl( 80 - 55 \nu + 4 \nu^2 \bigr) + \frac{8}{j} \bigl( 17 - 11 \nu \bigr) \biggr] \nonumber \\ &\qquad + \frac{\varepsilon^3}{192} \biggl[ - 768 - 344 \nu - 6 \pi^2 \nu - 216 \nu^2 + 3 j \bigl( - 1488 + 1556 \nu - 319 \nu^2 + 4 \nu^3 \bigr) \nonumber \\ &\qquad\qquad\quad\; - \frac{4}{j} \bigl( 588 - 8212 \nu + 177 \pi^2 \nu + 480 \nu^2 \bigr) \nonumber \\ &\qquad\qquad\quad\; + \frac{192}{j^2} \bigl( 134 - 281 \nu + 5 \pi^2 \nu + 16 \nu^2 \bigr) \biggr] + o(\varepsilon^3) \Biggr\}^{1/2} , \label{e_r_ADM} \\
		e_\phi^\text{ADM} &= \Biggl\{ 1 - j + \frac{\varepsilon}{4} \bigl[ 24 + j \bigl( - 15 + \nu \bigr) \bigr] \nonumber \\ &\qquad + \frac{\varepsilon^2}{16} \biggl[ - 32 + 176 \nu + 18 \nu^2 - j \bigl( 160 - 30 \nu + 3 \nu^2 \bigr) + \frac{1}{j} \bigl( 408 - 232 \nu - 15 \nu^2 \bigr) \biggr] \nonumber \\ &\qquad + \frac{\varepsilon^{3}}{384} \biggl[ - 16032 + 2764 \nu + 3 \pi^2 \nu + 4536 \nu^2 + 234 \nu^3 - 36 j \bigl( 248 - 80 \nu + 13 \nu^2 + \nu^3 \bigr) \nonumber \\ &\qquad\qquad\quad\; - \frac{6}{j} \bigl( 2456 - 26860 \nu + 581 \pi^2 \nu + 2689 \nu^2 + 10 \nu^3 \bigr) \nonumber \\ &\qquad\qquad\quad\; + \frac{3}{j^2} \bigl( 27776 - 65436 \nu + 1325 \pi^2 \nu + 3440 \nu^2 - 70 \nu^3
\bigr) \biggr] + o(\varepsilon^3) \Biggr\}^{1/2} \! , \label{e_phi_ADM} \\
		f_t^\text{ADM} &= - \frac{\varepsilon^2}{8} \, \frac{\sqrt{1-j}}{j^{1/2}} \, \nu \, \bigl( 4 + \nu \bigr) \nonumber \\ &\quad + \frac{\varepsilon^{3}}{64} \, \frac{j^{1/2}}{\sqrt{1-j}} \biggl[ \nu \, \bigl( - 64 - 4 \nu + 23 \nu^2 \bigr) + \frac{1}{j^2} \bigg( 576 - \frac{4148}{3} \, \nu + \pi^2 \nu + 200 \nu^2 + 11 \nu^3 \bigg) \nonumber \\ &\qquad\qquad\qquad\quad\; + \frac{1}{j} \biggl ( - 576 + \frac{4232}{3} \, \nu - \pi^2 \nu - 209 \nu^2 - 35 \nu^3 \biggr) \biggr] + o(\varepsilon^3) \, , \\
		g_t^\text{ADM} &= \frac{3\varepsilon^2}{2} \, \biggl( \frac{5 - 2 \nu}{j^{1/2}} \biggr) + \frac{\varepsilon^{3}}{192} \, \biggl[ \frac{1}{j^{3/2}} \bigl( 10080 - 13952 \nu + 123 \pi^2 \nu + 1440 \nu^2 \bigr) \nonumber \\ &\qquad\qquad\qquad\qquad\qquad\quad\; + \frac{1}{j^{1/2}} \bigl( - 3420 + 1980 \nu - 648 \nu^2 \bigr) \biggr] + o(\varepsilon^3) \, , \\
		i_t^\mathrm{ADM} &= \frac{\varepsilon^{3}}{32} \, \frac{1 - j}{j^{3/2}} \, \nu \, \bigl( 23 + 12 \nu + 6 \nu^2 \bigr) + o(\varepsilon^3) \, , \\
		h_t^\mathrm{ADM} &= \frac{13\varepsilon^{3}}{192} \, \biggl( \frac{1-j}{j} \biggr)^{3/2} \nu^3 + o(\varepsilon^3) \, , \\
		f_\phi^\text{ADM} &= \frac{\varepsilon^{2}}{8} \, \frac{1-j}{j^2} \, \nu \, \bigl( 1 - 3 \nu \bigr) \nonumber \\ &\quad + \frac{\varepsilon^{3}}{256} \biggl[ \frac{4\nu}{j} \bigl( - 11 - 40 \nu + 24 \nu^2 \bigr) + \frac{1}{j^2} \bigl( - 256 + 1192 \nu - 49 \pi^2 \nu + 336 \nu^2 - 80 \nu^3 \bigr) \nonumber \\ &\qquad\qquad\; + \frac{1}{j^3} \bigl( 256 - 1076 \nu + 49 \pi^2 \nu - 384 \nu^2 - 40 \nu^3 \bigr) \biggr] + o(\varepsilon^3) \, , \\
		g_\phi^\text{ADM} &= - \frac{3\varepsilon^2}{32} \, \frac{\nu^2}{j^2} \, (1 -j)^{3/2} \nonumber \\ &\quad - \frac{\varepsilon^3}{256} \, \frac{\sqrt{1-j}}{j} \, \nu \, \biggl[ \nu \bigl( 9 - 26 \nu \bigr) + \frac{1}{j} \biggl(
\frac{220}{3} + \pi^2 + 104 \nu + 50 \nu^2 \biggr) \nonumber \\ &\qquad\qquad\qquad\qquad\quad - \frac{1}{j^2} \biggl( \frac{220}{3} + \pi^2 + 32 \nu + 15 \nu^2 \biggr) \biggr] + o(\varepsilon^3) \, , \label{g_phi} \\
		i_\phi^\mathrm{ADM} &= \frac{\varepsilon^{3}}{128} \, \frac{(1-j)^2}{j^3} \, \nu \, \bigl( 5 + 28 \nu + 10 \nu^2 \bigr) + o(\varepsilon^3) \, , \\
		h_\phi^\mathrm{ADM} &= \frac{5\varepsilon^3}{256} \, \frac{\nu^3}{j^3} \, (1-j)^{5/2} + o(\varepsilon^3) \, .
	\end{align}
\end{subequations}
The eccentricities $e_t^\text{ADM}$, $e_r^\text{ADM}$ and $e_\phi^\text{ADM}$ are all such that $j = 1 - e^2 + \calO(c^{-2})$ at Newtonian order; they start differing from each other at leading 1PN order.

The expressions \eqref{orb_elts_ADM} are specific to the ADM coordinates. Before we give the corresponding expressions in MH coordinates, let us recall an important point related to the use of gauge-invariant variables. As shown in Ref.~\cite{DaSc.88}, the functional forms of $\Omega_r = 2\pi / T_r$ and $K = \Phi / (2\pi)$ as functions of gauge-invariant variables like $\varepsilon$ and $j$ are identical in different coordinate systems. In particular we have the exact same relations in MH coordinates as in ADM coordinates:
\begin{subequations}\label{Omega_r-K}
	\begin{align}
		\Omega_r^\text{MH} &= \Omega_r^\text{ADM} \equiv \Omega_r \, , \\
		K^\text{MH} &= K^\text{ADM} \equiv K \, .
	\end{align}
\end{subequations}
We may therefore use any combination of $\Omega_r$ and $K$ instead of the constants of the motion $\varepsilon$ and $j$ to parameterize in a physically meaningful way a given eccentric orbit (assuming a one-to-one relation). Following Ref.~\cite{Ar.al.08}, we introduce the frequency $\Omega_\phi \equiv K \, \Omega_r$, which is a natural generalization of the circular-orbit frequency $\Omega$,\footnote{Note that $\Omega_\phi$ coincides with the average angular frequency of the motion: $\Omega_\phi = \langle \dot{\phi} \rangle \equiv \frac{1}{T_r} \int_0^{T_r} \dot{\phi}(t) \, \ud t$.} and we define the dimensionless coordinate-invariant parameters (remember that $k = K - 1$)
\beq\label{x_iota}
	x \equiv \left( \frac{G m \Omega_\phi}{c^3} \right)^{2/3} , \qquad \iota \equiv \frac{3 x}{k} \, .
\eeq
The PN parameter $x$ is $\calO(c^{-2})$, while $\iota$ is merely Newtonian at leading order (the relativistic periastron advance first appears at 1PN order). The choice of variables \eqref{x_iota} is the obvious generalization of the gauge-invariant variable $x$ that is commonly used for circular orbits. It will thus facilitate checking the circular-orbit limit. In Sec.~\ref{sec:averaging_invariant}, we shall express our final results in terms of either of the two sets of gauge-invariant parameters $(\varepsilon,j)$ or $(x,\iota)$.

To compute the invariant relationships $\langle U \rangle(\varepsilon,j)$ and $\langle U \rangle(x,\iota)$ from the expressions \eqref{U_CM}--\eqref{U_CMbis} and \eqref{dxiU} for the redshift in MH coordinates, we shall also need expressions for the orbital elements $a_r$, $e_t$, $e_r$, $e_\phi$, $f_t$, $g_t$, $i_t$, $h_t$, $f_\phi$, $g_\phi$, $i_\phi$ and $h_\phi$ in these coordinates. They are given by Eqs.~\eqref{a_r}--\eqref{g_phi}, to which we must add the differences \cite{Me.al2.04}
\begin{subequations}\label{orb_elts_harm}
	\begin{align}
		a^\text{MH}_r - a^\text{ADM}_r &= \frac{G m \, \varepsilon}{c^2} \left\{ - \frac{5}{8} \nu + \frac{1}{j} \left( \frac{1}{4} + \frac{17}{4} \nu \right) \right\} \nonumber \\ &+ \frac{G m \, \varepsilon^2}{c^2} \left\{ \frac{\nu}{32} + \frac{\nu^2}{32} + \frac{1}{j} \left( - \frac{1}{2} + \left( - \frac{11499}{560} + \frac{21}{32} \pi^2 \right) \nu + \frac{19}{4} \nu^2 \right) \right. \nonumber \\ &\qquad\qquad\quad \left. + \, \frac{1}{j^2} \left( \frac{3}{2} + \left( \frac{14501}{420} - \frac{21}{16} \pi^2 \right) \nu - 5 \nu^2 \right) \right\} + o(\varepsilon^2) \, , \\
		e^\text{MH}_t - e^\text{ADM}_t &= \frac{\varepsilon^2}{\sqrt{1-j}} \left( \frac{1}{4} + \frac{17}{4} \nu \right) \! \left( 1 - \frac{1}{j} \right) \nonumber \\ &+ \frac{\varepsilon^3}{\sqrt{1-j}} \left\{ - \frac{19}{32} - \frac{52}{3} \nu + \frac{225}{32} \nu^2 + \frac{1}{j} \left( \frac{29}{16} + \left( \frac{79039}{1680} - \frac{21}{16} \pi^2 \right) \nu - \frac{201}{16} \nu^2 \right) \right. \nonumber \\ &\qquad\qquad\quad\; \left. + \, \frac{1}{j^2} \left( - \frac{3}{2} + \left( - \frac{14501}{420} + \frac{21}{16} \pi^2 \right) \nu + 5 \nu^2 \right) \right\} + o(\varepsilon^3) \, , \label{e_t_MH} \\
		e^\text{MH}_r - e^\text{ADM}_r &= \frac{\varepsilon^2}{\sqrt{1-j}} \left\{ \frac{1}{2} + \frac{73}{8} \nu - j \, \frac{5}{8} \nu - \frac{1}{j} \left( \frac{1}{2} + \frac{17}{2} \nu \right) \right\} \nonumber \\ &+ \frac{\varepsilon^3}{\sqrt{1-j}} \left\{ \frac{13}{16} + \left( - \frac{5237}{1680} + \frac{21}{32} \pi^2 \right) \nu + \frac{19}{16} \nu^2 + j \left( - \frac{143}{64} \nu + \frac{37}{64} \nu^2 \right) \right. \nonumber \\ &\qquad\qquad\quad\; + \frac{1}{j} \left( \frac{13}{8} + \left( \frac{3667}{56} - \frac{105}{32} \pi^2 \right) \nu - \frac{51}{4} \nu^2 \right) \nonumber \\ &\qquad\qquad\quad\; \left. + \, \frac{1}{j^2} \left( - 3 + \left( - \frac{14501}{210} + \frac{21}{8} \pi^2 \right) \nu + 10 \nu^2 \right) \right\} + o(\varepsilon^3) \, , \label{e_r_MH} \\
		e^\text{MH}_\phi - e^\text{ADM}_\phi &= \frac{\varepsilon^2}{\sqrt{1-j}} \left\{ - \frac{1}{4} - \frac{71}{16} \nu + j \, \frac{\nu}{32} + \frac{1}{j} \left( \frac{1}{4} + \frac{141}{32} \nu \right) \right\} \nonumber \\ &+ \frac{\varepsilon^3}{\sqrt{1-j}} \left\{ - \frac{13}{32} + \left( \frac{36511}{8960} - \frac{21}{128} \pi^2 \right) \nu - \frac{1723}{256} \nu^2 + j \left( \frac{17}{256} \nu + \frac{33}{256} \nu^2 \right) \right. \nonumber \\ &\qquad\qquad\quad\; + \frac{1}{j} \left( - \frac{13}{16} + \left( - \frac{21817}{480} + \frac{147}{64} \pi^2 \right) \nu + \frac{169}{8} \nu^2 \right) \nonumber \\ &\qquad\qquad\quad\; \left. + \, \frac{1}{j^2} \left( \frac{3}{2} + \left( \frac{621787}{13440} - \frac{273}{128} \pi^2 \right) \nu - \frac{1789}{128} \nu^2 \right) \right\} + o(\varepsilon^3) \, , \label{e_phi_MH} \\
		f^\text{MH}_t - f^\text{ADM}_t &= \frac{19 \varepsilon^2}{8} \left( \frac{1-j}{j} \right)^{1/2} \nu \nonumber \\ &+ \frac{\varepsilon^3}{\sqrt{j(1-j)}} \left\{ - 1 + \left( - \frac{296083}{6720} + \frac{21}{32} \pi^2 \right) \nu + \frac{989}{64} \nu^2 + j \left( \frac{361}{64} \nu - \frac{171}{64} \nu^2 \right) \right. \nonumber \\ &\qquad\qquad\qquad\;\; \left. + \, \frac{1}{j} \left( 1 + \left( \frac{276133}{6720} - \frac{21}{32} \pi^2 \right) \nu - \frac{799}{64} \nu^2 \right) \right\} + o(\varepsilon^3) \, , \\
		g^\text{MH}_t - g^\text{ADM}_t &= o(\varepsilon^3) \, , \\
		i^\text{MH}_t - i^\text{ADM}_t &=  \frac{11\varepsilon^3}{32} \frac{1-j}{j^{3/2}} \, \nu \, \bigl( 19 - 10 \nu \bigr) + o(\varepsilon^3) \, , \\
		h^\text{MH}_t - h^\text{ADM}_t &= \frac{\varepsilon^3}{192} \left( \frac{1-j}{j} \right)^{3/2} \nu \, \bigl( 23 - 73 \nu \bigr) + o(\varepsilon^3) \, , \\
		f^\text{MH}_\phi - f^\text{ADM}_\phi &= - \frac{\varepsilon^2}{8} \left( \frac{1}{j} - \frac{1}{j^2} \right) \bigl( 1 + 18 \nu \bigr) \nonumber \\ &\quad + \frac{\varepsilon^3}{j} \left\{ \frac{1}{32} + \frac{1045}{192} \nu - \frac{99}{32} \nu^2 + \frac{1}{j} \left( - \frac{5}{4} + \left( - \frac{139633}{3360} + \frac{21}{16} \pi^2 \right) \nu + \frac{117}{8} \nu^2 \right) \right. \nonumber \\ &\qquad\qquad \left. + \, \frac{1}{j^2} \left( \frac{3}{2} + \left( \frac{92307}{2240} - \frac{21}{16} \pi^2 \right) \nu - \frac{351}{32} \nu^2 \right) \right\} + o(\varepsilon^3) \, , \\
		g^\text{MH}_\phi - g^\text{ADM}_\phi &= \frac{\varepsilon^2}{32} \, \frac{{(1-j)^{3/2}}}{j^2} \, \nu \nonumber \\ &+ \varepsilon^3 \, \frac{\sqrt{1-j}}{j} \, \nu \left\{ \frac{7}{128} - \frac{5}{32} \nu + \frac{1}{j} \left( - \frac{49709}{13440} + \frac{21}{128} \pi^2 + \frac{445}{128} \nu \right) \right. \nonumber \\ &\qquad\qquad\qquad\; \left. + \, \frac{1}{j^2} \left( \frac{100783}{26880} - \frac{21}{128} \pi^2 - \frac{847}{256} \nu \right) \right\} + o(\varepsilon^3) \, , \\
		i^\text{MH}_\phi - i^\text{ADM}_\phi &= \frac{\varepsilon^3}{384} \frac{(1-j)^2}{j^3} \, \nu \, \bigl( 149 - 198 \nu \bigr) + o(\varepsilon^3) \, , \\
		h^\text{MH}_\phi - h^\text{ADM}_\phi &= \frac{\varepsilon^3}{256} \frac{(1-j)^{5/2}}{j^3} \, \nu \, \bigl( 1 - 5 \nu \bigr) + o(\varepsilon^3) \, .
	\end{align}
\end{subequations}
Notice, in agreement with the comment made earlier in Sec.~\ref{subsec:ADM}, that the MH coordinates differ from the ADM coordinates at leading 2PN order.

\subsection{Orbital average of the redshift}\label{sec:averaging}

We are finally in a position to compute the generalized redshift
\beq\label{def_<U>}
	{\langle U \rangle} \equiv \frac{1}{\Tau_r} \int_0^{\Tau_r} U(\tau) \, \ud \tau \, ,
\eeq
which coincides with the ratio $T_r / \Tau_r$ of the coordinate time period $T_r$ and the proper time period $\Tau_r$ of the radial motion.\footnote{Beware that, although we are using the same symbol to denote the generalized redshift in Eqs.~\eqref{eq:U} and \eqref{def_<U>}, the former definition is restricted to linear order in the mass ratio, while the latter holds for any $q$.} The averaged redshift \eqref{def_<U>} can be written in the convenient alternative forms\footnote{Notice the simple relation ${\langle U \rangle}_\tau {\langle 1 / U \rangle}_t = 1$, where ${\langle \cdot \rangle}_\tau$ (resp. ${\langle \cdot \rangle}_t$) denotes an averaging over one radial period with respect to the proper time $\tau$ (resp. the coordinate time $t$).}
\beq\label{<U>_QK}
	{\langle U \rangle}^{-1} = \frac{1}{T_r} \int_0^{T_r} \frac{\ud t}{U(t)} = \frac{1}{2\pi} \int_0^{2\pi} \frac{\ud \ell}{U(\ell)} = \frac{1}{2\pi} \int_0^{2\pi} \frac{\ell'(u)}{U(u)} \, \ud u \, ,
\eeq
where $\ell' \equiv \ud \ell / \ud u$ can be computed from Eqs.~\eqref{l_u} and \eqref{V_u}. We first perform the orbit averaging in MH coordinates.
 
\subsubsection{Orbital average in MH coordinates}

Using the generalized QK representation \eqref{l_t}--\eqref{V_u}, \eqref{orb_elts_ADM}--\eqref{Omega_r-K} and \eqref{orb_elts_harm}, the variables $r$, $\dot{r}$ and $v^2 = \dot{r}^2 + r^2 \dot{\phi}^2$ that enter the expression \eqref{U_CM}--\eqref{U_CMbis} and \eqref{dxiU_MH} for the redshift in MH coordinates can be expressed as functions of the binding energy $\varepsilon$, the time eccentricity $e_t \equiv e_t^\text{MH}$, and the eccentric anomaly $u$. The integrand in Eq.~\eqref{<U>_QK} then reads
\beq\label{integrant}
	\frac{\ell'}{U} = \sum_{N=-1}^6 \frac{\alpha_N(e_t,\varepsilon)}{(1-e_t \cos{u})^N} + \sum_{N=2}^4 \beta_N(e_t,\varepsilon) \, \frac{\ln{(1 - e_t \cos{u})}}{(1 - e_t \cos{u})^N} \, .
\eeq
The computation of the coefficients $\alpha_N$ and $\beta_N$ is straightforward, but the resulting expressions are too cumbersome to be reported here. The integral in \eqref{<U>_QK} is readily performed thanks to the following formulas, which are valid for all integers $N \geqslant 1$:
\begin{subequations}\label{integrals}
	\begin{align}
		I_N(e) &\equiv \frac{1}{2\pi} \int_0^{2\pi} \frac{\ud u}{(1 - e \cos{u})^N} = \frac{(-)^{N-1}}{(N-1)!} \, \frac{\ud^{N-1}}{\ud y^{N-1}} \left( \frac{1}{\sqrt{y^2-e^2}} \right) \bigg|_{y=1} \, , \\
		I_N^\text{log}(e) &\equiv \frac{1}{2\pi} \int_0^{2\pi} \frac{\ln{(1 - e \cos{u})}}{(1 - e \cos{u})^N} \, \ud u = \frac{(-)^{N-1}}{(N-1)!} \, \frac{\ud^{N-1} Y(y,e)}{\ud y^{N-1}} \bigg|_{y=1} ,
	\end{align}
\end{subequations}
where
\beq
	Y(y,e) \equiv \frac{1}{\sqrt{y^2-e^2}} \left\{ \ln{\left[ \frac{\sqrt{1-e^2} + 1}{2} \right]} + 2 \ln{\left[ 1 + \frac{\sqrt{1-e^2} - 1}{y + \sqrt{y^2-e^2}} \right]} \right\} .
\eeq

We note that the logarithmic contributions in \eqref{integrant} arise at 3PN order from those terms proportional to $\ln{\bigl(r/r_0\bigr)}$ and $\ln{\bigl(r/r'_1\bigr)}$ in Eq.~\eqref{Uln}. Indeed, combining Eqs.~\eqref{r_u}, \eqref{e_t_ADM}, \eqref{e_r_ADM}, \eqref{e_t_MH} and \eqref{e_r_MH}, one finds
\begin{subequations}
	\begin{align}
		\ln{\biggl(\frac{r}{r_0}\biggr)} &= \ln{\biggl(\frac{a_r}{r_0}\biggr)} + \ln{(1 - e_t \cos{u})} + \calO(c^{-2}) \, , \\
		\ln{\biggl(\frac{r}{r'_1}\biggr)} &= \ln{\biggl(\frac{a_r}{r'_1}\biggr)} + \ln{(1 - e_t \cos{u})} + \calO(c^{-2}) \, .
	\end{align}
\end{subequations}
Hence some coefficients $\alpha_N$ in \eqref{integrant} depend on the regularization constants $r_0$ and $r'_1$ through $\ln{\bigl(a_r/r_0\bigr)}$ and $\ln{\bigl(a_r/r'_1\bigr)}$. However, when averaged over one radial period, these terms cancel out from the final expression, because they appear only through the vanishing combination
\beq
	2 I_2(e) - 5 I_3(e) + 3 (1-e^2) I_4(e) = 0 \, .
\eeq
The final expression for ${\langle U \rangle}(\varepsilon,e_t^\text{MH})$ is thus free of the regularization constants $r_0$ and $r'_1$.

Implementing all the above integrations, the expression \eqref{U_CM}--\eqref{U_CMbis} and \eqref{dxiU_MH} for the redshift in MH coordinates can be averaged over an orbit. Up to 3PN order, the generalized redshift \eqref{def_<U>} then takes the form
\beq\label{U_et_eps_MH}
	{\langle U \rangle} = 1 + \calU_\text{N}^\text{MH} \, \varepsilon + \calU_\text{1PN}^\text{MH} \, \varepsilon^2 + \calU_\text{2PN}^\text{MH} \, \varepsilon^3 + \calU_\text{3PN}^\text{MH} \, \varepsilon^4 + o(\varepsilon^4) \, ,
\eeq
where the PN coefficients depend on the symmetric mass ratio $\nu$, the reduced mass difference $\Delta = \sqrt{1-4\nu}$, and the time eccentricity $e_t$ in MH coordinates (hence $e_t \equiv e_t^\text{MH}$). They read
\begin{subequations}\label{U_et_eps_MHbis}
	\begin{align}
		\calU_\text{N}^\text{MH}	&= \frac{3}{4} + \frac{3}{4} \Delta - \frac{\nu}{2} \, , \\
		\calU_\text{1PN}^\text{MH}  &=  - \frac{3}{4} - \frac{3}{4} \Delta - \frac{45}{16} \nu - \frac{9}{16} \Delta \, \nu + \frac{3 + 3 \Delta}{\sqrt{1 - e_t^2}} \, , \\
		\calU_\text{2PN}^\text{MH}  &= \frac{1}{8} + \frac{\Delta}{8} + \frac{111}{16} \nu - \frac{15}{16} \Delta \, \nu + \frac{75}{32} \nu^2 + \frac{3}{32} \Delta \, \nu^2 \nonumber \\ &\quad - \left( \frac{21}{2} + \frac{21}{2} \Delta + \frac{57}{4} \nu - \frac{15}{4} \Delta \, \nu + 3 \nu^2 \right) \frac{1}{\sqrt{1 - e_t^2}} \nonumber \\ &\quad + \left( \frac{41}{2} + \frac{41}{2} \Delta - \frac{37}{4} \nu - \frac{37}{4} \Delta \, \nu + 5 \nu^2 \right) \frac{1}{(1 - e_t^2)^{3/2}} \, , \\
		\calU_\text{3PN}^\text{MH}  &= - \frac{43}{4} \nu + \frac{119}{16} \Delta \, \nu - \frac{93}{32} \nu^2 + \frac{15}{8} \Delta \, \nu^2 - \frac{45}{64} \nu^3 \nonumber \\ &\quad + \left( \frac{405}{16} + \frac{405}{16} \Delta + \frac{1419}{16} \nu - \frac{525}{16} \Delta \, \nu - 3 \nu^2 - \frac{3}{4} \Delta \, \nu^2 - \frac{15}{4} \nu^3 \right) \frac{1}{\sqrt{1 - e_t^2}} \nonumber \\ &\quad\ - \left( 27 + 27 \Delta + 18 \nu - 18 \Delta \, \nu \right) \frac{1}{1 - e_t^2} + \left( \frac{45}{2} + \frac{45}{2} \Delta - 9 \nu - 9 \Delta \, \nu \right) \frac{1}{(1 - e_t^2)^2} \nonumber \\ &\quad+ \biggl( - \frac{1467}{8} - \frac{1467}{8} \Delta + \biggl[ \frac{3281}{48} - \frac{287}{256} \pi^2 \biggr] \nu + \biggl[ \frac{9185}{48} - \frac{287}{256} \pi^2 \biggr] \Delta \, \nu \nonumber \\ &\quad\quad\quad + \biggl[ - \frac{4193}{48} + \frac{41}{64} \pi^2 \biggr] \nu^2 - \frac{415}{16} \Delta \, \nu^2 + \frac{107}{4} \nu^3 \biggr) \, \frac{1}{(1 - e_t^2)^{3/2}} \nonumber \\ &\quad + \biggl( \frac{873}{4} + \frac{873}{4} \Delta + \biggl[ - 278 + \frac{861}{256} \pi^2 \biggr] \nu + \biggl[ - 278 + \frac{861}{256} \pi^2 \biggr] \Delta \, \nu \nonumber \\ &\quad\quad\quad + \biggl[ \frac{695}{4} - \frac{123}{64} \pi^2 \biggr] \nu^2 + \frac{135}{4} \Delta \, \nu^2 - \frac{51}{2} \nu^3 \biggr) \, \frac{1}{(1 - e_t^2)^{5/2}} \, .
	\end{align}
\end{subequations}
For notational simplicity we did not add a label on $e_t$ to indicate that it is the time eccentricity in MH coordinates. (No such label is required over $\varepsilon$, which is gauge invariant.) This point should be remembered when comparing expressions derived in different gauges, as we shall do next.

\subsubsection{Orbital average in ADM coordinates}

We shall now perform an independent calculation in ADM coordinates. We start from the expression for the redshift in ADM coordinates, as given by \eqref{U_CM}--\eqref{U_CMbis} and \eqref{dxiU_ADM}, employ the appropriate QK parametrization and perform the orbital averaging as outlined above. We find that the form \eqref{integrant} is obtained also in the ADM case, with the same coefficients $\beta_N$ but different coefficients $\alpha_N$ in general. The result for the generalized redshift in ADM coordinates is of the form
\beq\label{U_et_eps_ADM}
	{\langle U \rangle} = 1 + \calU_\text{N}^\text{ADM} \, \varepsilon + \calU_\text{1PN}^\text{ADM} \, \varepsilon^2 + \calU_\text{2PN}^\text{ADM} \, \varepsilon^3 + \calU_\text{3PN}^\text{ADM} \, \varepsilon^4 + o(\varepsilon^4) \, ,
\eeq
where the various coefficients depend on $\nu$, $\Delta$, and the time eccentricity in ADM coordinates (hence $e_t \equiv e_t^\text{ADM})$, and read
\begin{subequations}\label{U_et_eps_ADMbis}
	\begin{align}
		\calU_\text{N}^\text{ADM}	&= \frac{3}{4} + \frac{3}{4} \Delta - \frac{\nu}{2} \, , \\
		\calU_\text{1PN}^\text{ADM} &=  - \frac{3}{4} - \frac{3}{4} \Delta - \frac{45}{16} \nu - \frac{9}{16} \Delta \, \nu + \frac{3 + 3 \Delta}{\sqrt{1 - e_t^2}} \, , \\
		\calU_\text{2PN}^\text{ADM} &= \frac{1}{8} + \frac{\Delta}{8} + \frac{111}{16} \nu - \frac{15}{16} \Delta \, \nu + \frac{75}{32} \nu^2 + \frac{3}{32} \Delta \, \nu^2 \nonumber \\ &\quad - \left( \frac{21}{2} + \frac{21}{2} \Delta + \frac{57}{4} \nu - \frac{15}{4} \Delta \, \nu + 3 \nu^2 \right) \frac{1}{\sqrt{1 - e_t^2}} \nonumber \\ &\quad + \left( \frac{41}{2} + \frac{41}{2} \Delta - \frac{37}{4} \nu - \frac{37}{4} \Delta \, \nu + 5 \nu^2 \right) \frac{1}{(1 - e_t^2)^{3/2}} \, , \\
		\calU_\text{3PN}^\text{ADM} &= - \frac{43}{4} \nu + \frac{119}{16} \Delta \, \nu - \frac{93}{32} \nu^2 + \frac{15}{8} \Delta \, \nu^2 - \frac{45}{64} \nu^3 \nonumber \\ &\quad + \left( \frac{405}{16} + \frac{405}{16} \Delta + \frac{1419}{16} \nu - \frac{525}{16} \Delta \, \nu - 3 \nu^2 - \frac{3}{4} \Delta \, \nu^2 - \frac{15}{4} \nu^3 \right) \frac{1}{\sqrt{1 - e_t^2}} \nonumber \\ &\quad\ - \left( 27 + 27 \Delta + 18 \nu - 18 \Delta \, \nu \right) \frac{1}{1 - e_t^2} + \left( \frac{45}{2} + \frac{45}{2} \Delta - 9 \nu - 9 \Delta \, \nu \right) \frac{1}{(1 - e_t^2)^2} \nonumber \\ &\quad+ \biggl( - \frac{1461}{8} - \frac{1461}{8} \Delta + \biggl[ \frac{3893}{48} - \frac{287}{256} \pi^2 \biggr] \nu + \biggl[ \frac{9797}{48} - \frac{287}{256} \pi^2 \biggr] \Delta \, \nu \nonumber \\ &\quad\quad\quad + \biggl[ - \frac{4193}{48} + \frac{41}{64} \pi^2 \biggr] \nu^2 - \frac{415}{16} \Delta \, \nu^2 + \frac{107}{4} \nu^3 \biggr) \, \frac{1}{(1 - e_t^2)^{3/2}} \nonumber \\ &\quad + \biggl( \frac{435}{2} + \frac{435}{2} \Delta + \biggl[ - \frac{1163}{4} + \frac{861}{256} \pi^2 \biggr] \nu + \biggl[ - \frac{1163}{4} + \frac{861}{256} \pi^2 \biggr] \Delta \, \nu \nonumber \\ &\quad\quad\quad + \biggl[ \frac{695}{4} - \frac{123}{64} \pi^2 \biggr] \nu^2 + \frac{135}{4} \Delta \, \nu^2 - \frac{51}{2} \nu^3 \biggr) \, \frac{1}{(1 - e_t^2)^{5/2}} \, .
	\end{align}
\end{subequations}
Although the coefficients \eqref{U_et_eps_MHbis} and \eqref{U_et_eps_ADMbis} coincide through 2PN order, the 3PN coefficients $\calU_\text{3PN}^\text{MH}$ and $\calU_\text{3PN}^\text{ADM}$ are different. A useful internal check of the PN calculations of the generalized redshift in MH and ADM coordinates is the verification that the equality of Eqs.~\eqref{U_et_eps_MH}--\eqref{U_et_eps_MHbis} and \eqref{U_et_eps_ADM}--\eqref{U_et_eps_ADMbis} holds if and only if the time eccentricities $e^\text{MH}_t$ and $e^\text{ADM}_t$ are related by
\beq
	e_t^\text{MH} = e_t^\text{ADM} \left\{ 1 - \frac{1+17\nu}{1-(e^\text{ADM}_t)^2} \, \frac{\varepsilon^2}{4} + \calO(\varepsilon^3) \right\} .
\eeq
This relation is in perfect agreement with what is predicted from using different QK representations of the motion, namely Eq.~\eqref{e_t_ADM} together with \eqref{e_t_MH}.

\subsection{Gauge-invariant formulations}\label{sec:averaging_invariant}

To compare the analytical PN predictions with the numerical results of the GSF calculation (Sec.~\ref{sec:num_results}), it is best to use a coordinate-invariant relationship. We shall thus replace the coordinate-dependant time eccentricity $e_t$ in favor of the coordinate-invariant angular momentum variable $j$. Substituting the PN expansion \eqref{e_t_ADM} into Eq.~\eqref{U_et_eps_ADMbis}, or alternatively Eqs.~\eqref{e_t_ADM} and \eqref{e_t_MH} into \eqref{U_et_eps_MHbis}, we get
\beq\label{U_eps_j}
	{\langle U \rangle} = 1 + \calU_\text{N} \, \varepsilon + \calU_\text{1PN} \, \varepsilon^2 + \calU_\text{2PN} \, \varepsilon^3 + \calU_\text{3PN} \, \varepsilon^4 + o(\varepsilon^4) \, ,
\eeq
where
\begin{subequations}\label{U_eps_jbis}
	\begin{align}
		\calU_\text{N}	 &= \frac{3}{4} + \frac{3}{4} \Delta - \frac{\nu}{2} \, , \\
		\calU_\text{1PN} &= - \frac{3}{4} - \frac{3}{4} \Delta - \frac{45}{16} \nu - \frac{9}{16} \Delta \, \nu + \frac{3 + 3 \Delta }{\sqrt{j}} \, , \\
		\calU_\text{2PN} &= \frac{1}{8} + \frac{\Delta}{8} + \frac{111}{16} \nu - \frac{15}{16} \Delta \, \nu + \frac{75}{32} \nu^2 + \frac{3}{32} \Delta \, \nu^2 \nonumber \\ &\quad - \left( \frac{33}{8} + \frac{33}{8} \Delta + \frac{135}{8} \nu - \frac{9}{8} \Delta \, \nu + 3 \nu^2 \right) \frac{1}{\sqrt{j}} \nonumber \\ &\quad + \left( \frac{35}{2} + \frac{35}{2} \Delta - \frac{25}{4} \nu - \frac{25}{4} \Delta \, \nu + 5 \nu^2 \right) \frac{1}{j^{3/2}} \, , \\
		\calU_\text{3PN} &= - \frac{43}{4} \nu + \frac{119}{16} \Delta \, \nu - \frac{93}{32} \nu^2 + \frac{15}{8} \Delta \, \nu^2 - \frac{45}{64} \nu^3 \nonumber \\ &\quad + \left( \frac{297}{128} + \frac{297}{128} \Delta + \frac{3819}{64} \nu - \frac{1509}{64} \Delta \, \nu + \frac{453}{128} \nu^2 - \frac{459}{128} \Delta \, \nu^2 - \frac{9}{8} \nu^3 \right) \frac{1}{\sqrt{j}} \nonumber \\ &\quad - \left( \frac{9}{2} + \frac{9}{2} \Delta + 27 \nu - 9 \Delta \, \nu \right) \frac{1}{j} \nonumber \\ &\quad- \biggl( \frac{945}{16} + \frac{945}{16} \Delta + \biggl[ \frac{17}{96} + \frac{287}{256} \pi^2 \biggr] \nu + \biggl[ - \frac{10063}{96} + \frac{287}{256} \pi^2 \biggr] \Delta \, \nu \nonumber \\ &\quad\quad\quad + \biggl[ \frac{4471}{96} - \frac{41}{64} \pi^2 \biggr] \nu^2 + \frac{65}{32} \Delta \, \nu^2 - \frac{85}{8} \nu^3 \biggr) \, \frac{1}{j^{3/2}} \nonumber \\ &\quad + \biggl( \frac{693}{4} + \frac{693}{4} \Delta + \biggl[ - \frac{875}{4} + \frac{861}{256} \pi^2 \biggr] \nu + \biggl[ - \frac{875}{4} + \frac{861}{256} \pi^2 \biggr] \Delta \, \nu \nonumber \\ &\quad\quad\quad + \biggl[ \frac{271}{2} - \frac{123}{64} \pi^2 \biggr] \nu^2 + \frac{21}{2} \Delta \, \nu^2 - \frac{21}{2} \nu^3 \biggr) \, \frac{1}{j^{5/2}} \, .
	\end{align}
\end{subequations}

Since the relationship ${\langle U \rangle}(\varepsilon,j)$ is coordinate-invariant, it is physically meaningful. However, the binding energy $E$ and angular momentum $J$ are not easily accessible to perturbative GSF calculations, so a direct comparison is not obvious. Thanksfully, Eq.~\eqref{U_eps_j} can also be expressed using the invariant parameters \eqref{x_iota} defined with respect to the fundamental frequencies $\Omega_r$ and $\Omega_\phi$. Indeed, inverting the PN expansions \eqref{n} and \eqref{K} yields
\begin{subequations}\label{eps_j}
	\begin{align}
		\varepsilon &= x \, \biggl\{ 1 + \left( \frac{5}{4} - \frac{\nu}{12} - \frac{2}{\iota} \right) x + \left( \frac{5}{8} - \frac{5}{8} \nu - \frac{\nu^2}{24} + \frac{5 - 2\nu}{\sqrt{\iota}} - \frac{5 - \frac{\nu}{3}}{\iota} + \frac{5}{\iota^2} \right) x^2 \nonumber \\ &\quad\quad\quad\,\, + \bigg( - \frac{185}{192} - \frac{75}{64} \nu - \frac{25}{288} \nu^2 - \frac{35}{5184} \nu^3 + \left[ \frac{105}{8} - \frac{35}{6} \nu - \frac{7}{6} \nu^2 \right] \frac{1}{\sqrt{\iota}} \nonumber \\ &\qquad\qquad\quad\; + \left[ - \frac{15}{4} + \frac{15}{4} \nu + \frac{\nu^2}{4} \right] \frac{1}{\iota} - \left[ \frac{95}{8} + \left( \frac{211}{9} - \frac{41}{96} \pi^2 \right) \nu - \frac{5}{2} \nu^2 \right] \frac{1}{\iota^{3/2}} \biggr) \, x^3 + o(x^3) \biggr\} \, , \\
		j &= \iota + \left( \frac{27}{4} - \frac{5}{2} \nu + \frac{5}{12} \nu \, \iota \right) x + \biggl( - \frac{35}{8} + \left[ \frac{373}{16} - \frac{41}{128} \pi^2 \right] \nu - \frac{55}{24} \nu^2 + \left[ 5 - 2 \nu \right] \sqrt{\iota} \nonumber \\ &\quad\quad + \left[ - \frac{35}{16} + \frac{25}{48} \nu + \frac{\nu^2}{8} \right] \iota + \left[ \frac{115}{16} - \left( \frac{665}{12} - \frac{205}{128} \pi^2 \right) \nu - \frac{15}{8} \nu^2 \right] \frac{1}{\iota} \biggr) \, x^2 + o(x^2) \label{j_x_iota} \, .
	\end{align}
\end{subequations}
We thus have the leading-order relationships $x = \varepsilon + \calO(c^{-2})$ and $\iota = j + \calO(c^{-2})$. Introducing the expansions \eqref{eps_j} into Eq.~\eqref{U_eps_j}--\eqref{U_eps_jbis}, our main PN result reads
\beq\label{U_x_iota}
	{\langle U \rangle} = 1 + \calV_\text{N} \, x + \calV_\text{1PN} \, x^2 + \calV_\text{2PN} \, x^3 + \calV_\text{3PN} \, x^4 + o(x^4) \, ,
\eeq
where the various PN coefficients, which depend on the variable $\iota$ as well as on the particle's masses, read up to 3PN order
\begin{subequations}\label{U_x_iotabis}
	\begin{align}
		\calV_\text{N}	 &= \frac{3}{4} + \frac{3}{4} \Delta - \frac{\nu}{2} \, , \\
		\calV_\text{1PN} &= \frac{3}{16} + \frac{3}{16} \Delta - \frac{7}{2} \nu - \frac{5}{8} \Delta \, \nu + \frac{\nu^2}{24} + \frac{3 + 3 \Delta}{\sqrt{\iota}} - \left( \frac{3}{2} + \frac{3}{2} \Delta - \nu \right) \frac{1}{\iota} \, , \\
		\calV_\text{2PN} &= - \frac{41}{32} - \frac{41}{32} \Delta - \frac{3}{4} \nu - \frac{43}{16} \Delta \, \nu + \frac{99}{32} \nu^2 + \frac{5}{32} \Delta \, \nu^2 + \frac{\nu^3}{48} \nonumber \\ &\quad + \left( \frac{57}{8} + \frac{57}{8} \Delta - 22 \nu - \frac{3}{2} \Delta \, \nu - 2 \nu^2 \right) \frac{1}{\sqrt{\iota}} - \left( \frac{3}{4} + \frac{3}{4} \Delta - 14 \nu - \frac{5}{2} \Delta \, \nu + \frac{\nu^2}{6} \right) \frac{1}{\iota} \nonumber \\ &\quad - \left( \frac{37}{8} + \frac{37}{8} \Delta + \frac{5}{2} \nu + \frac{5}{2} \Delta \, \nu - 5 \nu^2 \right) \frac{1}{\iota^{3/2}} + \left( \frac{15}{4} + \frac{15}{4} \Delta - \frac{5}{2} \nu \right) \frac{1}{\iota^2} \, , \\
		\calV_\text{3PN} &= - \frac{605}{256} - \frac{605}{256} \Delta + \frac{385}{48} \nu + \frac{323}{128} \Delta \, \nu + \frac{565}{64} \nu^2 + \frac{419}{128} \Delta \, \nu^2 - \frac{895}{864} \nu^3 + \frac{25}{1728} \Delta \, \nu^3 + \frac{35}{10368} \nu^4 \nonumber \\ &\quad + \left( \frac{117}{128} + \frac{117}{128} \Delta - \frac{89}{2} \nu - \frac{499}{16} \Delta \, \nu + \frac{1267}{96} \nu^2 - \frac{91}{32} \Delta \, \nu^2 + \frac{5}{6} \nu^3 \right) \frac{1}{\sqrt{\iota}} \nonumber \\ &\quad + \left( \frac{411}{16} + \frac{411}{16} \Delta - \frac{63}{2} \nu + \frac{129}{8} \Delta \, \nu - \frac{297}{16} \nu^2 - \frac{15}{16} \Delta \, \nu^2 - \frac{\nu^3}{8} \right) \frac{1}{\iota} \nonumber \\ &\quad+ \biggl( - \frac{1755}{64} - \frac{1755}{64} \Delta + \biggl[ \frac{4339}{48} - \frac{41}{128} \pi^2 \biggr] \nu + \biggl[ \frac{497}{24} - \frac{41}{128} \pi^2 \biggr] \Delta \, \nu \nonumber \\ &\quad\quad\quad + \biggl[ - \frac{181}{144} + \frac{41}{96} \pi^2 \biggr] \nu^2 + \frac{115}{16} \Delta \, \nu^2 + \frac{5}{4} \nu^3 \biggr) \, \frac{1}{\iota^{3/2}} \nonumber \\ &\quad + \biggl( \frac{1797}{128} + \frac{1797}{128} \Delta + \biggl[ - \frac{355}{16} + \frac{123}{128} \pi^2 \biggr] \nu + \biggl[ - \frac{355}{16} + \frac{123}{128} \pi^2 \biggr] \Delta \, \nu \nonumber \\ &\quad\quad\quad + \biggl[ \frac{1321}{32} - \frac{123}{64} \pi^2 \biggr] \nu^2 - \frac{99}{32} \Delta \, \nu^2 + \frac{33}{4} \nu^3 \biggr) \, \frac{1}{\iota^{5/2}} \, .
	\end{align}
\end{subequations}
The noncircular nature of the motion only explicitly enters the result at leading 1PN order via the invariant parameter $\iota$. Since we have the qualitative behavior $\iota \sim 1 - e^2$, this suggests that the effect of the eccentricity on ${\langle U \rangle}$ will be moderate (at least in the weak-field regime).

\subsubsection{Circular-orbit limit}

Another key check of the results \eqref{U_eps_jbis} and \eqref{U_x_iotabis} is provided by the circular-orbit limit. For such orbits, the two constants of the motion are no longer independent variables. Indeed, the angular momentum variable, say $j_\odot$, is related to the energy $\varepsilon$ by the 3PN gauge-invariant expansion \cite{Da.al.00} 
\begin{align}\label{j_eps}
	j_\odot &= 1 + \left( \frac{9}{4} + \frac{\nu}{4} \right) \varepsilon + \left( \frac{81}{16} - 2 \nu + \frac{\nu^2}{16} \right) \varepsilon^2 \nonumber \\ &+ \left( \frac{945}{64} + \biggl[ - \frac{7699}{192} + \frac{41}{32} \pi^2 \biggr] \nu + \frac{\nu^2}{2} + \frac{\nu^3}{64} \right) \varepsilon^3 + o(\varepsilon^3) \, .
\end{align}
It can be checked that the eccentricities $e_t$, $e_r$, $e_\phi$ all vanish when $j$ is replaced by \eqref{j_eps} in Eqs.~\eqref{e_t_ADM}--\eqref{e_phi_ADM} and \eqref{e_t_MH}--\eqref{e_phi_MH}. The invariant result \eqref{U_eps_j}--\eqref{U_eps_jbis} then reduces to
\begin{align}\label{U_eps}
	U_\odot &= 1 +  \left( \frac{3}{4} + \frac{3}{4} \Delta - \frac{\nu}{2} \right) \varepsilon + \left( \frac{9}{4} + \frac{9}{4} \Delta - \frac{45}{16} \nu - \frac{9}{16} \Delta \, \nu \right) \varepsilon^2 \nonumber \\ &\quad\quad + \left( \frac{81}{8} + \frac{81}{8} \Delta - \frac{265}{16} \nu - \frac{103}{16} \Delta \, \nu + \frac{139}{32} \nu^2 + \frac{3}{32} \Delta \, \nu^2 \right) \varepsilon^3 \nonumber \\ &\quad\quad + \biggl( \frac{891}{16} + \frac{891}{16} \Delta - \biggl[ \frac{3809}{24} - \frac{287}{128} \pi^2 \biggr] \nu - \biggl[ \frac{4945}{48} - \frac{287}{128} \pi^2 \biggr] \Delta \, \nu \nonumber \\ &\quad\quad\quad\quad + \biggl[ \frac{7727}{96} - \frac{41}{32} \pi^2 \biggr] \nu^2 + \frac{143}{16} \Delta \, \nu^2 - \frac{205}{64} \nu^3 \biggr) \, \varepsilon^4 + o(\varepsilon^4) \, .
\end{align}
Setting $e_t \to 0$ in Eq.~\eqref{U_et_eps_MHbis} or \eqref{U_et_eps_ADMbis} yields the same expression.

We then replace the constant of the motion $\varepsilon$ in favor of the frequency-related parameter $x$ [recall Eq.~\eqref{x_iota}], using the well-known 3PN-accurate expression for the binding energy as a function of the circular-orbit frequency, namely [see, e.g., Eq.~(232) of Ref.~\cite{Bl.14}]
\begin{align}\label{eps_x}
	\varepsilon_\odot &= x \, \biggl\{ 1 + \left( - \frac{3}{4} - \frac{\nu}{12} \right) x + \left( - \frac{27}{8} + \frac{19}{8} \nu - \frac{\nu^2}{24} \right) x^2 \nonumber \\ &+ \left( - \frac{675}{64} + \biggl[ \frac{34445}{576} - \frac{205}{96} \pi^2 \biggr] \nu - \frac{155}{96} \nu^2 - \frac{35}{5184} \nu^3 \right) x^3 + o(x^3) \biggr\} \, .
\end{align}
Finally, replacing $\varepsilon$ in \eqref{U_eps} using \eqref{eps_x}, we recover the known 3PN result for the circular-orbit redshift (see Eq.~(4.10) of Ref. \cite{Bl.al.10}):
\begin{align}\label{U_x}
	U_\odot &= 1 + \left( \frac{3}{4} + \frac{3}{4} \Delta - \frac{\nu}{2} \right) x + \left( \frac{27}{16} + \frac{27}{16} \Delta - \frac{5}{2} \nu - \frac{5}{8} \Delta \, \nu + \frac{\nu^2}{24} \right) x^2 \nonumber \\ &\quad\quad + \left( \frac{135}{32} + \frac{135}{32} \Delta - \frac{37}{4} \nu - \frac{67}{16} \Delta \, \nu + \frac{115}{32} \nu^2 + \frac{5}{32} \Delta \, \nu^2 + \frac{\nu^3}{48} \right) x^3 \nonumber \\
			&\quad\quad + \biggl( \frac{2835}{256} + \frac{2835}{256} \Delta - \biggl[ \frac{2183}{48} - \frac{41}{64} \pi^2 \biggr] \nu - \biggl[ \frac{12199}{384} - \frac{41}{64} \pi^2 \biggr] \Delta \, \nu \nonumber \\ &\quad\quad\quad\quad + \biggl[ \frac{17201}{576} - \frac{41}{192} \pi^2 \biggr] \nu^2 + \frac{795}{128} \Delta \, \nu^2 - \frac{2827}{864} \nu^3 + \frac{25}{1728} \Delta \, \nu^3 + \frac{35}{10368} \nu^4 \biggr) \, x^4 \nonumber \\
			&\quad\quad + o(x^4) \, .
\end{align}
Interestingly, at Newtonian order, the averaged redshift ${\langle U \rangle}$ along an eccentric orbit has the same functional form as $U_\odot$ in the case of a circular orbit. This shows that the effect of the eccentricity cancels out at Newtonian order, because of the orbital averaging.

Alternatively, we can also combine Eqs.~\eqref{j_x_iota}, \eqref{j_eps}, \eqref{eps_x} to obtain the PN expansion of the invariant relation $\iota_\odot(x)$ in the circular-orbit limit, namely
\beq
	\iota_\odot = 1 + \left( - \frac{9}{2} + \frac{7}{3} \nu \right) x +  \left( - \frac{9}{4} + \biggl[ \frac{397}{12} - \frac{41}{32} \pi^2 \biggr] \nu + \frac{28}{9} \nu^2 \right) x^2 + o(x^2) \, ,
\eeq
and introduce this expression into \eqref{U_x_iota}--\eqref{U_x_iotabis} to recover \eqref{U_x}.

Our third and last check of the correctness of the formula \eqref{U_x_iotabis} will be to recover the known result in the test-particle limit.

\subsubsection{Extreme mass-ratio limit}

The 3PN result \eqref{U_x_iota}--\eqref{U_x_iotabis} is valid for \textit{any} mass ratio $q = m_1 / m_2$. To extract from this result the contribution due to the conservative piece of the GSF, we introduce an alternative set of dimensionless coordinate-invariant parameters, better suited than $(x,\iota)$ to the extreme mass-ratio limit $q \ll 1$:
\beq\label{y-lambda}
	y \equiv \left( \frac{G m_2 \Omega_\phi}{c^3} \right)^{2/3} , \qquad \lambda \equiv \frac{3 y}{k} \, .
\eeq
We substitute the relations $x = y \, (1+q)^{2/3}$ and $\iota = \lambda \, (1+q)^{2/3}$ in \eqref{U_x_iota}--\eqref{U_x_iotabis}, and expand in powers of the mass ratio $q$, neglecting terms of $\calO(q^3)$ or higher. The 3PN result for the sum of the test mass, GSF and post-GSF contributions reads
\beq
	{\langle U \rangle} = {\langle U \rangle}_0 + q \, \langle U \rangle_\text{gsf} + q^2 \, \langle U \rangle_\text{p-gsf} + \calO(q^3) \, ,
\eeq
where
\begin{subequations}
	\begin{align}
		{\langle U \rangle}_0 &= 1 + \frac{3}{2} \, y + \biggl( \frac{3}{8} + \frac{6}{\sqrt{\lambda}} - \frac{3}{\lambda} \biggr) \, y^2 + \biggl( - \frac{41}{16} + \frac{57}{4\sqrt{\lambda}} - \frac{3}{2\lambda} - \frac{37}{4\lambda^{3/2}} + \frac{15}{2\lambda^2} \biggr) \, y^3 \nonumber \\ &+ \biggl( - \frac{605}{128} + \frac{117}{64\sqrt{\lambda}} + \frac{411}{8\lambda} - \frac{1755}{32\lambda^{3/2}} + \frac{21}{4\lambda^2}  + \frac{1797}{64\lambda^{5/2}} - \frac{20}{\lambda^3} \biggr) \, y^4 + o(y^4) \, , \label{U_Schw} \\
		\langle U \rangle_\text{gsf}  &= - y - \biggl( 4 - \frac{2}{\lambda} \biggr) \, y^2 - \biggl( 6 + \frac{14}{\sqrt{\lambda}} - \frac{16}{\lambda} + \frac{5}{\lambda^{3/2}} + \frac{5}{\lambda^2} \biggr) \, y^3 + \biggl( \frac{8}{3} - \frac{293}{4\sqrt{\lambda}} + \frac{36}{\lambda} \nonumber \\ &+ \left[ \frac{1789}{24} - \frac{41}{64} \pi^2 \right] \frac{1}{\lambda^{3/2}} - \frac{56}{\lambda^2} + \left[ - \frac{355}{8} + \frac{123}{64} \pi^2 \right] \frac{1}{\lambda^{5/2}} + \frac{40}{3\lambda^3} \biggr) \, y^4 + o(y^4) \, , \label{U_SF} \\
		\langle U \rangle_\text{p-gsf}  &= y + \biggl( 4 - \frac{2}{\lambda} \biggr) \, y^2 + \biggl( \frac{69}{8} + \frac{29}{4\sqrt{\lambda}} - \frac{16}{\lambda} + \frac{15}{\lambda^{3/2}} + \frac{5}{\lambda^2} \biggr) \, y^3 + \biggl( \frac{275}{24} + \frac{1533}{32\sqrt{\lambda}} - \frac{207}{4\lambda} \nonumber \\ &+ \left[ - \frac{6377}{96} + \frac{41}{32} \pi^2 \right] \frac{1}{\lambda^{3/2}} + \frac{56}{\lambda^2} + \left[ \frac{2031}{16} - \frac{369}{64} \pi^2 \right] \frac{1}{\lambda^{5/2}} - \frac{40}{3\lambda^3} \biggr) \, y^4 + o(y^4) \, . \label{U_PSF}
	\end{align}
\end{subequations}
In the test-particle limit $q = 0$, we recover the 3PN expansion of the fully relativistic result \eqref{eq:U0} for a geodesic orbit, as derived in App.~\ref{app:test-mass}. The 3PN prediction \eqref{U_PSF} could be compared with future calculations of the second-order GSF \cite{De.12,Gr.12,Po.12,Po2.12,Po.14,PoMi.14}.

Finally, we may express the result \eqref{U_SF} for the 3PN expansion of the GSF contribution to the generalized redshift by means of the usual parametrization of bound timelike geodesic orbits in Schwarzschild in terms of the semi-latus rectum $p$ and eccentricity $e$ (see Sec.~\ref{sec:Schwarzschild_geodesics}). Substituting for $y$ and $\lambda$ from Eqs.~\eqref{y_lambda} into \eqref{U_SF}, we find
\begin{align}\label{<U>_gsf_3PN}
	\langle U \rangle_\text{gsf} &= - \frac{j_e}{p} \, \bigg\{ 1 + \frac{2 j_e}{p} + \left( 5 \sqrt{j_e} - 4 j_e + 9 j_e^{3/2} - 5 j_e^2 \right) \frac{1}{p^2} + \biggl( \, \left[ 95 - \frac{123}{64} \pi^2 \right] \sqrt{j_e} \nonumber \\ &\qquad\qquad - 16 j_e + \left[ - \frac{79}{6} + \frac{41}{64} \pi^2 \right] j_e^{3/2} - 16 j_e^2 - \frac{27}{2} j_e^{5/2} + 4 j_e^3 \biggr) \, \frac{1}{p^3} + o(p^{-3}) \biggr\} \, ,
\end{align}
where $j_e \equiv 1 - e^2$. For small eccentricities, we may write
\beq\label{<U>_gsf_3PNbis}
	\langle U \rangle_\text{gsf} = a + b \, e^2 + c \, e^4 + d \, e^6 + \calO(e^8) \, ,
\eeq
where the weak-field expansions of the coefficients $a(p)$, $b(p)$, $c(p)$ and $d(p)$ read
\begin{subequations}\label{a-b-c-d}
	\begin{align}
		a &= - \frac{1}{p} - \frac{2}{p^2} - \frac{5}{p^3} - \left( \frac{121}{3} - \frac{41}{32} \pi^2 \right) \frac{1}{p^4} + o(p^{-4}) \, , \\
		b &= \frac{1}{p} + \frac{4}{p^2} + \frac{7}{p^3} - \left( \frac{5}{3} + \frac{41}{32} \pi^2 \right) \frac{1}{p^4} + o(p^{-4}) \, , \\
		c &= - \frac{2}{p^2} + \frac{1}{4p^3} + \left( \frac{705}{8} - \frac{123}{256} \pi^2 \right) \frac{1}{p^4} + o(p^{-4}) \, , \\
		d &= - \frac{5}{2p^3} + \left( - \frac{475}{12} + \frac{41}{128} \pi^2 \right) \frac{1}{p^4} + o(p^{-4}) \, ,
	\end{align}
\end{subequations}
and higher-order terms in the eccentricity all contribute at leading 2PN order.

\section{Comparison of post-Newtonian and self-force results}\label{sec:comparison}

\begin{figure}[h]
	\begin{subfigure}[b]{0.5\textwidth}
		\includegraphics[width=\textwidth]{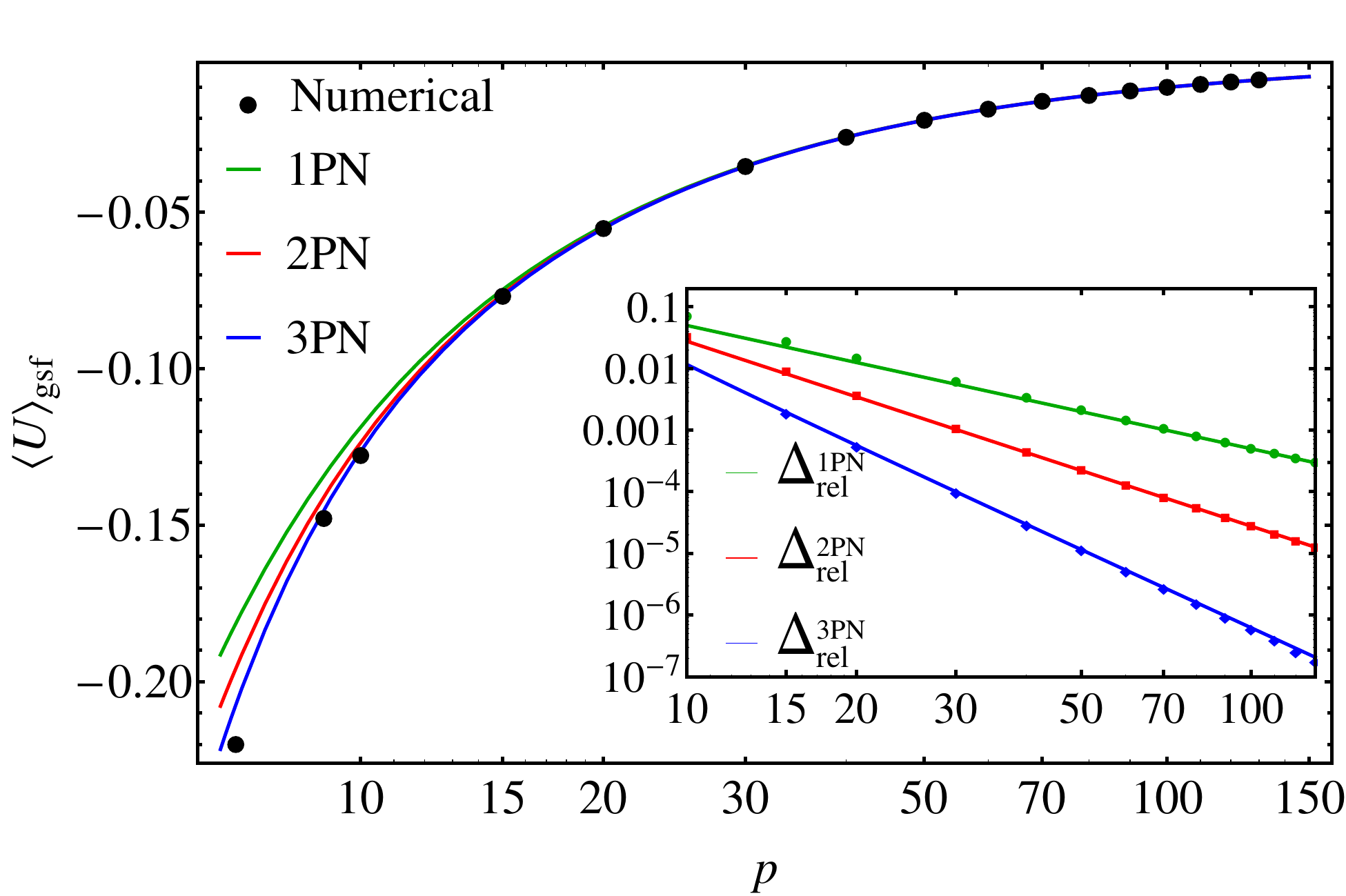}
		\caption{$ e = 0.1 $}
		\label{fig:e0.1_plot}
	\end{subfigure}%
	\hfill
	\begin{subfigure}[b]{0.5\textwidth}
		\includegraphics[width=\textwidth]{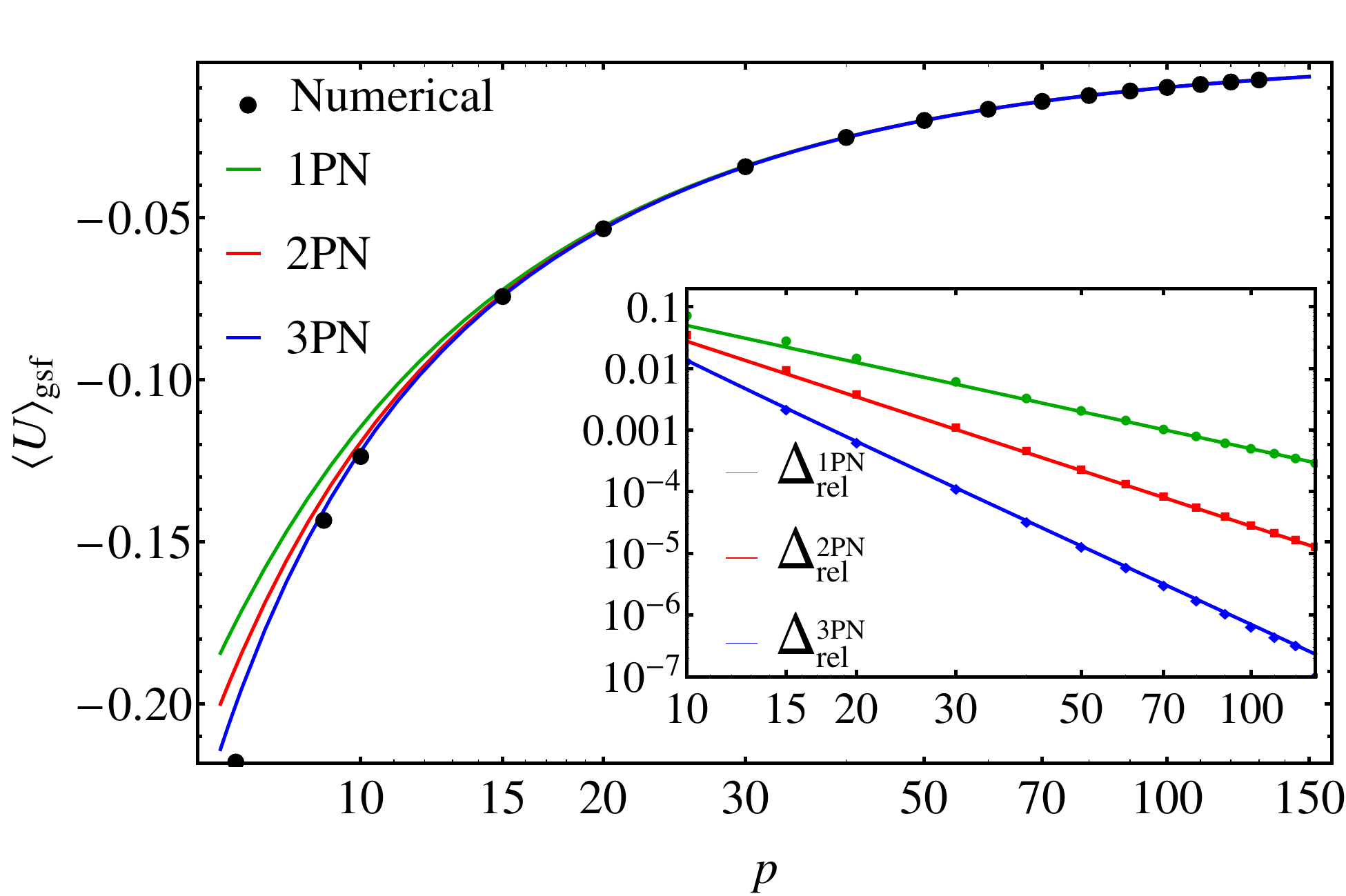}
		\caption{$ e = 0.2 $}
		\label{fig:e0.2_plot}
	\end{subfigure}
	\begin{subfigure}[b]{0.5\textwidth}
		\includegraphics[width=\textwidth]{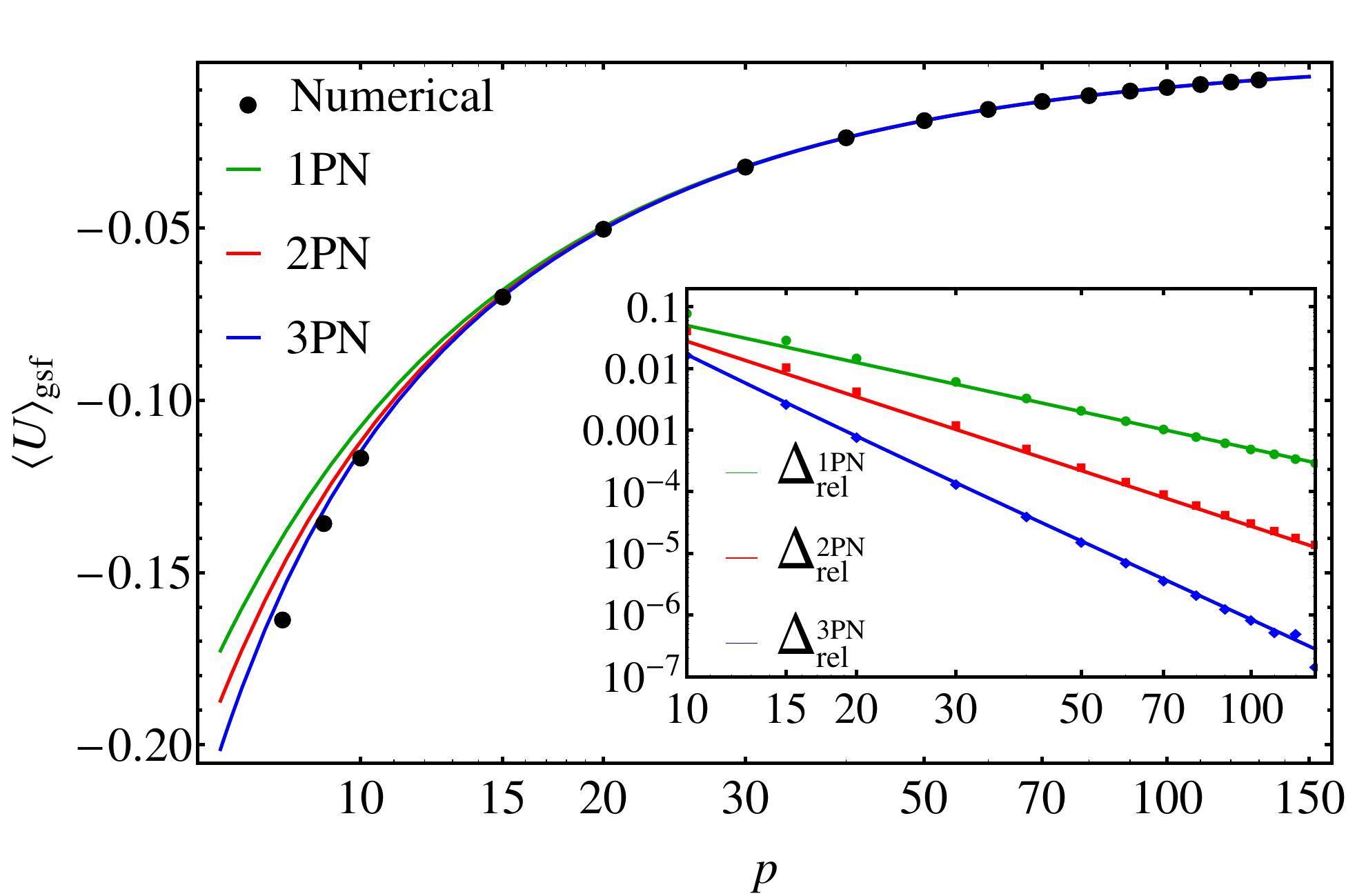}
		\caption{$ e = 0.3 $}
		\label{fig:e0.3_plot}
	\end{subfigure}%
	\hfill
	\begin{subfigure}[b]{0.5\textwidth}
		\includegraphics[width=\textwidth]{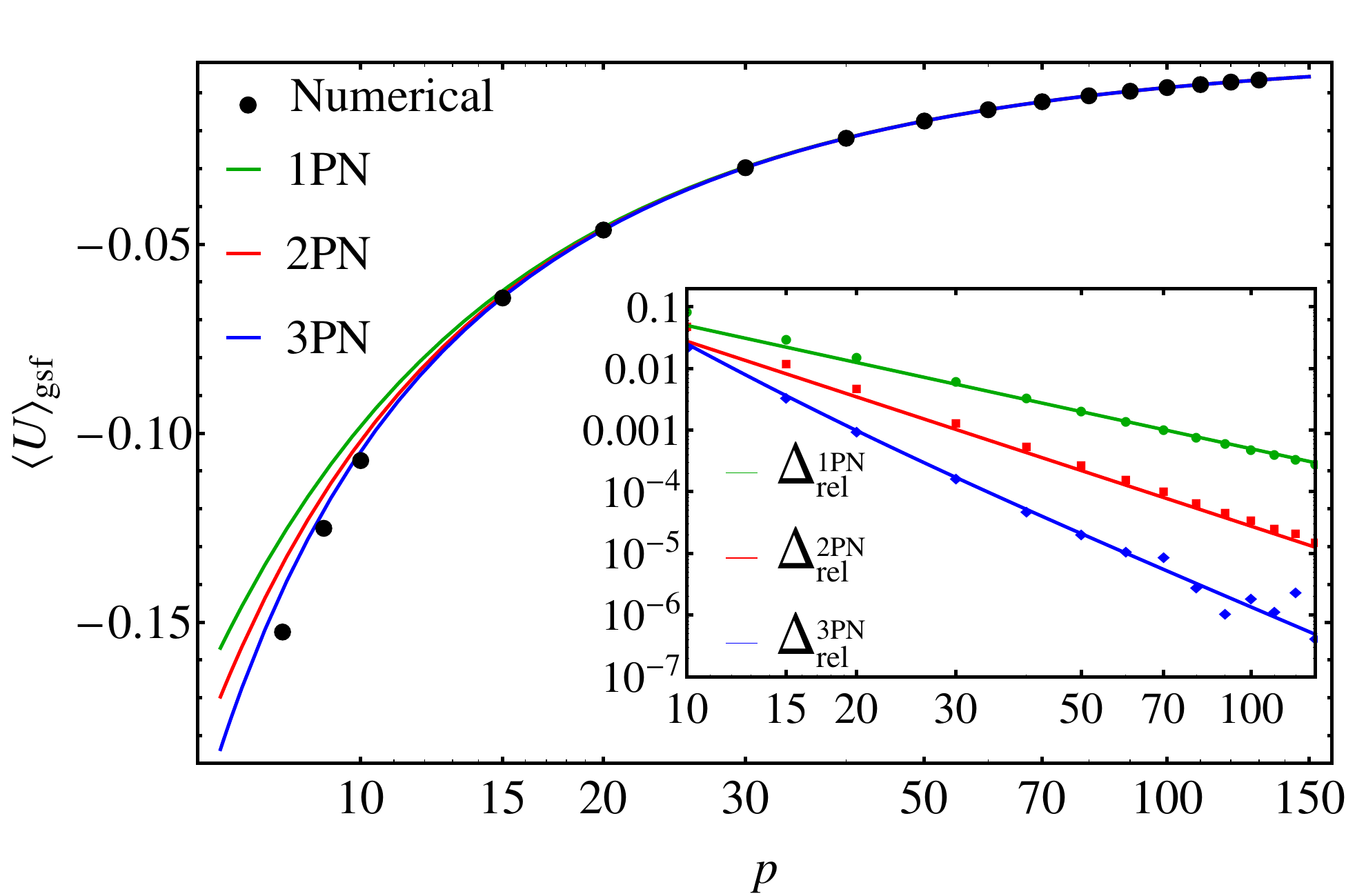}
		\caption{$ e = 0.4 $}
		\label{fig:e0.4_plot}
	\end{subfigure}
	\caption{Numerical GSF output for $ \langle U \rangle_{\text{gsf}}$ (black data points) versus analytical PN approximations (solid curves). Each panel shows $\langle U\rangle_{\text{gsf}}$ as a function of semi-latus rectum $p$ for a fixed eccentricity $e$. Insets display, on a log-log scale, the relative differences  $\Delta_\text{rel}^{n\text{PN}} \equiv |1-U_{n{\rm PN}}/\langle U \rangle_{\rm gsf}|$, where $U_{n{\rm PN}}$ is the PN approximation through $n$PN order. In both the main plots and the insets, the three curves correspond, top to bottom, to the 1PN, 2PN and 3PN approximations. Solid curves in the insets are the analytical PN residues $1-U_{\rm 1PN}/U_{\rm 3PN}$ (upper curve) and $1-U_{\rm 2PN}/U_{\rm 3PN}$ (middle curve); for the lower curve we have fitted the simple model $1-U_{\rm 3PN}/\langle U \rangle_{\rm gsf} = p^{-4} \, (\alpha_1 + \alpha_2 \ln p + \alpha_3 / p)$.}
	\label{fig:GSF_data_all}
\end{figure}

In Fig.~\ref{fig:GSF_data_all} we plot our data for $\langle U \rangle_\text{gsf}$ as a function of $p$ for a sample $e=\{0.1, 0.2, 0.3,0.4\}$ of eccentricities. We show, superposed, the corresponding 1PN, 2PN and 3PN predictions from Eq.~\eqref{<U>_gsf_3PN}. The insets display the relative differences between the GSF data and the successive PN approximations. We make the following observations:
\begin{itemize}
	\item[(i)] There is an excellent agreement between the numerical GSF results and the analytical PN prediction at ``large'' $p$, in what should be considered a very strong test of both calculations. This is a first demonstration of such an agreement for noncircular orbits.
	\item[(ii)] The PN series appears to converge uniformly to the GSF result at any $p$ for any fixed $e$ in our survey, at least through 3PN order.
	\item[(iii)] The 3PN formula reproduces the GSF results extremely well even in what might be considered a ``strong-field'' regime: at $p=10$ it does so to within $\sim 1\%$ for $e=0.1$ and to within a few percent for $e=0.4$; at $p=20$ the agreement is already at the level of one part in a thousand.
\end{itemize}

We can make the comparison more quantitative by attempting to extract the large-$p$ behavior of the numerically computed function $\langle U \rangle_{\rm gsf}(p,e)$. Our strategy will be to fit the numerical data against the PN model \eqref{<U>_gsf_3PN}, leaving the numerical coefficients as unknown fitting parameters, later to be compared with the analytically known values. Given the relative sparseness of data available, we shall not attempt a simultaneous fit over $p$ and $e$, but rather fit over each of the two dimensions separately, as described below.  We will follow a ``marginalization'' procedure, whereby each of the PN orders is fitted for in turn, assuming the analytic values of all terms at lower PN order. Since the circular limit of $\langle U \rangle_{\rm gsf}$ has been computed previously at great accuracy \cite{Bl.al2.10,Sh.al.11,Ak.al.12,Sh.al.14}, we are able to accurately ``remove'' the circular ($e$-independent) part of $\langle U \rangle_{\rm gsf}$ from the data, fitting only for the $e$-dependent residue. This should allow to fit the eccentricity-related terms of interest here with greater accuracy.

Let us now describe this procedure in more detail. We assume the $e$-expanded form \eqref{<U>_gsf_3PNbis} of the full PN expression \eqref{<U>_gsf_3PN}. The term $a(p)$ is the circular-orbit limit of $\langle U \rangle_{\rm gsf}$, which has been computed to at least ten significant figures in Refs.~\cite{Bl.al2.10,Sh.al.11,Ak.al.12,Sh.al.14}. By subtracting off these numerical data from ours, we construct a new data set for the difference
\be\label{U^e}
	\langle U \rangle^{(e)}_{\rm gsf}\equiv \langle U \rangle_{\rm gsf}-a(p) = b(p) \, e^2 + c(p) \, e^4 + \cdots \p
\ee
We assume that the functions $b(p), c(p), \ldots$ admit expansions in $p^{-1}$ as in Eqs.~\eqref{a-b-c-d}, but pretend that the PN coefficients are unknown:
\begin{align}\label{bc_model}
	b &= p^{-1} + b_1 \, p^{-2} + b_2 \, p^{-3} + \cdots \c \nonumber \\
	c &= c_1 \, p^{-2} + c_2 \, p^{-3} + \cdots \c
\end{align}
where subscripts are mnemonics for the PN order at which coefficients occur, and we have fixed the ``Newtonian,'' $1/p$ term of $b(p)$ at its known value of unity. Our goal is to determine the coefficient $b_n$, $c_n$, $\ldots$ from the numerical data for $\langle U \rangle ^{(e)}_{\rm gsf}$. To this end, we first prepare subsets of data where in each subset $p$ is fixed and $e$ varies. We fit each subset with respect to $e$ using the model (\ref{U^e}), including terms through $\calO(e^6)$. This yields three one-dimensional data sets, representing $b(p)$, $c(p)$ and $d(p)$. 

Focusing first on the data set for $b(p)$, we fit it against the PN model
\be\label{eq:fits_for_PN_terms}
	b(p) = p^{-1} + \sum_{n=1}^N  p^{-(n+1)} \left( b_n + b_n^{\rm log} \ln{p} \right) ,
\ee
in which $b_1^{\rm log}=b_2^{\rm log}=b_3^{\rm log}=0$, since logarithmic terms are known not to occur before the 4PN order \cite{An.al.82,BlDa.88}.\footnote{The form of the circular-orbit limit, in which the PN expansion of $\langle U \rangle_{\rm gsf}$ is known analytically up to a very high order \cite{Bl.al2.10,Bl.al.14,Bl.al2.14,Sh.al.14,BiDa.15}, suggests that the function $b(p)$ could also involve powers of $\ln{p}$. However, those would contribute at even higher orders than we consider here, so we do not include them in the PN model \eqref{eq:fits_for_PN_terms}.} The truncation order $N$ is left as a control parameter; by varying it we obtain a rough estimate of the numerical uncertainty in the fitted values of the parameters. We apply a marginalization procedure, whereby to determine $b_n$ we set all $b_{n'<n}$ at their known analytic values. We use this procedure to estimate the values of $b_1$, $b_2$ and $b_3$, and we later similarly determine $c_1$. Our results are shown in Table~\ref{table:numerical_values}, alongside the known analytic values for these parameters. We see a good agreement through 3PN order in the $\calO(e^2)$ term, and at 1PN order in the $\calO(e^4)$ term.

\begin{table}[h]
	\begin{tabular}{cclcl}
		\toprule
		Coefficient & & Estimate & & Exact result \\
		\hline                       
		$b_{1}$ & & $+4.0002(8)$ & & $+4$ \\
		$b_{2}$ & & $+7.02(3)$   & & $+7$ \\
		$b_{3}$ & & $-14.5(4)$	 & & $-14.312\dots$ \\
		$c_{1}$ & & $-2.00(1)$	 & & $-2$ \\
		\botrule
	\end{tabular}
	\caption{Best-fit values for the PN coefficients $b_n$ and $c_n$ [Eqs.\ (\ref{U^e}) and (\ref{bc_model})] as extracted from the numerical data, compared to their known exact values. Parenthetical figures are estimated fitting uncertainties in the last displayed decimals, obtained by varying the value of the truncation index $N$ in the fitting model [e.g., Eq.\ (\ref{eq:fits_for_PN_terms}) for $b(p)$]. The exact value of $b_3$ is $- \left( 5/3 + 41\pi^2/32 \right)$.}
	\label{table:numerical_values}
\end{table}

\begin{figure}[h]
	\begin{subfigure}[b]{0.48\textwidth}
		\includegraphics[width=\textwidth]{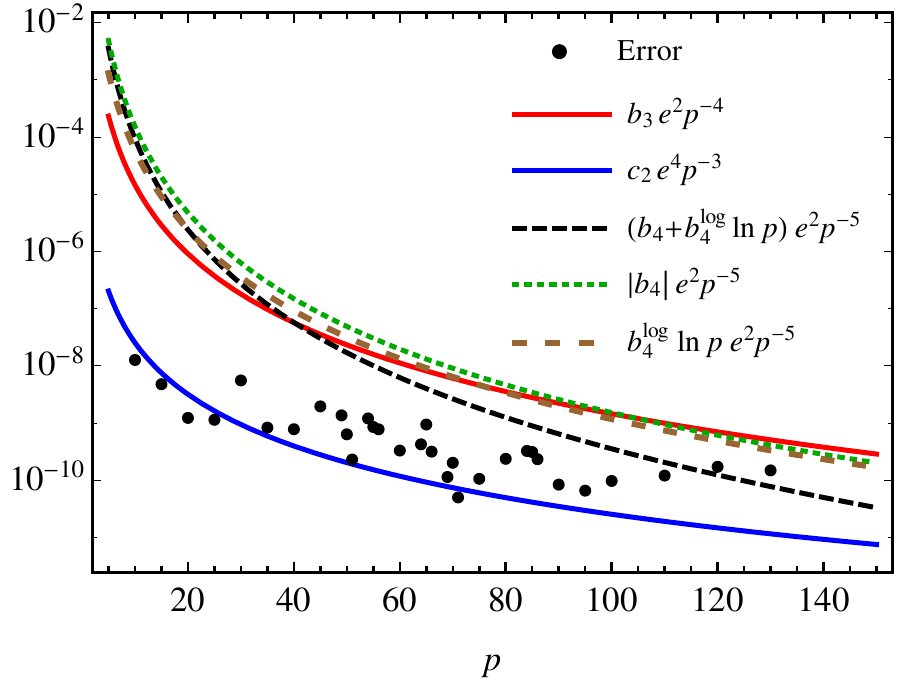}
		\label{fig:e0.1_error_plot}
	\end{subfigure}%
	\quad
	\begin{subfigure}[b]{0.48\textwidth}
		\includegraphics[width=\textwidth]{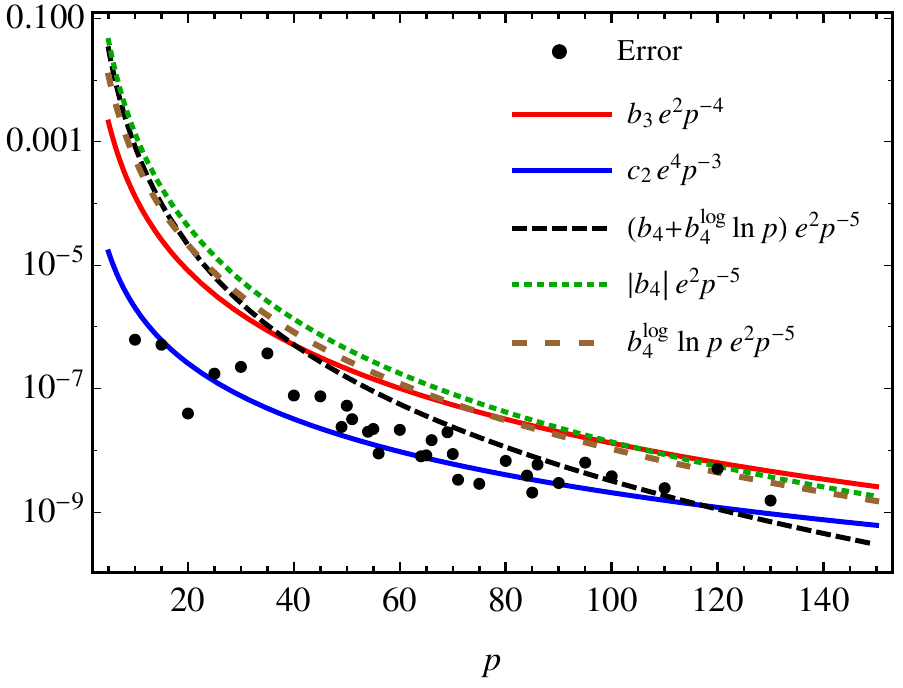}
		\label{fig:e0.3_error_plot}
	\end{subfigure}
	\caption{Absolute magnitude of various PN terms (``signal'') compared to the magnitude of numerical error in the data (``noise''), shown as a function of $p$ for $e=0.1$ (left panel) and $e=0.3$ (right panel). The red (upper solid) curve shows the 3PN term of the $\calO(e^2)$ piece of $\langle U \rangle_{\rm gsf}$, and the blue (lower solid) curve shows the 2PN term of the $\calO(e^4)$ piece. Comparison with the magnitude of numerical noise (black dots) suggests that $b_3$ should be easily discernible while $c_2$ might not. This is confirmed by attempting to fit the data against PN models, as detailed in the text. The dashed curve estimates the amplitude of the 4PN term of the $\calO(e^2)$ piece of $\langle U \rangle_{\rm gsf}$, which is not known analytically. We used here the values $b_4=-1500$ and $b_4^\text{log} = 250$ chosen from the middle of the estimated range shown in Eqs.~\eqref{b4}. This 4PN signal appears to lie just over the noise and is detectable. However, as it can be seen from the near overlap of the densely dashed (green) and sparsely dashed (brown) curves, the $|b_4|$ and $b_4^\text{log}$ terms become almost equal in magnitude as $p$ increases, hence making it very difficult to extract the individual values of $b_4$ and $b_4^\text{log}$.}
	\label{fig:b3PN_c2PN_b4PN_errors}
\end{figure}

Unfortunately, the accuracy of our current code (and its limited utility at $e\gtrsim 0.4$) does not seem to allow us an accurate extraction of $b_{n\geq 4}$, $c_{n\geq 2}$, or any of the $b_n^\text{log}$'s. The reason for this can be appreciated from Fig.~\ref{fig:b3PN_c2PN_b4PN_errors}, where we compare the amplitudes of the 3PN and 4PN terms with the amplitude of numerical noise in our $\langle U \rangle^{(e)}_{\rm gsf}$ data. Note that, while the ``signal'' from the $b_3$ term lies well above the noise, the $c_2$ signal is buried deep inside it. Since our data is limited to relatively small eccentricities, it is clear why we have less ``handle'' on the $c_n$ [$\calO(e^4)$] terms than on the $b_n$ [$\calO(e^2)$] terms.

Figure \ref{fig:b3PN_c2PN_b4PN_errors} also suggests that we might have just enough ``signal'' coming from the $\calO(e^2)$ terms at 4PN to allow a rough estimation of the coefficients $b_4$ and $b_4^{\rm log}$, which are not known analytically. We have experimented fitting to a large number of models of the form (\ref{eq:fits_for_PN_terms}), where all the analytically known coefficients are pre-specified, and varying both the cutoff $N$ and the number of nonzero logarithmic terms. We find that fitting uncertainties are almost as large as the fitted values themselves. However, we are able to confidently constrain the values of $b_4$ and $b_4^{\rm log}$ to lie within the ranges
\begin{subequations}\label{b4}
	\begin{align}
		-2000 &\lesssim b_4 \lesssim -1000 \c \\
		+150 &\lesssim b_4^{\rm log} \lesssim +350 \p
	\end{align}
\end{subequations}
Future analytic calculations of the 4PN terms may be checked against these predictions. 

Our current code does not allow the determination of unknown PN coefficients related to eccentricity with any greater accuracy. To improve on our predictions would require (i) to push the reach of the computation to higher eccentricities and larger $p$, and at the same time (ii) to reduce the numerical error in the calculation of $\langle U \rangle_{\rm gsf}$. Some improvement may be achieved using the method of Ref.~\cite{Os.al.14}, which is a slightly more advanced variant of our method.  More significant improvements may have to await the development of eccentric-orbit GSF codes based on the Teukolsky equation \cite{Po.al.14,MeSh.15}. We expect such codes to start delivering accurate numerical results in the very near future.

\section*{Acknowledgements}

We are grateful to Maarten van de Meent for providing us with unpublished comparison data generated by a new GSF code now being developed by him. SA also thanks Maarten van de Meent and Haris Markakis for many useful discussions. The research leading to these results received funding from the European Research Council under the European Union's Seventh Framework Programme (FP7/2007-2013)/ERC grant agreement no.~304978. SA and LB acknowledge additional support from STFC through grant number PP/E001025/1. SA's work is further supported by the Irish Research Council, funded under the National Development Plan for Ireland. ALT acknowledges support from a Marie Curie FP7 Integration Grant under the 7th European Union Framework Programme (PCIG13-GA-2013-630210). NS acknowledges the support of the Grand-in-Aid for Scientific Research (No. 25800154). NW gratefully acknowledges support from a Marie Curie International Outgoing Fellowship (PIOF-GA-2012-627781) and the Irish Research Council, which is funded under the National Development Plan for Ireland.

\appendix

\section{Equivalence between our expression for $\langle U \rangle_{\rm gsf}$ and BS2011's}\label{app:proof}

BS2011 give their formula for $\langle U \rangle_{\rm gsf}$ in their Eq.~(84). Adjusting notation and rearranging the terms, their expression reads
\be\label{eq:Ugsf_breakdown}
q \, \langle U \rangle_{\rm gsf} = \f{1}{2} {\langle U \rangle}_0 \langle \hat{h}_{uu}^{R} \rangle + \left(\alpha +\frac{\Delta T_r}{T_{r0}}\right)\left({\langle U \rangle}_0+C_r \Omr+C_\phi\Omph\right) - C_{\phi}\Omph\frac{\Delta\Phi}{\Phi_0} - {\langle U \rangle}_0 \frac{\Delta\Tau_{r}}{\Tau_{r0}} \, ,
\ee
where
\be
C_r \equiv \f{\partial {\langle U \rangle}_0}{\partial \Omr} \c \qquad C_\phi \equiv \f{\partial {\langle U \rangle}_0}{\partial \Omph} \c
\ee
and $\Delta X$ denotes the GSF correction to a quantity $X$, holding fixed $p$ and $e$, rather than the invariant frequencies $\Omega_i$. BS2011 give explicit expressions for $\Delta T_r$, $\Delta\Phi$ and $\Delta\Tau_r$ in terms of GSF quantities, but these will not be needed here. We observe that the expression \eqref{eq:Ugsf_breakdown} is much more complicated than our result, Eq.\ (\ref{Ugsf}). Our goal here is to show that the two expressions are, in fact, identical.

To this end, we note the two key relations
\begin{subequations}
	\begin{align}
		C_\phi &= \ang \, {\langle U \rangle}_0^2 \c \label{eq:C_phi_identity} \\
		C_r &= \f{{\langle U \rangle}_0}{\Omega_{r}} \left[ \left(\en - \ang \Omph\right) {\langle U \rangle}_0 -1 \right] , \label{eq:C_r_identity}	\end{align}
\end{subequations}
which shall be derived below. Substituting these $C_r$ and $C_\phi$ into Eq.~\eqref{eq:Ugsf_breakdown} and using $\Tau_{r0} = T_{r0} / {\langle U \rangle}_0$, we obtain
\be\label{eq:Ugsf_breakdown2}
q \, \langle U \rangle_{\rm gsf} = \f{1}{2} {\langle U \rangle}_0 \langle \hat{h}_{uu}^{R} \rangle + \alpha {\cal E} {\langle U \rangle}_0^2 + \frac{{\langle U \rangle}_0^2}{T_{r0}} \left({\cal E} \, \Delta T_r - {\cal L} \, \Delta\Phi - \Delta\Tau_r \right) .
\ee
For this to be identical to Eq.~\eqref{Ugsf}, the sum of three term in brackets on the right-hand side should vanish. Indeed, writing $(\ud\tau/\ud\chi)^2=-g_{\alpha\beta}(\ud x^{\alpha}/\ud\chi)(\ud x^{\beta}/\ud\chi)$ and perturbing linearly with $\Delta$, holding $p$, $e$ and $r_{\rm p}$ (or, equivalently, $p$, $e$ and $\chi$) fixed, we find 
\be\label{eq:DeltaTau_integrand_equality}
\Delta(\ud\tau/\ud\chi) = \en \, \Delta(\ud t/\ud\chi) - \ang \, \Delta(\ud\phi/\ud\chi) \c 
\ee
which, upon integrating over a radial period, gives
\be 
\Delta \Tau_r = \en \, \Delta T_r - \ang \, \Delta \Phi \p
\ee
Hence Eq.~\eqref{eq:Ugsf_breakdown2} reduces to our Eq.\ (\ref{Ugsf}) for $\langle U \rangle_{\rm gsf}$.

It remains to establish the relations \eqref{eq:C_phi_identity} and \eqref{eq:C_r_identity}. This can be achieved by manipulating the explicit elliptic-integral representations of $\Omph,\Omr$ and ${\langle U \rangle}_0$, given in BS2011, but this approach involves much ungainly algebra and will not be presented here. A much neater derivation uses general results derived from the Hamiltonian formulation of geodesic motion in Kerr spacetime \cite{Le.14}. Start by averaging $u_0^{\alpha}u_{0\alpha}=-1$ with respect to $t$ over a radial period of the geodesic orbit, to obtain 
\be\label{Av_norm}
{\langle U \rangle}_0^{-1} = {\cal E} - \Omph {\cal L} - \Omr J_r \c
\ee
where $J_r \equiv {(2\pi)}^{-1} \oint u_{0r} \, \ud r = {(2\pi {\cal E}_0)}^{-1} \int_0^{T_r} {(u_0^r)}^2 \, \ud t$ is the invariant action variable (per mass $m_1$) associated with the radial motion \cite{Sc.02}. This relation is the Schwarzschild reduction of Eq.~(3.4) of Ref.~\cite{Le.14}. In addition, we require a relation between the partial derivatives of ${\cal E}$, ${\cal L}$ and $J_r$ with respect to $\Omega_i$. The necessary relation follows most directly from the general variational formula (``first law'')
\be 
\delta {\cal E} = \Omph \, \delta {\cal L} + \Omr \, \delta J_r 
\ee
established in \cite{Le.14} [this form is the reduction of Eq.~(3.5) therein to Schwarzschild spacetime, with a fixed black-hole mass $m_2$, and with suitable notational adjustments]. Here $\delta {\cal E}$, $\delta {\cal L}$ and $\delta J_r$ correspond to an arbitrary variation of a geodesic with frequencies $\Omega_i$ onto a nearby geodesic. If we regard ${\cal E}$, ${\cal L}$ and $J_r$ as functions of $\Omega_i$, we obtain
\be \label{partials}
\frac{\partial {\cal E}}{\partial\Omega_i} - \Omph \frac{\partial {\cal L}}{\partial\Omega_i} - \Omr \frac{\partial J_r}{\partial\Omega_i} = 0 \p
\ee
Taking the partial derivative of (\ref{Av_norm}) with respect to $\Omph$ and using (\ref{partials}) immediately leads to \eqref{eq:C_phi_identity}. Equation (\ref{eq:C_r_identity}), in turn, is obtained by taking the derivative of (\ref{Av_norm}) with respect to $\Omr$, then using Eq.\ (\ref{partials}), and finally substituting for $J_r$ from \eqref{Av_norm}. 

The above establishes the equivalence of our simple expression (\ref{Ugsf}) and the BS2011 result \eqref{eq:Ugsf_breakdown}. The simplification obtained here owes itself primarily to the two key relations (\ref{eq:C_phi_identity}) and (\ref{eq:C_r_identity}), which have unfortunately gone unnoticed (by two of us) in BS2011.

\section{Post-Newtonian expansion of ${\langle U \rangle}_0$}\label{app:test-mass}

Here we consider a test mass on a bound geodesic orbit around a nonspinning black hole of mass $m_2$ and obtain the PN expansion of the relationship ${\langle U \rangle}_0(\Omega_r,\Omega_\phi)$. This calculation provides a powerful check of our PN result \eqref{U_x_iota}--\eqref{U_x_iotabis}, because it is based on a different formalism, it makes use of an alternative parametrization of the motion, and it is performed using a different coordinate system.

Since the relationships \eqref{eq:dt0-dphi0}--\eqref{eq:geod_frequencies} cannot be inverted analytically to yield the expressions for the parameters $p$ and $e$ as functions of the frequencies $\Omr = 2 \pi / T_{r 0}$ and $\Omph = \Phi_0 / T_{r0}$, we shall work perturbatively, expanding all quantities in powers of the small parameter $1/p$. From $T_{r0}$ and $\Phi_0$ we define the invariant parameters $y \!\equiv (m_2 \Omph)^{2/3}$ and $\lambda \!\equiv 3 y \, {(\Phi_0/2\pi-1)}^{-1}$ [recall Eq.~\eqref{y-lambda}]. Expanding the formulas \eqref{eq:dt0-dphi0}--\eqref{eq:geod_frequencies} up to 3PN order, we obtain
\begin{subequations}\label{y_lambda}
	\begin{align}
		y &= \frac{j_e}{p} \left\{ 1 + \frac{2 \left( 1 - j_e \right)}{p} + \left( \frac{17}{2} \left[ 1 - j_e \right] + 5 j_e \left[ j_e - \sqrt{j_e} \right] \right) \frac{1}{p^2} \right. \nonumber \\ &\left. + \left( \frac{133}{3} - 48 j_e - 35 j_e^{3/2} + 27 j_e^2 + 25 j_e^{5/2} - \frac{40}{3} j_e^3 \right) \frac{1}{p^3} + o(p^{-3}) \right\} , \\
		\lambda &= j_e \left\{ 1 - \left( \frac{11}{4} + \frac{7}{4} j_e \right) \frac{1}{p} + \left( - \frac{75}{16} + \frac{23}{8} j_e - 5 j_e^{3/2} + \frac{73}{16} j_e^2 \right) \frac{1}{p^2} \right. \nonumber \\ &\left. + \left( - \frac{1849}{96} + \frac{849}{64} j_e - \frac{45}{4} j_e^{3/2} - \frac{17}{16} j_e^2 + \frac{95}{4} j_e^{5/2} - \frac{2341}{192} j_e^3 \right) \frac{1}{p^3} + o(p^{-3}) \right\} ,
	\end{align}
\end{subequations}
where we introduced the notation $j_e \equiv 1 - e^2$. In the limit of vanishing eccentricity, $e \rightarrow 0$, we have the simple relation $y = p^{-1} + o(p^{-4})$. Actually, we know that for circular orbits the relation $y = 1 / p$ holds exactly, such that in Schwarzschild coordinates the semi-major axis coincides with an invariant measure of the orbital radius. [This, however, is no longer true at $\calO(q)$ in the GSF approximation.] For circular orbits, the 3PN-accurate relationship between the invariants $y$ and $\lambda$ then reads
\beq\label{lambda_y}
	\lambda = 1 - \frac{9}{2} y - \frac{9}{4} y^2 - \frac{27}{4} y^3 + o(y^3) \, .
\eeq
Inverting the relations \eqref{y_lambda} yields expressions for the semi-latus rectum $p$ and eccentricity $e$ (or equivalently $j_e = 1 - e^2$) as functions of the invariant parameters $y$ and $\lambda$. Up to 3PN order, we find
\begin{subequations}\label{p_e}
	\begin{align}
		\frac{1}{p} &= \frac{y}{\lambda} \left\{ 1 + \left( \frac{1}{4} - \frac{19}{4\lambda} \right) y + \left( \frac{9}{16} - \frac{5}{16\lambda} + \frac{151}{8\lambda^2} \right) y^2 \right. \nonumber \\ &\left. + \left( \frac{65}{64} + \frac{5}{4\sqrt{\lambda}} - \frac{25}{64\lambda} + \frac{1}{4\lambda^2} - \frac{2255}{32\lambda^3} \right) y^3 + o(y^3) \right\} , \\
		j_e &= \lambda \left\{ 1 + \left( \frac{7}{4} + \frac{11}{4\lambda} \right) y + \left( 2 + \frac{5}{\sqrt{\lambda}} + \frac{63}{16\lambda} - \frac{13}{16\lambda^2} \right) y^2 \right. \nonumber \\ &\left. + \left( \frac{5}{6} + \frac{145}{8\sqrt{\lambda}} + \frac{221}{32\lambda} + \frac{95}{8\lambda^{3/2}} - \frac{289}{64\lambda^2} + \frac{263}{192\lambda^3} \right) y^3 + o(y^3) \right\} .
	\end{align}
\end{subequations}

We now have all the pieces required to compute the relation ${\langle U \rangle}_0(y,\lambda)$ up to the required PN order. The generalized redshift is defined as
\begin{align}
	{\langle U \rangle}_0 \equiv \frac{1}{\Tau_{r0}} \int_0^{\Tau_{r0}} u_0^t(\tau_0) \, \ud \tau_0  = {\biggl( \frac{1}{T_{r0}} \int_0^{2\pi} \frac{\ud t_0}{\ud \chi} \, \frac{\ud \chi}{u_0^t(\chi)} \biggr)}^{-1} \, ,
\end{align}
where $\Tau_{r0}$ is the proper time period of the radial motion. From the expressions \eqref{eq:E0_L0},  \eqref{eq:r_of_chi}, \eqref{eq:dt0_dchi} and \eqref{eq:radial_period}, we find
\begin{align}\label{U_e_p}
	{\langle U \rangle}_0 &= 1 + \frac{j_e}{p} \left\{ \frac{3}{2} + \left( 6 \sqrt{j_e} - \frac{21}{8} j_e \right) \frac{1}{p} + \left( 23 \sqrt{j_e} - 6 j_e - 12 j_e^{3/2} + \frac{55}{16} j_e^2 \right) \frac{1}{p^2} \right. \nonumber \\ & \left. + \left( \frac{249}{2} \sqrt{j_e} - 24 j_e - 105 j_e^{3/2} + 12 j_e^2 + \frac{75}{4} j_e^{5/2} - \frac{525}{128} j_e^3 \right) \frac{1}{p^3} + o(p^{-3}) \right\} .
\end{align}
Finally, substituting for $(p,e)$ in terms of $(y,\lambda)$ in Eq.~\eqref{U_e_p}, using \eqref{p_e}, we obtain the 3PN-accurate coordinate-invariant relation
\begin{align}\label{U_y_lambda}
	{\langle U \rangle}_0 &= 1 + \frac{3}{2} \, y + \biggl( \frac{3}{8} + \frac{6}{\sqrt{\lambda}} - \frac{3}{\lambda} \biggr) \, y^2 + \biggl( - \frac{41}{16} + \frac{57}{4\sqrt{\lambda}} - \frac{3}{2\lambda} - \frac{37}{4\lambda^{3/2}} + \frac{15}{2\lambda^2} \biggr) \, y^3 \nonumber \\ &+ \biggl( - \frac{605}{128} + \frac{117}{64\sqrt{\lambda}} + \frac{411}{8\lambda} - \frac{1755}{32\lambda^{3/2}} + \frac{21}{4\lambda^2} + \frac{1797}{64\lambda^{5/2}} - \frac{20}{\lambda^3} \biggr) \, y^4 + o(y^4) \, .
\end{align}
In the circular-orbit limit, we may introduce the PN expansion \eqref{lambda_y} for $\lambda(y)$ in \eqref{U_y_lambda}, expand in powers of $y$ up to the appropriate PN order, and recover the 3PN expansion of the fully relativistic result $U_\odot = {(1-3y)}^{-1/2}$. Although the result \eqref{U_y_lambda} can in principle be extended up to an arbitrarily high PN order, we only need here the 3PN approximation to the exact result. Comparing with the formula \eqref{U_Schw} derived from our 3PN calculation valid for any mass ratio, we find perfect agreement.

\bibliography{}

\end{document}